\documentclass[pdftex,twoside,11pt]{elsarticle}

\usepackage{epsfig,psfrag,amssymb,amsmath,fancyhdr,psfrag,graphics,subfigure}
\usepackage{dcolumn,bm,float,epstopdf,hyperref,setspace}
\hypersetup{colorlinks=true,linktoc=all,bookmarks=true,pdftoolbar=true}
\usepackage[english]{babel}
\usepackage{color}

\def\fermi{{\it Fermi }}

\journal{Physics Reports}

\begin{document}

\begin{frontmatter}

\title{The nature of the Diffuse Gamma-Ray Background}

\author{Mattia Fornasa}
\ead{fornasam@gmail.com}
\address{School of Physics and Astronomy, University of Nottingham, University Park NG7 2RD, UK}

\author{Miguel A. S\'{a}nchez-Conde}
\ead{sanchezconde@fysik.su.se}
\address{Oskar Klein Centre for Cosmoparticle Physics, Department of Physics, Stockholm University, SE-10691 Stockholm, Sweden}

\begin{abstract}
We review the current understanding of the diffuse gamma-ray background 
(DGRB). The DGRB is what remains of the total measured gamma-ray emission 
after the subtraction of the resolved sources and of the diffuse Galactic 
foregrounds. It is interpreted as the cumulative emission of sources that are 
not bright enough to be detected individually. Yet, its exact composition 
remains unveiled. Well-established astrophysical source populations (e.g. 
blazars, misaligned AGNs, star-forming galaxies and millisecond pulsars) all 
represent guaranteed contributors to the DGRB. More exotic scenarios, such as 
dark matter annihilation or decay, may contribute as well. In this review, we 
describe how these components have been modeled in the literature and how the 
DGRB can be used to provide valuable information on each of them. We summarize 
the observational information currently available on the DGRB, paying 
particular attention to the most recent measurement of its intensity energy 
spectrum by the \fermi LAT Collaboration. We also discuss the novel analyses 
of the auto-correlation angular power spectrum of the DGRB and of its 
cross-correlation with tracers of the large-scale structure of the Universe. 
New data sets already (or soon) available are expected to provide further 
insight on the nature of this emission.  By summarizing where we stand on the 
current knowledge of the DGRB, this review is intended both as a useful 
reference for those interested in the topic and as a means to trigger new
ideas for further research.
\end{abstract}


\end{frontmatter}

\newpage
\tableofcontents

\section*{Table of acronyms}
\label{sec:acronyms}
\addcontentsline{toc}{section}{Table of acronyms}
\hspace{-2em}
\begin{tabular}{p{2cm}l}
1FGL & \fermi LAT First Source Catalog \\
2FGL & \fermi LAT Second Source Catalog \\ 
2FPC & \fermi LAT Second Catalog of gamma-ray pulsars \\
2LAC & \fermi LAT Second Catalog of Active Galactic Nuclei \\
2MASS & 2 Micron All-Sky Survey \\
3FGL & \fermi LAT Third Source Catalog \\ 
AGN & Active Galactic Nucleus \\
APS & Angular Power Spectrum \\
CFHTLenS & Canada-France-Hawaii Telescope Lensing Survey \\ 
CMB & Cosmic Microwave Background \\
CR & Cosmic Ray \\
CTA & Cherenkov Telescope Array \\
DES & Dark Energy Survey \\
DESI & Dark Energy Spectroscopic Instrument \\
DGRB & Diffuse Gamma-Ray Background \\
DM & Dark Matter \\
\end{tabular}
\newpage
\hspace{-2em}
\begin{tabular}{p{2cm}l}
eBOSS & extended Baryon Oscillation Spectroscopy Survey \\
EBL & Extragalactic Background Light \\
FRI & Fanaroff-Riley Class I \\
FRII & Fanaroff-Riley Class II \\
FSRQ & Flat-Spectrum Radio Quasar \\
HMF & Halo Mass Function \\
HOD & Halo Occupation Distribution \\
HSP & High-Synchrotron Peak \\
IC & Inverse Compton \\
IR & InfraRed \\
ISP & Intermediate-Synchrotron Peak \\
JWST & James Webb Space Telescope \\
LF & Luminosity Function \\
LOFAR & Low-Frequency Array \\
LSP & Low-Synchrotron Peak \\
LSS & Large-Scale Structure \\
MAGN & Misaligned Active Galactic Nucleus \\
MSP & MilliSecond Pulsar \\
MW & Milky Way \\
NFW & Navarro-Frenk-White \\
NuSTAR & Nuclear Spectroscopic Telescope Array \\
NVSS & NRAO VLA Sky Survey \\
PDF & Probability Distribution Function \\
PSF & Point Spread Function \\
SDSS & Sloan Digital Sky Survey \\
SED & Spectral Energy Distribution \\ 
SFG & Star-Forming Galaxy \\
SFR & Star-Forming Rate \\
SKA & Square Kilometer Array \\ 
UHECR & Ultra-High-Energy Cosmic Ray \\
UV & UltraViolet \\
WIMP & Weakly Interacting Massive Particle \\
WISE & Wide-field Infrared Survey Explorer \\
WMAP & Wilkinson Microwave Anisotropy Probe \\
\end{tabular}

\section{Introduction}
\label{sec:introduction}
The first full-sky image of gamma-ray emission was obtained in 1972 by the 
OSO-3 satellite. It consisted of 621 events detected above 50 MeV 
\cite{Kraushaar:1972}. Since then, telescopes with lower sensitivities and 
better angular and energy resolutions have significantly improved our 
understanding of the gamma-ray Universe. The all-sky maps produced by the 
Fermi Large Area Telescope\footnote{http://fermi.gsfc.nasa.gov/} (\fermi LAT 
from now on) after more than 6 years of data taking contain more than 5 
million events above 1 GeV. These maps exhibit a rich morphology: along the 
Galactic plane, the diffuse Galactic foreground is the most evident feature 
and, overall, it accounts for $\sim 80\%$ of the detected gamma rays. This 
diffuse radiation is produced by the interaction of cosmic rays (CRs) with the 
Galactic interstellar radiation field and with the nuclei of the Galactic 
interstellar medium. Also, the Third \fermi LAT catalog (3FGL) reported the 
detection of 3033 sources throughout the sky \cite{TheFermi-LAT:2015hja}. 
Extended gamma-ray emitters are discussed in Ref.~\cite{Lande:2012xn}. Other 
structures, e.g. the Fermi bubbles \cite{Fermi-LAT:2014sfa}, represent more 
complex phenomena, whose emission is not fully understood yet.

In order to reproduce the data from the \fermi LAT it is necessary to include
one additional contribution, i.e. a diffuse and {\it nearly} isotropic 
emission called the Diffuse Gamma-Ray Background (DGRB) \cite{Kraushaar:1972,
Fichtel:1975,Sreekumar:1997un,Strong:2004ry,Abdo:2010nz,Ackermann:2014usa}.
The DGRB is thought to be predominantly of extragalactic origin: gamma-ray 
sources with a flux smaller than the sensitivity of \fermi LAT are not 
detected individually, producing instead a cumulative diffuse glow that 
contributes to the DGRB. Unresolved blazars \cite{Stecker:1993ni,Stecker:1996ma,
Muecke:1998cs,Narumoto:2006qg,Dermer:2007fg,Pavlidou:2007dv,Inoue:2008pk,
Ajello:2009ip,Collaboration:2010gqa,Abazajian:2010pc,Stecker:2010di,
Singal:2011yi,Ajello:2011zi,Ajello:2013lka,DiMauro:2013zfa,Ajello:2015mfa}, 
misaligned Active Galactic Nuclei (MAGNs) \cite{Stawarz:2005tq,Massaro:2011ww,
Inoue:2011bm,DiMauro:2013xta}, star-forming galaxies (SFGs) 
\cite{Bhattacharya:2009yv,Fields:2010bw,Makiya:2010zt,Ackermann:2012vca,
Lacki:2012si,Chakraborty:2012sh} and millisecond pulsars (MSPs) 
\cite{FaucherGiguere:2009df,SiegalGaskins:2010mp,Calore:2014oga} are 
guaranteed components to the DGRB. More uncertain source classes, e.g. 
galaxy clusters \cite{Zandanel:2014pva} or Type Ia supernovae 
\cite{Horiuchi:2010kq,Lien:2012gz}, may also play a role, together 
blue with diffuse phenomena, e.g. the radiation produced by the interaction of 
Ultra-High-Energy Cosmic Rays (UHECRs) with the Extragalactic Background Light 
(EBL) \cite{Kalashev:2007sn,Ahlers:2011sd}.\footnote{Different names have been 
used in the literature to denote the DGRB, e.g. Extragalactic Gamma-Ray 
Background or Isotropic Gamma-Ray Background. We believe the denomination used 
in this review is more precise since, as we will see in the following sections, 
the DGRB may be not entirely extragalactic and since it exhibits a certain 
amount of anisotropy. We also note that, in Ref.~\cite{Ackermann:2014usa}, the 
\fermi LAT Collaboration uses the name Extragalactic Gamma-Ray Background to 
characterize the cumulative emission of {\it all} sources (both resolved and 
unresolved), while Isotropic Gamma-Ray Background refers to the unresolved 
component only, i.e. what we call DGRB (see Sec.~\ref{sec:energy_spectrum}).} 
It is possible to estimate how much different source classes contribute to the 
DGRB, but its exact composition remains one of the main unanswered questions 
of gamma-ray astrophysics. Finding a definitive answer would constrain the 
faint end of the luminosity function of the DGRB contributors. Indeed, the 
study of the DGRB may represent the only source of information about those 
objects that are too faint to be detected individually. 

The DGRB may also shed some light on exotic Physics as, e.g., on the nature of
Dark Matter (DM): a huge experimental effort is currently devoted to the 
so-called indirect detection of DM, i.e. the search for particles (e.g. 
gamma rays, neutrinos, positrons or anti-protons) produced by the annihilations 
or decays of DM. Such a signal would constitute the first evidence that DM 
can interact non-gravitationally and it would represent an enormous step 
forward in our understanding of its nature \cite{Bertone:2004pz,Cirelli:2010xx,
Ibarra:2012yra,Bringmann:2012ez}. Targets like the center of the Milky Way 
(MW), local satellite galaxies or nearby galaxy clusters are considered 
optimal, thanks to the intensity of the expected DM signal and/or to the 
absence of significant competing backgrounds. Yet, no signal has been robustly 
associated with DM. Then, if DM annihilates or decays producing gamma rays and
such a signal has not been detected up to now, it is most probably unresolved 
and it contributes to the DGRB. Looking for the features of a DM component in 
the DGRB, one can, then, hope to finally unravel the long-standing mystery of 
the nature of DM \cite{Ullio:2002pj,Taylor:2002zd,Ando:2005hr,Ando:2005xg,
Ando:2006cr,SiegalGaskins:2008ge,Ando:2009fp,Fornasa:2009qh,Zavala:2009zr,
Ibarra:2009nw,Abdo:2010dk,Hutsi:2010ai,Cirelli:2010xx,Zavala:2011tt,
Calore:2011bt,Fornasa:2012gu,DiMauro:2014wha,Ackermann:2015gga,
DiMauro:2015tfa}. 

The most recent measurement of the intensity of the DGRB has been performed by 
the \fermi LAT, in the range between 100 MeV and 820 GeV 
\cite{Ackermann:2014usa} (see also Refs.~\cite{Kraushaar:1972,Fichtel:1975,
Sreekumar:1997un,Strong:2004ry,Abdo:2010nz} for the previous measurements). 
Valuable information on the nature of the DGRB can be extracted from its 
intensity energy spectrum. Its steepness could indicate whether a particular 
class of sources dominates the emission. Moreover, the transition between two 
energy regimes dominated by different classes could, in principle, give rise 
to breaks and features in the energy spectrum. Yet, since the intensity of
the DGRB is only sensitive to the sum of its contributions, ultimately there 
is only a limited amount of information that can be extracted from it.

Fortunately, due to its excellent sensitivity and exemplary angular resolution, 
the \fermi LAT marked the beginning of an era in which the intensity energy 
spectrum is no longer the only observational data available for the study of 
the DGRB. In 2012, the first measurement of anisotropies in the DGRB was 
reported \cite{Ackermann:2012uf}, and the detection of a non-null 
auto-correlation angular power spectrum (APS) provided complementary 
constraints on the composition of the DGRB \cite{Cuoco:2012yf,Harding:2012gk,
DiMauro:2014wha}. Moreover, in Ref.~\cite{Malyshev:2011zi}, the authors used 
the photon-count probability distribution measured after 11 months of \fermi 
LAT data to constrain the contribution of unresolved blazars to the DGRB. More
recently, the cross-correlation of the DGRB anisotropies with observables 
tracing the Large Scale Structure (LSS) of the Universe has also been 
considered. In Ref.~\cite{Xia:2015wka} the authors measured the 2-point 
correlation of the DGRB with 5 different galaxy catalogs and they reported a 
signal with 4 of them. Ref.~\cite{Shirasaki:2014noa}, instead, cross-correlated 
the DGRB with the cosmic shear observed by the Canada France Hawaii Telescope 
Lensing Survey (CFHTLenS) and found no significant detection. These works, 
together with Refs.~\cite{Camera:2012cj,Ando:2013xwa,Ando:2014aoa,
Camera:2014rja}, have proved that the study of the cross-correlation of the 
DGRB with the LSS is a very powerful strategy that may provide access to 
components of the DGRB that are only subdominant in the intensity energy 
spectrum or in the auto-correlation APS. In particular, the technique has the 
potential to deliver the first detection of DM-induced gamma-ray emission.

In the near future, the data on the DGRB is expected to further increase, due 
to the extended run of the \fermi LAT and to the fore-coming Cherenkov 
Telescope Array (CTA) \cite{Doro:2012xx,Acharya:2013sxa}. Other frequencies and 
other messengers will also be crucial to improve our modeling of the DGRB and
to extract complementary information about its nature. The scenario is rapidly 
evolving and the study of the DGRB is quickly becoming a standard tool for the 
characterization of unresolved astrophysical sources and of a potential 
DM-induced gamma-ray signal.

Therefore, we believe that this is the right moment to summarize where we stand
in our understanding of the DGRB. In this article we will survey the data 
available on the DGRB at present and we will discuss how these observations 
have been used to constrain the nature of the emission. We will also enumerate 
the classes of sources or emission mechanisms that have been proposed as 
contributors to the DGRB. By sketching a snapshot of the state-of-art on the 
DGRB {\it circa} 2015, we intend to provide the community with a reference 
point from which to build on.

We end by noting that the DGRB is intrinsically an analysis- and 
time-dependent quantity. Indeed, its intensity depends on the sensitivity of 
the telescope employed to detect it and on its instrumental capability to 
resolve sources. Even with the same detector, an increase in statistics or, in 
general, any improvement in the detection sensitivity will result in a 
different DGRB. In the following sections, every time we mention the DGRB we 
will make sure to specify which measurement of the DGRB we refer to.

The paper is organized as follows: in Sec.~\ref{sec:energy_spectrum} we focus 
on the intensity of the DGRB. Sec.~\ref{sec:intensity_measurement} reviews the 
recent measurement of the DGRB energy spectrum by the \fermi LAT, while 
Secs.~\ref{sec:components} and \ref{sec:DM} are devoted to the description of 
the sources that have been proposed as contributors to the emission: 
astrophysical objects are studied in Sec. \ref{sec:components}, while 
Sec.~\ref{sec:DM} discusses the case of the DM-induced emission. 
Sec.~\ref{sec:angular_spectrum} reviews the \fermi LAT measurement of the 
auto-correlation APS of anisotropies and its impact on our understanding of 
the DGRB. The topic of Sec.~\ref{sec:1-point} is the measurement and the 
interpretation of the photon count probability distribution, while in 
Sec.~\ref{sec:cross_correlation} we investigate the cross-correlation with 
probes of LSS, namely galaxy catalogs in Sec.~\ref{sec:correlation_galaxies}, 
cosmic shear in Sec.~\ref{sec:correlation_shear} and other observables in 
Sec.~\ref{sec:correlation_others}. Finally, we present our conclusions in 
Sec.~\ref{sec:conclusions}.

\section{The intensity energy spectrum}
\label{sec:energy_spectrum}
The subject of this section is the intensity energy spectrum of the DGRB. The 
first subsection (Sec.~\ref{sec:intensity_measurement}) summarizes the most 
recent measurement of this observable, i.e. the one performed by the \fermi 
LAT in Ref.~\cite{Ackermann:2014usa}. The following subsections 
(Sec.~\ref{sec:components} and Sec.~\ref{sec:DM}) present the different source 
classes and emission mechanisms that have been proposed to interpret the
observed emission. For each population we summarize how the DGRB energy 
spectrum has been used to learn about the properties of the gamma-ray emitters.

\subsection{The new \fermi LAT measurement of the Diffuse Gamma-Ray Background}
\label{sec:intensity_measurement}
The \fermi LAT \cite{Atwood:2009ez} on board the NASA \fermi satellite started 
its scientific operation on August 2008 and, since then, it has revolutionized 
our knowledge of the most violent phenomena in the Universe. The \fermi LAT 
covers 4 decades in energy, from few dozens of MeV up to the TeV regime. It 
has an unprecedented sensitivity ($\sim 30$ times better than its predecessor 
EGRET) and an extremely large field of view reaching almost one fifth of the 
sky. Its angular resolution is of about $0.8^\circ$ at 1 GeV and better than 
$0.2^\circ$ above 10 GeV.\footnote{These values refer to the 68\% containment 
radius.} Such superb capabilities allowed the discovery of hundreds of new 
gamma-ray sources and an impressive cartography of the Galactic CR-induced 
gamma-ray diffuse emission that reaches, for the first time, energies greater 
than 10 GeV \cite{FermiLAT:2012aa}.

In addition to these achievements, the analysis of 10 months of data delivered 
the first \fermi LAT measurement of the DGRB energy spectrum at Galactic 
latitudes, $b$, greater than 10 degrees \cite{Abdo:2010nz}. Such a measurement 
was performed between 200 MeV and 102 GeV and it represents the third 
independent observation of the DGRB, after the one by the SAS-2 satellite in 
1978 between 40 and 300 MeV \cite{Fichtel:1978} and that by EGRET between 40 
MeV and 10 GeV in 1998 \cite{Sreekumar:1997un}. The energy spectrum of the DGRB 
as measured by the \fermi LAT in Ref.~\cite{Abdo:2010nz} is featureless and it 
can be well described by a single power law with a spectral index of 
$2.41 \pm 0.05$. This is significantly softer than both the DGRB initially 
reported by EGRET \cite{Sreekumar:1997un} and the revised estimate of 
Ref.~\cite{Strong:2003ex} from the same data set.\footnote{However, note that, 
in the re-analysis of Ref.~\cite{Strong:2003ex}, the energy spectrum is well 
described by a single power law {\it only} up to 2 GeV.}

\begin{figure}[h]
\begin{center}
\includegraphics[width=0.95\linewidth]{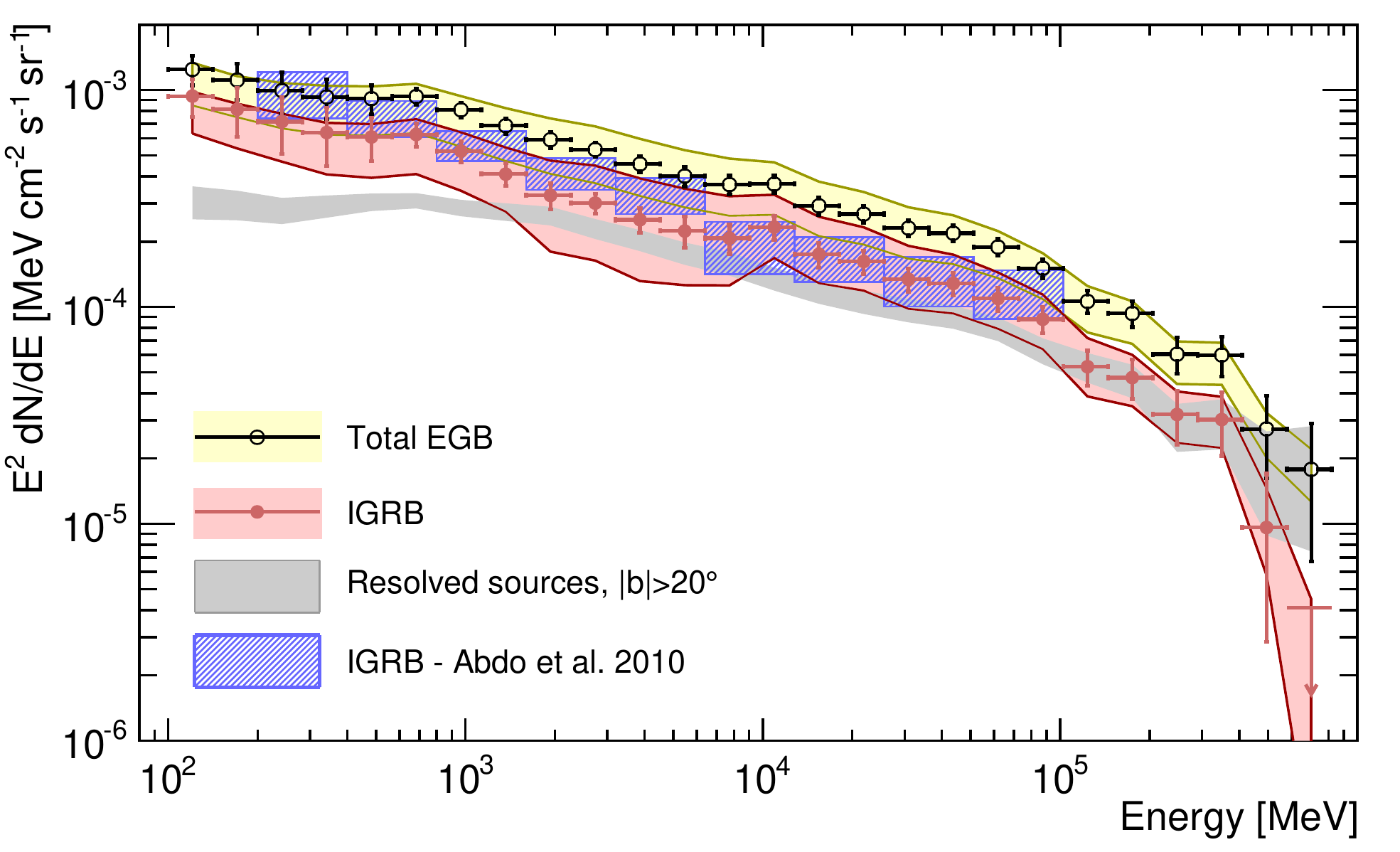}
\caption{\label{fig:DGRB_intensity} In red, the DGRB intensity energy spectrum, labeled ``Isotropic Gamma-Ray Background (IGRB)'' in the figure. Model A from Ref.~\cite{Ackermann:2014usa} is used for the diffuse Galactic foreground. The spectrum is compared to the previous \fermi LAT measurement of the DGRB from Ref.~\cite{Abdo:2010nz} (blue bands) and to the total ``Extragalactic Gamma-ray Background (EGB)'' (black data points). The latter is defined here as the sum of the DGRB and of the resolved sources at $|b|>20^\circ$ (shown in gray). The comparison shows that, above 100 GeV, $\sim 50\%$ of the Extragalactic Gamma-ray Background is now resolved into individual LAT sources. Yellow and red shaded bands represent the systematic error associated with the uncertainties in the modeling of the Galactic diffuse emission. Taken from Ref.~\cite{Ackermann:2014usa}.}
\end{center}
\end{figure}

More recently, a new measurement of the DGRB energy spectrum has been performed 
by the \fermi LAT \cite{Ackermann:2014usa}. The analysis used 50 months of data 
with $|b|>20^\circ$, it employed a dedicated event selection and it took 
advantage of improvements in the determination of the CR background and of the 
diffuse Galactic foreground. The measurement, denoted by red data points in 
Fig.~\ref{fig:DGRB_intensity}, now extends down to 100 MeV and up to 820 GeV. 
It represents the most complete and accurate picture that we currently possess 
of the DGRB intensity. Interestingly, the DGRB now exhibits a high-energy 
exponential cut-off at $279 \pm 52$ GeV (for the baseline model of Galactic 
diffuse foreground used in Ref.~\cite{Ackermann:2014usa}), and it is well 
described by a single power law with a spectral index of $2.31 \pm 0.02$ at 
lower energies. The cut-off is compatible with the attenuation expected from 
the interaction of high-energy photons with the EBL \cite{Franceschini:2008tp,
Dominguez:2010bv,Ackermann:2012sza,Abramowski:2012ry,Gilmore:2011ks,
Dominguez:2013lfa,Khaire:2014xqa} over cosmological distances 
\cite{Ajello:2015mfa}. The largest systematic uncertainty (represented by the 
red shaded region in Fig.~\ref{fig:DGRB_intensity}) ranges between a factor of 
$\sim 15\%$ and 30\% (depending on the energy range considered) and it comes 
from the modeling of the Galactic diffuse emission.

For the purposes of this review, it is convenient to briefly summarize here 
the main steps followed in Ref.~\cite{Ackermann:2014usa} to measure the DGRB. 
Two different sets of selection cuts are applied to the gamma-ray data, 
optimized over different energy ranges, namely below and above 12.8 GeV. 
Different selection criteria are required because of the energy dependence in
the composition of the CR-induced backgrounds and the way CRs interact with 
the detector. For the low-energy data sample, the all-sky Galactic 
diffuse emission is modeled as a sum of templates obtained with 
{\ttfamily GALPROP}\footnote{A CR propagation code available from 
http://galprop.stanford.edu. See also Ref.~\cite{FermiLAT:2012aa}.} and the 
point-like sources are modeled following the information in the Second \fermi 
LAT catalog (2FGL). Additional templates are used to model subdominant 
contributions from the Loop I large-scale Galactic structure and from 
electrons interacting with the Solar radiation field through Inverse Compton 
(IC). A fully isotropic template is also included, in order to describe the 
DGRB and the residual CR contamination. The fit to the data determines the 
normalizations of the different templates and it is performed independently 
in each energy bin considered. At the highest energies, where the statistic is 
scarce, the normalizations of the templates for the Galactic diffuse emission 
are fixed to the best-fit values obtained at intermediate energies (between 
6.4 and 51.2 GeV). The spectral shapes of their emission are also fixed to 
those predicted by {\ttfamily GALPROP}. Above 12.8 GeV, then, the 
normalizations of the point-like sources and of the isotropic template are the 
only free parameters in the fit.

Monte Carlo simulations are used to estimate the residual CR contamination. The 
energy spectrum of the DGRB is, finally, obtained by subtracting the CR 
contamination from the isotropic component determined in the template fitting. 
The systematic error induced by the imperfect knowledge of the Galactic 
diffuse emission is estimated by repeating the whole procedure for three 
benchmark models of Galactic foreground and for different values of the 
parameters controlling the propagation of CRs in the MW (see 
Ref.~\cite{Ackermann:2014usa} for further details).

\subsection{The astrophysical components of the Diffuse Gamma-Ray Background}
\label{sec:components}
In this section we describe the classes of astrophysical sources and emission 
mechanisms that have been proposed as contributors to the DGRB over the years. 
Well-established astrophysical populations, whose brightest members have
been robustly detected, represent ``guaranteed'' components to the DGRB. They 
are discussed in Sec.~\ref{sec:blazars}, Sec.~\ref{sec:MAGNs}, 
Sec.~\ref{sec:SFGs} and Sec.~\ref{sec:MSPs}, which are devoted, respectively, 
to blazars, MAGNs, SFGs and MSPs. Then, in Sec.~\ref{sec:other_astrophysics},
we turn to more speculative scenarios. We do not discuss the possibility of
a DM-induced contribution to the DGRB, since that is the subject of the 
following section (Sec.~\ref{sec:DM}).

We start by presenting a formalism that will be adopted throughout the
manuscript for the description of a generic population of sources. The sources
are supposed to be characterized by a measurable quantity $Y$. In most cases
$Y$ will be the blue gamma-ray luminosity $L_\gamma$ of the source but, in some 
instances, it will indicate another parameter such as, e.g., the mass of the 
galaxy or of the DM halo hosting the gamma-ray emitter. The differential 
gamma-ray flux $d\Phi/dEd\Omega$ (i.e. the number of photons per unit area, 
time, energy and solid angle) expected from the {\it unresolved} objects in 
such a population can be written as follows:
\begin{eqnarray}
\frac{d\Phi}{dE d\Omega}(E_0) & = & \int_{z_{\rm max}}^{z_{\rm min}} dz 
\int_{Y_{\rm min}}^{Y_{\rm max}} dY \int_{\Gamma_{\rm min}}^{\Gamma_{\rm max}} d\Gamma 
\, \frac{dN}{dVdYd\Gamma}(z,Y,\Gamma) \, \frac{dV}{dzd\Omega} 
\, F_{E_0}(z,Y,\Gamma) \times
\nonumber \\
& & \left( 1 - \frac{\Omega(z,Y,\Gamma)}{\Omega_{\rm max}} \right) 
e^{-\tau_{\rm EBL}(E_0,z)},
\label{eqn:unresolved_sources}
\end{eqnarray}
where $z$ is the redshift and $\Gamma$ is a parameter that characterizes the 
shape of the energy spectrum of the sources. The domain of integration will 
depend on the particular population considered. The factor $dN/dVdYd\Gamma$ 
indicates the comoving number density of sources per unit $Y$ and $\Gamma$, 
while $dV/dzd\Omega$ is the comoving volume per unit redshift and solid angle. 
The factor $F_{E_0}(z,Y,\Gamma)$ is the gamma-ray flux (at energy $E_0$) 
produced by the source identified by the value $Y$, located at redshift $z$ 
and with an energy spectrum characterized by $\Gamma$. If sources are 
described by their gamma-ray luminosity ($Y=L_\gamma$), $F_{E_0}$ can be written 
as follows:
\begin{equation}
F_{E_0} = \frac{(1+z)^2 L_\gamma}{4 \pi D_L(z)^2} g_{E_0}(\Gamma),
\end{equation}
where $D_{L}(z)$ is the luminosity distance at redshift $z$ and $g_{E_0}(\Gamma)$
parametrizes the gamma-ray energy spectrum. For other choices of $Y$, the
relation between $Y$ and $F_{E_0}$ needs to be specified case per case.

Thus, the first line of Eq.~(\ref{eqn:unresolved_sources}) indicates the 
cumulative emission expected from all the sources located between $z_{\rm min}$ 
and $z_{\rm max}$ and with characteristics between $(Y_{\rm min},\Gamma_{\rm min})$ 
and $(Y_{\rm max},\Gamma_{\rm max})$. The quantity $\Omega(z,Y,\Gamma)$ is 
called sky coverage and it is defined as the area $\Omega_{\rm max}$ surveyed by 
the telescope multiplied by the detection efficiency. It represents the 
probability for a source characterized by parameters $(z,Y,\Gamma)$ to be 
detected and it encodes the sensitivity of the instrument, i.e. 
$\Omega(z,Y,\Gamma)$ is equal to zero if the source is too faint to be 
detected. It also accounts for any potential selection effect. An estimate of 
$\Omega(z,Y,\Gamma)$ for the high-latitude blazars in the first \fermi LAT 
catalog (1FGL) is provided in Ref.~\cite{Collaboration:2010gqa}. Then, the 
factor $(1-\Omega(z,Y,\Gamma)/\Omega_{\rm max})$ in 
Eq.~(\ref{eqn:unresolved_sources}) has the effect of selecting only the 
sources that remain undetected. Finally, the exponential $e^{-\tau_{\rm EBL}(E_0,z)}$ 
accounts for the effect of the EBL attenuation.

It is also convenient to define a generic expression for the differential 
source count distribution $dN/dS$, i.e. the number of sources per unit solid 
angle and per unit flux $S$:
\begin{equation}
\frac{dN}{dS} = \int_{z_{\rm max}}^{z_{\rm min}} dz 
\int_{\Gamma_{\rm min}}^{\Gamma_{\rm max}} d\Gamma \, \frac{dN}{dVdYd\Gamma}(z,Y,\Gamma) 
\, \frac{dV}{dzd\Omega} \, \frac{dY}{dS} \, 
\frac{\Omega(z,Y,\Gamma)}{\Omega_{\rm max}},
\label{eqn:differential_source_count}
\end{equation}
where $S$ is the gamma-ray flux associated with the source characterized by 
$(z,Y,\Gamma)$.\footnote{Normally $S$ is the gamma-ray flux above some energy 
$E_{\rm min}$ and, therefore, it is related to $F_{E_0}(z,Y,\Gamma)$ as 
$S=\int_{E_{\rm min}} F_{E_0}(z,Y,\Gamma) dE_0$.} Note that the factor 
$\Omega(z,Y,\Gamma)/\Omega_{\rm max}$ has now the effect of selecting only the 
resolved objects. The number of sources (per solid angle) with a flux larger 
than $\bar{S}$, i.e. the {\it cumulative} source count distribution 
$N(>\bar{S})$, can be obtained by integrating 
Eq.~(\ref{eqn:differential_source_count}) above $\bar{S}$.

\subsubsection{Blazars}
\label{sec:blazars}
Blazars are among the brightest gamma-ray sources in the sky. They are 
interpreted as Active Galactic Nuclei (AGNs) with the relativistic jet 
directed towards the observer \cite{Urry:1995mg,Blandford:1978}. The emission 
coming from the jet normally outshines the radiation associated with the 
accretion onto the supermassive Black Hole at the center of the nucleus. The 
Spectral Energy Distribution (SED) is bimodal: the low-energy peak is located 
between ultraviolet (UV) and radio frequencies and it corresponds to the 
synchrotron emission produced by the electrons that are accelerated in the jet. 
Instead, the high-energy peak reaches the gamma-ray band and it is produced by 
IC of the same population of synchrotron-emitting electrons. The seed photons 
for the IC emission come either from the synchrotron radiation itself (i.e., 
the so-called synchrotron self-Compton) or from the accretion disk (external 
Compton). Blazars can be divided in two categories: BL Lacs and Flat Spectrum 
Radio Quasars (FSRQs). Sources belonging to the first class have a nuclear 
non-thermal emission so strong that the rest-frame equivalent width of the 
strongest optical emission line is narrower than 5 $\dot{\mbox{A}}$, while 
FSRQs have broader lines and a spectral index $\alpha_r<0.5$ in the radio band. 
Blazars have also been classified according to the frequency of their 
synchrotron peak, $\nu_{\rm S}$ \cite{Ghisellini:2009,Meyer:2012}: 
low-synchrotron-peaked (LSP) blazars have a peak in the infrared (IR) or 
far-IR band, ($\nu_{\rm S}<10^{14}$ Hz), intermediate-synchrotron-peaked (ISP) 
blazars for $\nu_{\rm S}$ in the near-IR or UV 
($10^{14} \mbox{ Hz} \leq \nu_{\rm S} < 10^{15} \mbox{ Hz}$) and 
high-synchrotron-peaked (HSP) blazars for $\nu_{\rm S} \geq 10^{15}$ Hz, i.e. 
in the UV band or higher. From the study of the 886 blazars present in the 
Second \fermi LAT AGN catalog (2LAC) \cite{Ackermann:2011} it was confirmed 
that FSRQs (which are almost entirely LSPs) are generally brighter than BL 
Lacs. Also, it was possible to determine a correlation between the steepness of 
their energy spectrum (in the gamma-ray band) and the position of the 
synchrotron peak, suggesting that FSRQs (with a low $\nu_{\rm S}$) have softer 
spectra than BL Lacs \cite{Ackermann:2011}.

Since the EGRET era, blazars were suspected to play a significant role in 
explaining the DGRB emission, with estimates ranging from 20\% to 100\% 
\cite{Padovani:1993,Stecker:1993ni,Stecker:1996ma,Muecke:1998cs,
Narumoto:2006qg,Pavlidou:2007dv,Kneiske:2007jq,Bhattacharya:2009yv,
Inoue:2008pk}. Fewer AGNs were known before the \fermi LAT (66 in the third 
EGRET catalog \cite{Hartman:1999}, mainly FSRQs), preventing reliable 
population studies to be performed entirely at gamma-ray energies. Predictions 
for the emission of unresolved blazars were obtained by means of 
well-established correlations between the gamma-ray luminosity $L_\gamma$ and 
the luminosity at lower frequencies, either in the radio or in the X-ray band
\cite{Padovani:1993,Stecker:1996ma}. Thus, the strategy followed in these 
early works was to characterize the blazar population in those low-frequency
regimes, where the statistical sample was much larger, and then export the
information up to the gamma-ray range by means of the aforementioned 
correlation between luminosities.

Ref.~\cite{Inoue:2008pk} considers the correlation between $L_\gamma$ and the
luminosity in X-rays, $L_X$. The $L_\gamma-L_X$ connection is a consequence of 
the relation between the emission of the jet (which can be linked to $L_\gamma$) 
and the mass accretion onto the supermassive Black Hole \cite{Falcke:1994eb,
Falcke:2003ia,Merloni:2003aq,Kording:2006sa}. This, in turn, correlates with 
the X-ray luminosity of the accretion disk $L_X$. Ref.~\cite{Inoue:2008pk} 
considers $L_\gamma$ as the $Y$-parameter in Eqs.~(\ref{eqn:unresolved_sources}) 
and (\ref{eqn:differential_source_count}). The dependence on $\Gamma$ in the 
factor $dN/dV dL_\gamma d\Gamma$ is taken care of by the so-called 
``blazar sequence'' \cite{Fossati:1997vu,Fossati:1998zn,Donato:2001}, which 
predicts how the shape of the blazar SED changes as a function of their 
luminosity. Then, integrating Eq.\ref{eqn:unresolved_sources} over $\Gamma$, 
the blazar sequence selects, for each $L_\gamma$, the only SED compatible with 
$L_\gamma$. On the other hand, $dN/dV dL_\gamma$ is the gamma-ray luminosity 
function (LF), $\Phi_\gamma(z,L_\gamma)$. Given the $L_\gamma-L_X$ relation, 
$\Phi_\gamma(z,L_\gamma)$ can be inferred from the X-ray LF $\Phi_X(L_X,z)$:
\begin{equation}
\Phi_\gamma(L_\gamma,z) = \kappa \, \frac{dL_X}{dL_\gamma} \, \Phi_X(L_X,z).
\label{eqn:L_gamma_L_X}
\end{equation}
The factor $\kappa$ indicates the fraction of AGNs observed as blazars and 
a parametrization for $\Phi_X(z,L_X)$ is available in Refs.~\cite{Ueda:2003yx,
Hasinger:2005sb,Gilli:2006zi}. X-ray AGNs are found to evolve positively (i.e. 
they are more abundant as redshift increases) until a certain redshift peak 
$z_c$, above which the X-ray LF decreases \cite{Ueda:2003yx,Hasinger:2005sb,
Gilli:2006zi}. It was found that allowing $z_c$ to vary with $L_X$ (i.e., what 
is called a luminosity-dependent density evolution) provides a better 
description of the blazars observed by EGRET than other evolution schemes, 
e.g., pure luminosity evolution or pure density evolution 
\cite{Narumoto:2006qg}.

Parameter $\kappa$ in Eq.~(\ref{eqn:L_gamma_L_X}) is left free in the analysis
of Ref.~\cite{Inoue:2008pk}, together with the faint-end slope $\gamma_1$ of 
the X-ray LF and the proportionality coefficient between $L_\gamma$ and 
$L_X$.\footnote{More precisely, the assumed proportionality is between $L_X$ 
and the so-called ``bolometric luminosity'', i.e. a measure of the total 
emission power of the source. However, under the hypothesis of the blazar 
sequence, $L_\gamma$ can be uniquely derived from the bolometric luminosity.} 
These parameters are determined by a maximum-likelihood fit to the differential 
source count $dN/dS$ of the blazars detected by EGRET. The best-fit model is, 
then, used to determine the contribution of unresolved blazars to the DGRB 
through Eq.~(\ref{eqn:unresolved_sources}). Ref.~\cite{Inoue:2008pk} finds that 
unresolved blazars can explain approximately 45\% of the DGRB measured by
EGRET in Ref.~\cite{Strong:2004ry} at 100 MeV.

Similar results are obtained in other works that follow the same formalism:
Refs.~\cite{Narumoto:2006qg,Ando:2006mt,Ando:2006cr} also consider a 
luminosity-dependent density evolution for the X-ray LF but they assume a 
common power-law energy spectrum at gamma-ray frequencies, instead of the 
SED predicted by the blazar sequence. In Ref.~\cite{Abazajian:2010pc} the 
formalism outlined before is fitted to both the blazar $dN/dS$ computed from 
the 1FGL \cite{Abdo:2010ru} {\it and} to the first \fermi LAT measurement of 
the DGRB in Ref.~\cite{Abdo:2010nz} (see also Ref.~\cite{Cavadini:2011ig}). The 
result of the fit proves that it is possible to explain both observables with 
one single population of blazars.\footnote{The best-fit point obtained in 
Ref.~\cite{Abazajian:2010pc} favors a $\gamma_1$ much larger than the one 
found in Ref.~\cite{Inoue:2008pk} for EGRET blazars.} Their predictions are
re-calibrated in Ref.~\cite{Harding:2012gk}, where the model is fitted against 
the 1FGL $dN/dS$ and the \fermi LAT measurement of the DGRB auto-correlation 
APS of anisotropies from Ref.~\cite{Ackermann:2012uf} (see 
Sec.~\ref{sec:angular_spectrum}). Ref.~\cite{Harding:2012gk} finds that the 
DGRB intensity energy spectrum and the auto-correlation APS cannot be 
{\it simultaneously} explained in terms of unresolved blazars, since a model 
that fits both the abundance of resolved blazars (i.e., their $dN/dS$) and the 
DGRB intensity energy spectrum would exceed the measured auto-correlation 
APS. Alternatively, unresolved blazars can reproduce the measured anisotropies
but, then, their emission only accounts for a maximum of 4.3\% of the DGRB 
intensity in Ref.~\cite{Abdo:2010nz}, above 1 GeV.

Instead of using the $L_\gamma-L_X$ correlation discussed above, 
Ref.~\cite{Stecker:1996ma} relates $L_\gamma$ to the radio luminosity, $L_r$. 
The gamma-ray LF is inferred from the radio LF, similarly to what done in
Eq.~(\ref{eqn:L_gamma_L_X}). The radio LF is taken from 
Ref.~\cite{Dunlop:1990kf}. A power-law energy spectrum is assumed at 
gamma-ray frequencies, while the distribution of spectral indexes $dN/d\Gamma$ 
is calibrated to reproduce the sources detected by EGRET. 
Ref.~\cite{Stecker:1996ma} finds that unresolved blazars can fit reasonably 
well the EGRET DGRB energy spectrum reported in Ref.~\cite{Sreekumar:1997un}. 
Other works exploit the $L_\gamma-L_r$ relation with similar findings: 
Ref.~\cite{Chiang:1998} determines the properties of the blazar population by 
fitting their measured radio LF (assuming it follows a pure luminosity 
evolution), while in Ref.~\cite{Stecker:2010di} the fit is performed with the 
cumulative source distribution $N(>S)$ of 1FGL blazars. The authors of 
Ref.~\cite{Stecker:2010di} also caution about the use of the sky coverage 
determined in Ref.~\cite{Collaboration:2010gqa}, since it may be affected by 
systematic uncertainties in the low-flux regime due to low statistics. 

\begin{figure}[h]
\begin{center}
\includegraphics[width=0.95\linewidth]{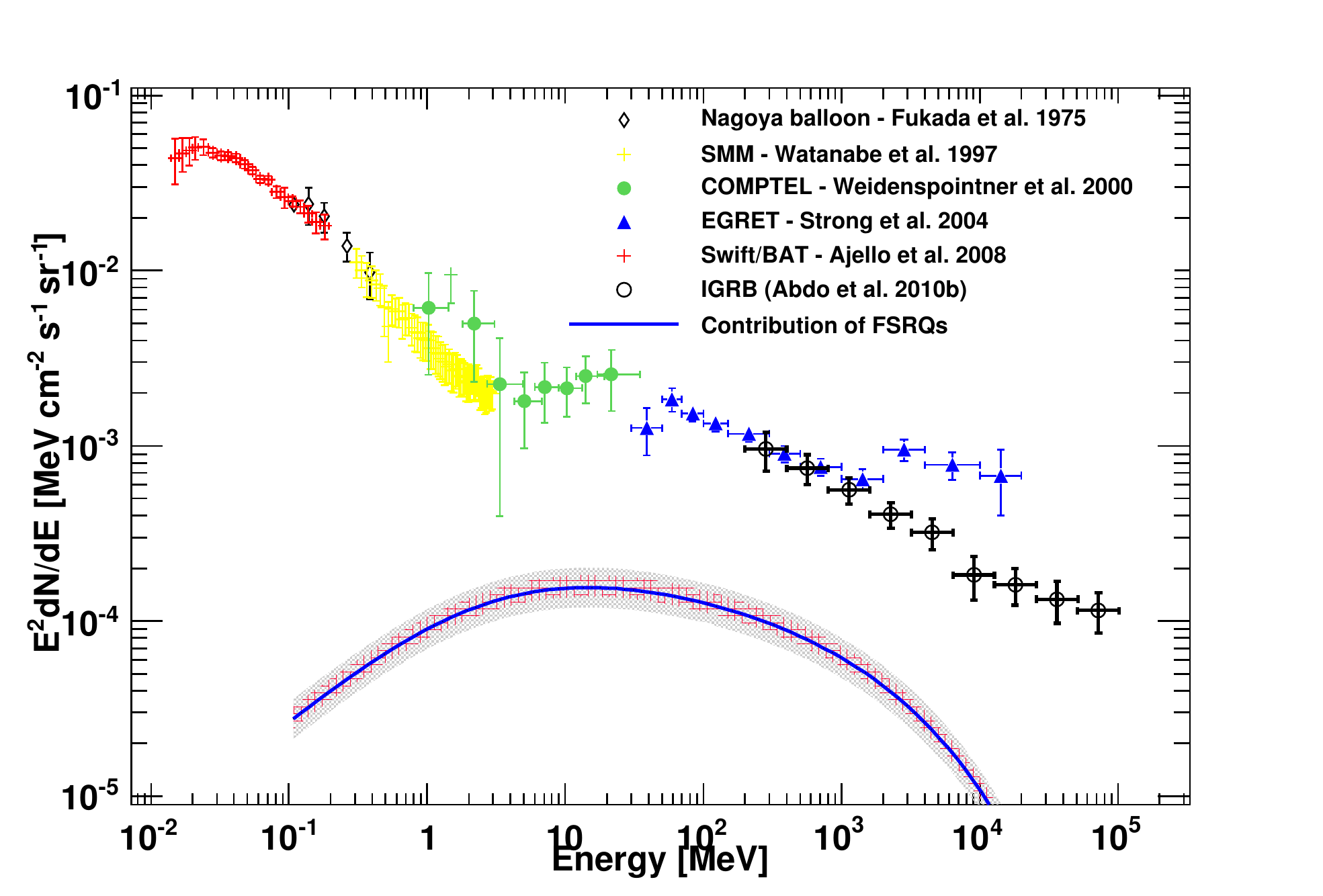}
\caption{\label{fig:Energy_spectrum_FSRQs} Contribution of unresolved FSRQs to the DGRB as determined by integrating the LF coupled to the SED model derived in Ref.~\cite{Ajello:2011zi}. The red hatched band around the best-fit point prediction (solid blue line) shows the $1\sigma$ statistical uncertainty, while the gray band represents the systematic uncertainty. Note that, in the figure, the DGRB measured by the \fermi LAT in Ref.~\cite{Abdo:2010nz} (black data points) is refereed to as ``IGRB''. Taken from Ref.~\cite{Ajello:2011zi}.}
\end{center}
\end{figure}

With the advent of the \fermi era, direct population studies at gamma-ray
frequencies became possible, without the need to rely on correlations with 
lower frequencies. Following the formalism of 
Eqs.~(\ref{eqn:unresolved_sources}) and (\ref{eqn:differential_source_count}), 
in Ref.~\cite{Collaboration:2010gqa} the $Y$ parameter is taken to be the 
flux $S$ above 100 MeV. The energy spectra of the blazars are assumed to be 
power laws with indexes $\Gamma$ characterized by a Gaussian probability 
distribution independent of $S$. The gamma-ray LF $dN/dV dS$ is combined with 
the factor $dV/dz d\Omega$ in Eq.~(\ref{eqn:differential_source_count}) into 
$dN/dz dS d\Omega$, which is assumed to be the same for all the sources 
considered and to depend on $S$ as a broken power law. No correlation with 
other frequencies is required. The parameters of the broken power law (as well 
as the mean and standard deviation of the Gaussian distribution for $\Gamma$) 
are inferred through a maximum-likelihood fit to the blazars in the 1FGL with 
$|b|>20^\circ$. The best-fit point has a break at 
$S=(5.99 \pm 0.91) \times 10^{-8} \mbox{cm}^{-2} \mbox{s}^{-1} $ and slopes of
$1.58 \pm 0.08$ and $2.44 \pm 0.11$ above and below the break 
\cite{Collaboration:2010gqa}. The model is, then, used to determine the 
cumulative emission of unresolved sources. Results indicate that unresolved 
blazars (with $S_{\rm min}=0$) account for 23\% of the DGRB detected by \fermi 
LAT in Ref.~\cite{Abdo:2010nz}.

\begin{figure}[h]
\begin{center}
\includegraphics[width=0.95\linewidth]{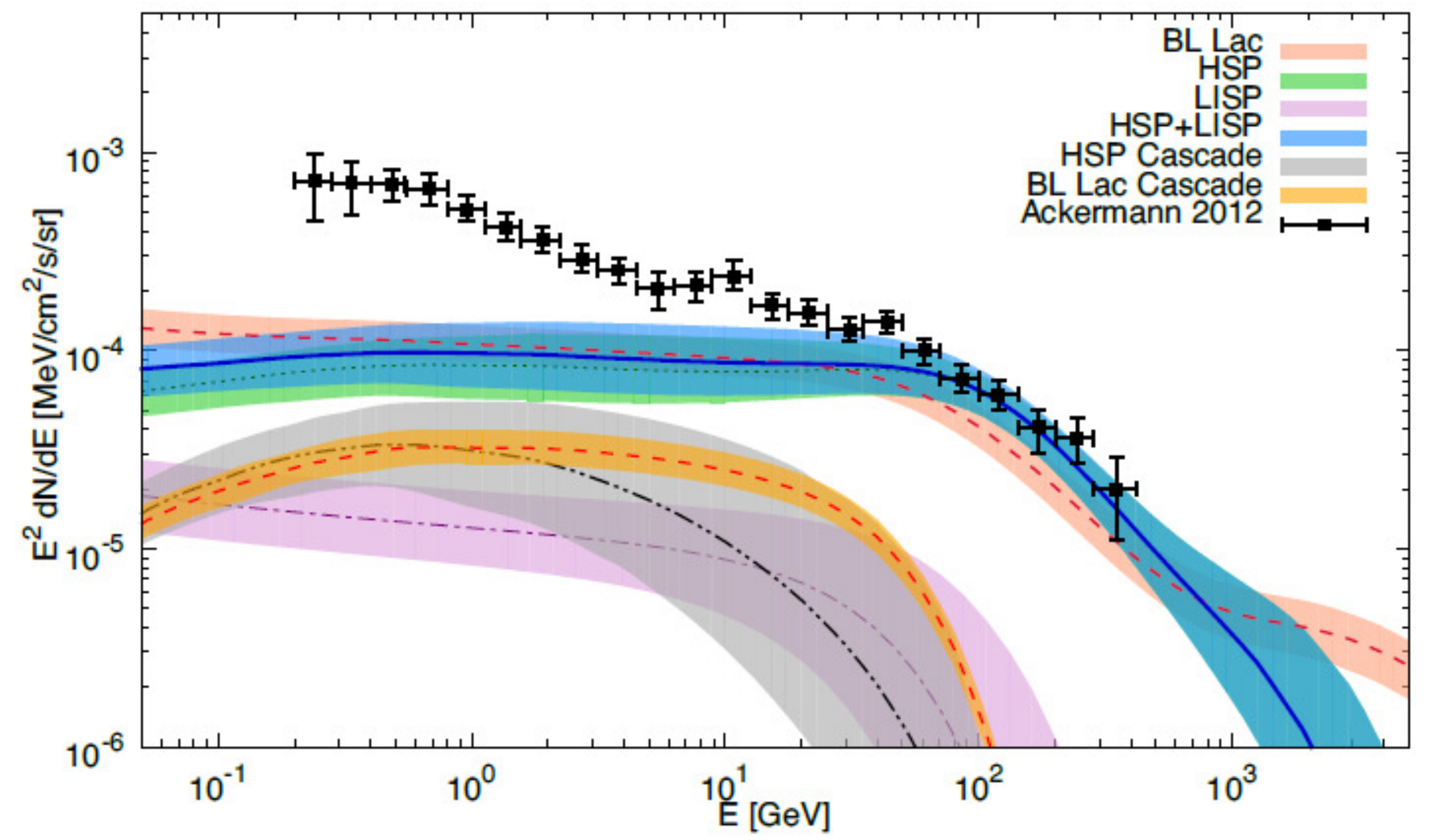}
\caption{\label{fig:Energy_spectrum_BLLacs} Diffuse gamma-ray emission from unresolved BL Lacs. Predictions for the best-fit model in Ref.~\cite{DiMauro:2013zfa} are shown embedded in their $1\sigma$ uncertainty bands: the summed contribution from LSPs and ISPs is plotted by means of the purple dot-dashed line and the purple band, the one from HSP BL Lacs by the green band and green dotted line (almost overlapped with the blue band), the one from the sum of the two by blue band and solid blue line and the contribution from BL Lacs considered as a unique population is depicted by the pink band and dashed pink line. The DGRB data are taken from Ref.~\cite{Ackermann:2012} and are displayed as black points. The gray band and double-dot-dashed black line (orange band and dashed red line) represent the cascade emission from the HSP BL Lacs (the whole BL Lac population). Taken from Ref.~\cite{DiMauro:2013zfa}.}
\end{center}
\end{figure}

Ref.~\cite{Ajello:2011zi} improves the analysis in 
Ref.~\cite{Collaboration:2010gqa}, attacking one of its main limitations,
i.e. the fact that the broken power law considered for $dN/dz dS d\Omega$ is 
assumed {\it a priori} and it is not inferred from a specific evolution scheme. 
In Ref.~\cite{Ajello:2011zi}, the analysis is restricted to FSRQs: sources are 
identified by their $L_\gamma$ (i.e., $Y \equiv L_\gamma)$, their redshift and 
the slope of the power-law fit to their energy spectrum. 
$dN/dV dL_\gamma d\Gamma$ is split into the gamma-ray LF and the spectral index 
distribution $dN/d\Gamma$. The former is described by a parametric expression 
inspired by the results in the radio and X-ray bands, while a Gaussian 
distribution is assumed for $dN/d\Gamma$, with no dependence on $L_\gamma$. A 
maximum-likelihood fit is performed to determine the parameters of the model. 
186 FSRQs in the 1FGL and with $|b| \geq 15^\circ$ are considered in the fit. 
Results establish that FSRQs evolve positively until a $z_c$ that depends on 
$L_\gamma$, i.e. a luminosity-dependent density evolution performs better than 
other evolution formalisms. The best-fit model corresponds to an unresolved 
emission which accounts for $9.3_{-1.0}^{+1.6}\%$ of the \fermi LAT DGRB in 
Ref.~\cite{Abdo:2010nz}, between 0.1 and 100 GeV (see 
Fig.~\ref{fig:Energy_spectrum_FSRQs}).\footnote{In Ref.~\cite{Ajello:2011zi} 
only sources with $L_\gamma \geq 10^{44} \mbox{erg s}^{-1}$ are considered when 
computing the emission of unresolved FSRQs. Such a value corresponds 
approximately to the fainter blazar in the 1FGL. Thus, the quoted contribution 
of FSRQs to the DGRB implicitly assumes that no sources are present below 
$10^{44} \mbox{erg s}^{-1}$ and, therefore, it should be considered as a lower 
limit to the total emission of unresolved FSRQs.} The contribution peaks 
below the GeV scale and it becomes more subdominant at higher energies.

\begin{figure}[h]
\begin{center}
\includegraphics[width=0.95\linewidth]{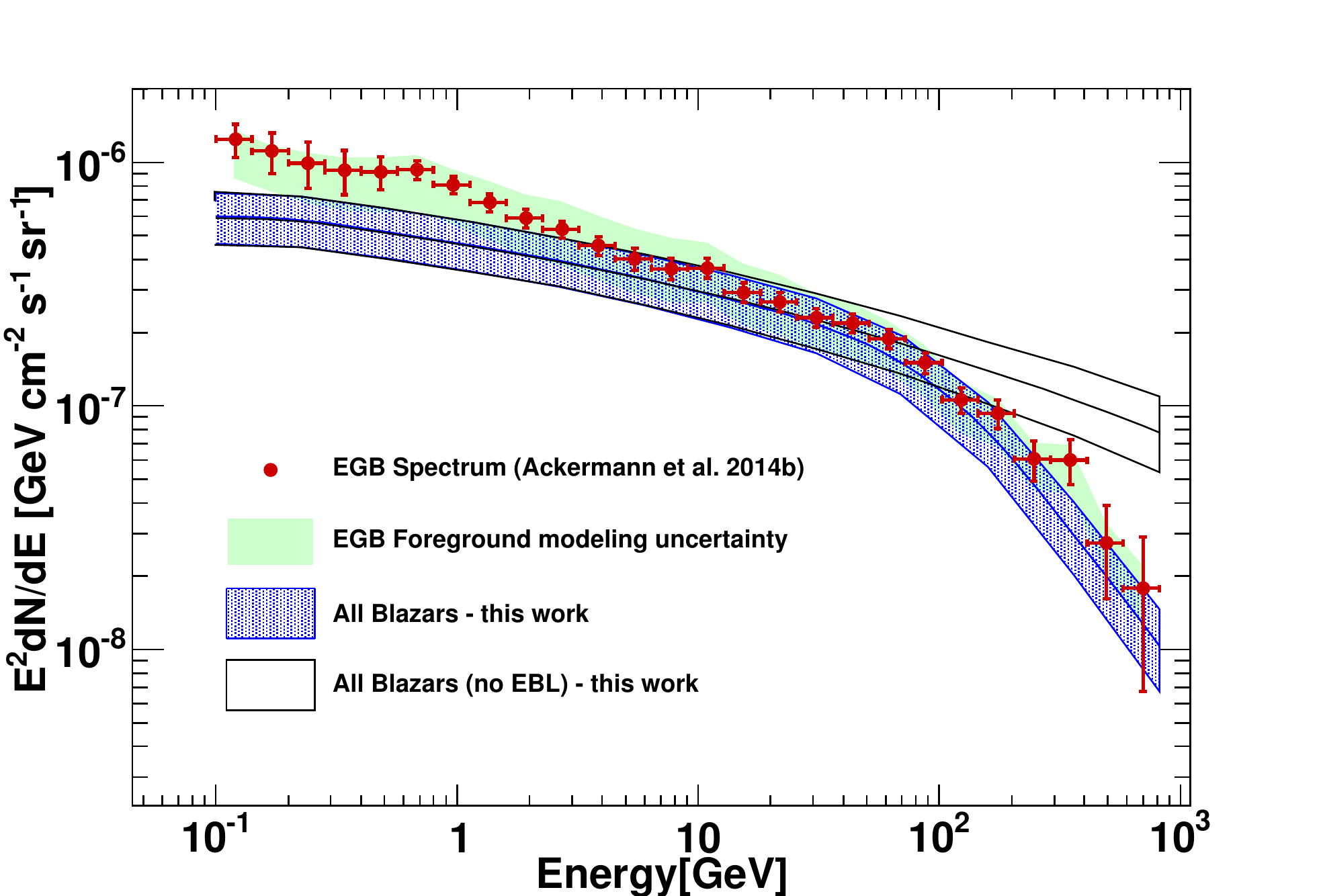}
\caption{\label{fig:Energy_spectrum_blazars} Diffuse emission arising from blazars (with or without EBL absorption), in comparison with the intensity of the total emission from sources (both resolved and unresolved), called here ``EGB'' (red data points, from Ref.~\cite{Ackermann:2014usa}). Taken from Ref.~\cite{Ajello:2015mfa}}.
\end{center}
\end{figure}

Performing a similar population study for BL Lacs is hampered by the fact 
that it is more difficult to obtain a measure of spectroscopic redshift for 
these objects due to the lack of strong emission lines: indeed, approximately 
55\% of the BL Lacs in the 2LAC do not have an associated $z$. This issue is 
somehow alleviated in Ref.~\cite{Ajello:2013lka} by considering photometric 
redshift estimates \cite{Rau:2012}, lower \cite{Shaw:2013pp} or upper limits 
on $z$ \cite{Rau:2012,Shaw:2013pp} and host-galaxy spectral fitting 
\cite{Shaw:2013pp}. The 211 BL Lacs studied in Ref.~\cite{Ajello:2013lka}
are taken from Ref.~\cite{Collaboration:2010gqa} and analyzed by means of
the same pipeline applied in Ref.~\cite{Ajello:2011zi} to FSRQs: the $Y$ 
parameter in Eq.~(\ref{eqn:unresolved_sources}) is again $L_\gamma$ but the 
mean of the Gaussian distribution of spectral indexes depends now linearly on 
$\log_{10}(L_\gamma)$. Only sources with 
$L_\gamma \geq 7 \times 10^{43} \mbox{erg s}^{-1}$ are considered. The best-fit
model suggests that the number density of faint BL Lacs (probably HSPs) 
decreases with redshift, while BL Lacs with 
$L_\gamma \geq 10^{45.8} \mbox{erg s}^{-1}$ are more numerous at large redshift,
i.e. more similar to the positive evolution FSRQs. The redshift estimates 
adopted for the BL Lacs without spectroscopic information are crucial to 
constrain the evolution of their LF. Results indicate that unresolved BL Lacs 
contribute to $7.7_{-1.3}^{+2.0}\%$ of the \fermi LAT DGRB in 
Ref.~\cite{Abdo:2010nz}, between 0.1 and 100 GeV. Due to the large density of 
low-luminosity hard sources at low redshift, the emission of unresolved BL 
Lacs is expected to be harder than that of FSRQs and, thus, BL Lacs may play 
a more significant role at higher energies.

This hypothesis was tested by Ref.~\cite{DiMauro:2013zfa}, where the authors
considered a set of 148 BL Lacs with redshift and synchrotron peak frequency
$\nu_{\rm S}$ obtained from the 2FGL \cite{Fermi-LAT:2011iqa}. Their model for 
the gamma-ray LF is fitted to the observed cumulative source distribution 
$N(>S)$, the gamma-ray LF and the redshift distribution $dN/dz$ of the 
detected sources. The energy spectra of the sources are obtained from 
a combination of \fermi LAT and Imaging Air Cherenkov Telescopes data. A 
luminosity-dependent density evolution is found to provide the best fit to 
the data. Pure power laws, log-parabolae and power laws with exponential 
cut-offs are considered as possible SEDs, with the last one corresponding to 
the most accurate description of the BL Lacs in the sample. The sources were 
considered as either one single population, or split into HSPs and a second 
sub-class including ISPs and LSPs. In their best-fit model, HSPs dominates 
the $dN/dS$ below $S=5 \times 10^{-9} \mbox{cm}^{-2} \mbox{s}^{-1}$ and their 
SED extends to much higher energies than in the ISP+LSP class (the best-fit 
cut-off energy is 910 GeV for HSPs and 37 GeV for the class of ISPs and LSPs). 
That is the reason why the cumulative emission from HSPs (computed from 
Eq.~(\ref{eqn:unresolved_sources}) above 
$L_\gamma \geq 10^{38} \mbox{erg s}^{-1}$) can extend up to very high energies and 
it is able to explain the whole DGRB emission reported in 
Ref.~\cite{Ackermann:2012} above few tens of GeV (see 
Fig.~\ref{fig:Energy_spectrum_BLLacs}). Between 0.1 and 100 GeV, unresolved BL 
Lacs account for $\sim$ 11\% of the \fermi LAT DGRB in 
Ref.~\cite{Ackermann:2012}, in agreement with Ref.~\cite{Ajello:2013lka}.

Ref.~\cite{Ajello:2015mfa} repeated the analysis of Ref.~\cite{Ajello:2013lka} 
on a sample of 403 blazars from 1FGL, this time considering both FSRQs and BL
Lacs as one single population by allowing the spectral index distribution to
depend on $L_\gamma$. A double power-law energy spectrum, proportional to 
$[(E_0/E_b)^{1.7}+(E_0/E_b)^{2.6}]^{-1}$, is assumed and the energy scale $E_b$ 
is found to correlate with the index $\Gamma$ obtained when the SED is fitted 
by a single power law. The same LF used in Ref.~\cite{Ajello:2013lka} and 
based on a luminosity-dependent density evolution is implemented in 
Ref.~\cite{Ajello:2015mfa}, together with other evolution schemes. They all 
provide an acceptable description of the blazar population, even if the 
luminosity-dependent density evolution is the one corresponding to the
largest log-likelihood. The predicted cumulative emission of blazars (FSRQs 
and BL Lacs, resolved and unresolved) can be seen in the 
Fig.~\ref{fig:Energy_spectrum_blazars} as a dotted blue band, compared to the 
total emission from resolved and unresolved sources taken from 
Ref.~\cite{Ackermann:2014usa} (labeled ``EGB'' here, red data points). 
Blazars (both resolved and unresolved) accounts for the $50^{+12}_{-11}\%$ of the 
total emission from resolved and unresolved sources, above 100 MeV. Unresolved 
blazars, on the other hand, are responsible for approximately 20\% of the new
\fermi LAT measurement of the DGRB in Ref.~\cite{Ackermann:2014usa}, in the 
0.1-100 GeV energy band.

\subsubsection{Misaligned Active Galactic Nuclei}
\label{sec:MAGNs}
Under the AGN unification scenario, the parameter that discriminates among
different classes of AGNs is the viewing angle \cite{Urry:1995mg}. A value of 
$14^\circ$ separate blazars (with the jet pointing towards the observer) from 
non-blazars, i.e. MAGNs. Among MAGNs it is possible to further distinguish 
between radio galaxies (with a viewing angle larger than $44^\circ$) and radio 
quasars \cite{Barthel:1989gf}. Radio galaxies are classified either as 
Fanaroff-Riley Type I or Type II (FRI and FRII, respectively) according to
their morphology \cite{Fanaroff:1974}. The emission of FRIs peaks at the 
center of the AGN and it is dominated by two-sided decelerating jets. On the 
other hand, FRIIs are brighter and they are characterized by edge-brightened 
radio lobes with bright hotspots, while jets and core (when detected) are 
subdominant. FRIs and FRIIs are normally interpreted as the misaligned 
counterparts of BL Lacs and FSRQs, respectively. Indeed, the detection of 
MAGNs in the 1FGL and 2FGL confirms that, similar to the case of BL Lacs and 
FSRQs, FRIs have harder spectra than FRIIs \cite{Abdo:2010}.

The mechanisms of gamma-ray production in MAGNs are less clear than for 
blazars: the SED exhibits a bimodal structure, with the low-frequency peak due 
to synchrotron radiation, while at higher energies, the bulk of gamma-ray 
emission may be due to synchrotron self-Compton \cite{Maraschi:1992iz}. 
External IC may also contribute \cite{Dermer:1993cz}. If present, external IC 
would be localized within one parsec from the center of the AGN, while 
synchrotron self-Compton emission is produced outside that region 
\cite{Maraschi:1992iz,Dermer:1993cz}. Furthermore, spatially coincident with 
the radio lobes, there may also be emission from IC off the Cosmic Microwave
Background (CMB) \cite{Massaro:2011ww,Stawarz:2005tq} (see the discussion at 
the end of this section).

With no Doppler boost, MAGNs are expected to be less bright but more abundant 
than blazars, making them potentially important contributors to the DGRB 
\cite{Inoue:2011bm}. 15 MAGNs were reported in Ref.~\cite{Abdo:2010}, 
precluding the possibility of deriving their gamma-ray LF directly from 
gamma-ray observations, as done with blazars. Yet, the LF of MAGNs is well 
known at radio frequencies. As done in the previous section, once a correlation
between their radio and gamma-ray luminosities is established, it is possible 
to deduce the gamma-ray properties of MAGNs by studying the sources in the 
radio band. 

Ref.~\cite{Inoue:2011bm} considers $L_\gamma$ (between 0.1 and 10 GeV) for 10 
MAGNs detected by the \fermi LAT in Ref.~\cite{Abdo:2010}. The author studies
the possibility of a linear correlation between $\log_{10}(L_\gamma)$ and 
$\log_{10}(L_r)$, where $L_r$ is the radio luminosity. The slope of the observed 
correlation is very similar to the one found by Ref.~\cite{Ghirlanda:2010qw} 
for blazars. Then, $\Phi_\gamma(z,L_\gamma)$ is determined in terms of the radio 
LF, $\Phi_r(z,L_r)$, as 
$\Phi_\gamma(z,L_\gamma) = \kappa \Phi_r(z,L_r) dL_r/dL_\gamma$. This is similar to 
Eq.~(\ref{eqn:L_gamma_L_X}) for blazars. Ref.~\cite{Inoue:2011bm} takes 
$\Phi_r$ from Ref.~\cite{Willott:2000dh} and the value of $\kappa$ is tuned to 
reproduce the number of sources observed by the \fermi LAT. All MAGNs are 
assumed to share the same energy spectrum, i.e. a power law with an index of 
2.39. Such a value is the mean spectral index among the 10 MAGNs considered. 
The emission from unresolved MAGNs is, finally, estimated from 
Eq.~(\ref{eqn:unresolved_sources}), for $L_\gamma \geq 10^{39} \mbox{erg s}^{-1}$. 
The result indicates that $\sim 25\%$ of the DGRB measured by \fermi LAT in 
Ref.~\cite{Abdo:2010nz} above 100 MeV can be explained in terms of unresolved 
MAGNs.

\begin{figure}[h]
\begin{center}
\includegraphics[width=0.95\linewidth]{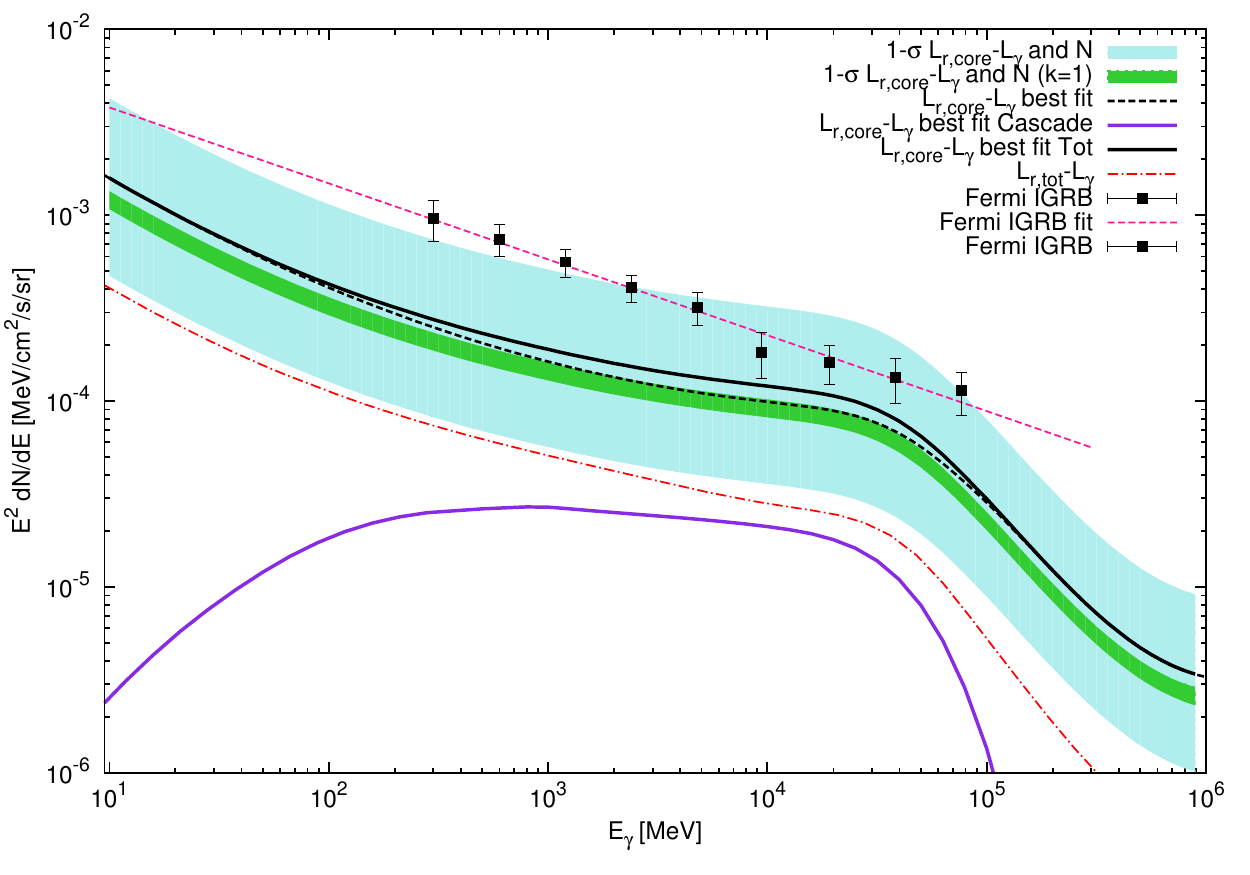}
\caption{\label{fig:Energy_spectrum_MAGNs} Diffuse gamma-ray flux due to MAGNs as a function of the gamma-ray energy. The dashed black line is obtained from the best-fit to the $L_{r,\rm core}-L_\gamma$ correlation. The cyan shaded area is derived by allowing for the uncertainty on the $L_{r,\rm core}-L_\gamma$ correlation and by including all the models that are less than $1\sigma$ away from the best-fit point. The green band corresponds to the case with $\kappa=1$ at $1\sigma$ confidence level. The red dot-dashed curve shows the diffuse flux obtained when assuming a correlation between the total radio luminosity and the gamma-ray luminosity. The black squares corresponds to the DGRB measured by the \fermi LAT in Ref.~\cite{Abdo:2010nz} and the magenta dashed curve represents the best-fit to the data. Taken from Ref.~\cite{DiMauro:2013xta}.}
\end{center}
\end{figure}

Ref.~\cite{DiMauro:2013xta} improves the analysis in Ref.~\cite{Inoue:2011bm}
by properly estimating the uncertainties involved: the authors consider 
$L_\gamma$ (defined above 100 MeV) of 12 MAGNs in the 2FGL and the correlation 
with $L_r$ at 5 GHz. They make use of the total radio luminosity, $L_{r,\rm tot}$, 
and the so-called ``radio core luminosity'' $L_{r,\rm core}$, defined as the 
emission from the central arcsecond-scale region of the source. Both the 
$\log_{10}(L_\gamma)-\log_{10}(L_{r,\rm core})$ and 
$\log_{10}(L_\gamma)-\log_{10}(L_{r,\rm tot})$ relations are considered. The 
gamma-ray LF is inferred from the radio LF as follows:
\begin{equation}
\Phi_\gamma(z,L_\gamma) = \kappa_i \Phi_{r,i} 
\frac{\log_{10}(L_{r,i})}{\log_{10}(L_\gamma)},
\label{eqn:L_gamma_L_r}
\end{equation}
where $i$ stands for ``total'' or ``core''.\footnote{The luminosity functions 
in Ref.~\cite{DiMauro:2013xta} are defined per units of $\log(L)$, hence the 
factor $\log_{10}(L_{r,i})/\log_{10}(L_\gamma)$ in Eq.~(\ref{eqn:L_gamma_L_r}) as 
compared to Eq.~(\ref{eqn:L_gamma_L_X}).} If $L_{r,\rm tot}$ is used, 
$\Phi_{r,\rm tot}$ is taken from Ref.~\cite{Willott:2000dh}.On the other hand, 
it is not possible to build the core radio LF directly from radio observation, 
due to the scarce data available \cite{Yuan:2011hi}. In 
Ref.~\cite{DiMauro:2013xta}, a relation is assumed between $L_{r,\rm core}$ and 
$L_{r,\rm tot}$ \cite{Lara:2004ee}, so that $\Phi_{r,\rm core}$ can be derived from 
$\Phi_{r,\rm tot}$. The factor $\kappa_i$ in Eq.~(\ref{eqn:L_gamma_L_r}) is tuned 
to reproduce the number of observed MAGNs. Different values for $\kappa$ are 
obtained in Ref.~\cite{DiMauro:2013xta}, depending on which LF is considered 
(total radio or core radio) and on how the uncertainties in the 
$\log_{10}(L_\gamma)-\log_{10}(L_{r,i})$ correlations are treated. Finally, the 
emission from unresolved MAGNs is computed as in 
Eq.~(\ref{eqn:unresolved_sources}), assuming a Gaussian distribution for the 
spectral indexes. The results (for $L_\gamma>10^{41} \mbox{erg s}^{-1}$) are 
summarized in Fig.~\ref{fig:Energy_spectrum_MAGNs}. For the best-fit model, 
the contribution of MAGNs accounts for $\sim 25\%$ of the \fermi LAT DGRB in 
Ref.~\cite{Abdo:2010nz} above 100 MeV. The value agrees very well with that 
given in Ref.~\cite{Inoue:2011bm}. Yet, in Ref.~\cite{DiMauro:2013xta}, the 
prediction is embedded in an uncertainty band with a size of almost one order 
of magnitude. Such large uncertainty mainly comes from the possible different 
values for $\kappa$.

Given the complex morphology of the emission in radio galaxies, it is possible
that different regions in the source emit gamma rays which are not accounted 
for by the analyses presented above. Ref.~\cite{Massaro:2011ww} considers a 
scenario in which the non-thermal electrons responsible for the synchrotron 
emission in the radio lobes of FRIIs could emit gamma rays by IC off the 
CMB. Indeed, gamma-ray emission spatially coincident with the radio lobes of 
Centaurus A has been recently observed by the \fermi LAT \cite{Abdo:2010a}.
Such a component would peak at around 1 MeV and could explain up to 10\% of 
the DGRB in Ref.~\cite{Abdo:2010nz} below 1 GeV.

Similarly, Ref.~\cite{Stawarz:2005tq} considers the X-ray-emitting kpc-scale
jets of FRIs as possible contributors to the DGRB. The established synchrotron 
origin of those X-rays suggests that the same non-thermal electrons could also 
emit gamma rays through IC with the ambient photons of the host galaxy. 
However, even assuming this is common for all FRIs, the contribution adds up 
to only $\sim 1\%$ of the DGRB reported in Ref.~\cite{Strong:2004ry}.

\subsubsection{Star-forming galaxies}
\label{sec:SFGs}
The majority of the gamma-ray emission detected by the \fermi LAT is associated 
with the MW \cite{FermiLAT:2012aa}. This diffuse Galactic foreground is 
produced by the interaction of Galactic CRs (mainly protons and electrons) 
with the Galactic interstellar medium and interstellar radiation field. Other 
SFGs similar to the MW are expected to shine in gamma rays thanks to the 
same emission mechanisms. Up to now, the \fermi LAT has only detected a few 
SFGs other than the MW: M31 and M33 \cite{Fermi-LAT:2010zba}, the Large 
and Small Magellanic Clouds \cite{Abdo:2010pq,Fermi-LAT:2010nva} in the Local
Group and the Circinus Galaxy \cite{Hayashida:2013wha}, M82, NGC 253 
\cite{Abdo:2009aa,Abdo:2010ge}, NGC 1068 and NGC 4945 \cite{Ackermann:2012vca}. 
Ref.~\cite{Tang:2014dia} also reported the detection of gamma-ray emission 
from NGC 2146. Being intrinsically faint but numerous, SFGs are expected to 
contribute significantly to the DGRB.

Massive stars in SFGs emit the majority of their light in the UV band. The
emission is then absorbed by interstellar dust and re-emitted as IR light. 
The IR luminosity can be used as a good tracer of the star-forming rate (SFR) 
$\psi$, i.e. the amount of mass converted in stars per unit time 
\cite{Kennicutt:1998zb}. The same massive stars finally explode into 
core-collapse supernovae, leaving behind supernova remnants which are 
considered to be the main sources of accelerated CRs on galactic scales 
\cite{Ginzburg:1964}. The leptonic component of these CRs is responsible for 
the synchrotron radio emission observed from SFGs and they may also contribute 
at gamma-ray frequencies when interacting with the interstellar radiation 
field of the host galaxy (through IC or bremsstrahlung). However, such 
leptonic gamma-ray emission is expected to be subdominant with respect to the 
hadronic one \cite{Chakraborty:2012sh}, since CR protons have an intrinsically 
larger injection rate than CR electrons \cite{Strong:2010pr,Lacki:2012si}. 
Hadronic gamma-ray emission mainly comes from inelastic interactions of CR 
protons with the nuclei of the interstellar medium, producing neutral pions 
that decay into gamma rays with an energy spectrum that peaks at around 
300-400 MeV. 

The so-called ``initial mass function'' determines the relative number of 
stars produced in any star-formation event in a galaxy, as a function of the
stellar mass \cite{Salpeter:1955it,Bastian:2010di}. Under the assumption of an 
universal high-mass end of the initial mass function, the SFR of a galaxy is 
proportional to the rate of supernova explosions and, thus, to the CR 
abundance and the CR-induced emission \cite{Thompson:2006qd,Lacki:2012si}. 
Then, it is expected that, in a generic galaxy, the gamma-ray and radio 
luminosities (both associated with CRs) are correlated with the IR emission,
which depends on the SFR.

The cumulative gamma-ray flux produced by IC of CR electrons in {\it all} SFGs 
at {\it all} redshifts is computed in Ref.~\cite{Chakraborty:2012sh}. In this 
work, the emission of a generic galaxy is modeled based on a template which 
is tuned to reproduce the emission of the MW. The cumulative gamma-ray flux 
depends on the abundance of SFGs. Ref.~\cite{Chakraborty:2012sh} assumes that, 
at a certain redshift, the total emission scales as the so-called cosmic SFR 
$\dot{\rho}_\star(z)$, i.e. the mass converted to stars per unit time and 
comoving volume:
\begin{equation}
\dot{\rho}_\star(z) = \int d\psi \, \psi \frac{dN}{dVd\psi},
\label{eqn:cosmic_SFR}
\end{equation}
where $dN/dVd\psi$ is the comoving density of galaxies per unit SFR. 
Parametric fits for $\dot{\rho}_\star(z)$ are available from the observation 
of the total luminosity density at various wavelengths \cite{Cole:2000ea,
Hopkins:2006bw,Horiuchi:2008jz}. Ref.~\cite{Chakraborty:2012sh} finds that the 
IC-induced emission is always subdominant with respect to the hadronic one. 
The two become comparable only above 100 GeV. 

Indeed, the majority of the studies on the SFGs focus on their hadronic 
emission and they neglect the contribution from primary electrons. The 
hadronic gamma-ray luminosity of a SFG, as a function of the observed energy 
$E_0$, can be written as follows:
\begin{equation}
L_\gamma(E_0) = 
\int \Gamma_{\pi^0 \rightarrow \gamma\gamma}(E_{\rm em}) \, n_{\rm H} \, dV =
\Gamma_{\pi^0 \rightarrow \gamma\gamma}(E_{\rm em}) \, N_{\rm H},
\label{eqn:gamma_ray_luminosity_SFG}
\end{equation}
where $E_{\rm em}$ is the energy in the emitter frame, 
$\Gamma_{\pi^0 \rightarrow \gamma\gamma}$ is the pionic gamma-ray production rate per 
interstellar hydrogen atom and $n_{\rm H}$ is density of hydrogen atoms in the 
interstellar medium \cite{Fields:2010bw,Stecker:2010di}. Integrated over the 
volume of the medium, $N_{\rm H}$ gives the total number of hydrogen atoms, 
which can be expressed in terms of the total interstellar gas mass $M_{\rm gas}$ 
in the galaxy as $X_{\rm H} M_{\rm gas}/m_p$. $X_{\rm H} \sim 0.7$ is the hydrogen 
mass fraction and $m_{\rm p}$ is the mass of the proton. As for 
$\Gamma_{\pi_0 \rightarrow \gamma\gamma}$ in Eq.~(\ref{eqn:gamma_ray_luminosity_SFG}), 
it is the product of the flux $\Phi_p$ of CR protons (averaged over the volume 
of the galaxy) and the cross section for the production of gamma rays 
\cite{Pohl:1994,Persic:2009nj}.

The distribution and propagation of CR protons is governed by the diffuse-loss 
equation, which depends on their injection rate, diffusion, energy losses and 
possible inelastic interactions (see Ref.~\cite{Lacki:2012si} for a recent 
review). Different scenarios are possible depending on the strength of those 
terms. Here we only mention two possibilities that provide good descriptions 
to different typologies of galaxies. The first is the so-called 
{\it escape regime} and it corresponds to a situation in which the energy 
losses of CR protons are dominated by their escape from the diffuse region of 
the galaxy. An equilibrium is reached between the energy losses and the 
acceleration of CR protons so that their flux is proportional to the product 
of the SFR of the galaxy and the CR escape path-length, $\Lambda_{\rm esc}$: 
$\Phi_p \propto \Lambda_{\rm esc} \psi$. Consequently, 
$L_\gamma \propto M_{\rm gas} \psi$ \cite{Pavlidou:2002va,Ando:2009nk,
Fields:2010bw}. The escape regime provides a good description of the diffuse 
foreground of the MW. This motivates the use of the gamma-ray production rate 
of the MW, $\Gamma_{\pi^0 \rightarrow \gamma\gamma}^{\rm MW}$, to normalize 
the proportionality relation between $L_\gamma$ and $M_{\rm gas} \psi$. Then
Eq.~(\ref{eqn:gamma_ray_luminosity_SFG}) becomes:
\begin{equation}
L_\gamma(\psi,M_{\rm gas}) = X_{\rm H} \Gamma_{\pi^0 \rightarrow \gamma\gamma}^{\rm MW}
\frac{M_{\rm gas}}{m_p} \frac{\psi}{\psi_{\rm MW}},
\label{eqn:luminosity_escape}
\end{equation}
where $\psi_{\rm MW}$ is the SFR of the MW. This relation provides a good 
description of the so-called {\it quiescent} or {\it normal} SFGs, 
characterized by properties similar to those of the MW.

The second scenario considered for the modeling of SFGs is the so-called 
{\it calorimetric regime}: in this case the energy losses of CR protons is
mainly due to their inelastic interactions. This means that protons lose all 
their energy into pions and SFGs act effectively as calorimeters. 
Their gamma-ray lumonisity, then, can be computed from the amount of energy 
available to CRs. Under the paradigm that supernova remnants are the primary 
source of CRs, the energy available in the form of CRs for a calorimetric SFG 
is proportional to the supernova rate, multiplied by the energy $E_{\rm SN}$ 
released per supernova and by the fraction $\eta$ of that energy going into CR 
protons (i.e. the so-called acceleration efficiency) \cite{Thompson:2006qd,
Ackermann:2012vca}. In turn, the supernova rate is proportional to the SFR, so 
that, finally,
\begin{equation}
L_\gamma(\psi) \propto \psi \, E_{\rm SN} \, \eta.
\label{eqn:luminosity_calorimeter}
\end{equation}
Starburst galaxies are well modeled as proton calorimeters: normally brighter 
and less numerous than quiescent galaxies, starburst ones are characterized by 
at least one region undergoing intense star formation. This is often induced 
by major merger events or by bar instabilities, leading to a large gas density 
\cite{Lacki:2010vs}.

Several works analyze quiescent and starburst SFGs separately. 
Refs.~\cite{Pavlidou:2002va,Ando:2009nk,Fields:2010bw}, e.g., focus on
quiescent galaxies. Their contribution to the DGRB is computed by using a 
formalism similar to Eq.~(\ref{eqn:unresolved_sources}), where $L_\gamma$ is
considered as the $Y$-parameter. Then, by means of 
Eq.~(\ref{eqn:luminosity_escape}), $L_\gamma$ is written as a function of 
$M_{\rm gas}$ and $\psi$. In Refs.~\cite{Pavlidou:2002va,Ando:2009nk}, the former 
is factorized out of the integral in $dL_\gamma$ in 
Eq.~(\ref{eqn:unresolved_sources}), assuming that $M_{\rm gas}$ only depends on 
redshift. The integration in $dL_\gamma$ now only depends on the SFR and it 
amounts to the cosmic SFR $\dot{\rho}_\star(z)$ in Eq.~(\ref{eqn:cosmic_SFR}). 
The net result is that the cumulative emission of quiescent SFGs scales with 
redshift as the product of $\dot{\rho}_\star(z)$ and the average amount of gas. 
Refs.~\cite{Pavlidou:2002va,Ando:2009nk} take $\dot{\rho}_\star(z)$ from 
Refs.~\cite{Cole:2000ea,Hopkins:2006bw}, respectively, and derive the average
gas mass assuming that the total baryonic mass of a galaxy (stars and gas) 
remains constant time and that it can be computed de-evolving backwards the 
cosmic history of the SFR density \cite{Pavlidou:2002va,Ando:2009nk}. They 
find that quiescent SFGs can contribute significantly to the DGRB, especially 
below 1 GeV, but they are only subdominant at higher energies.

\begin{figure}[h]
\begin{center}
\includegraphics[width=0.95\linewidth]{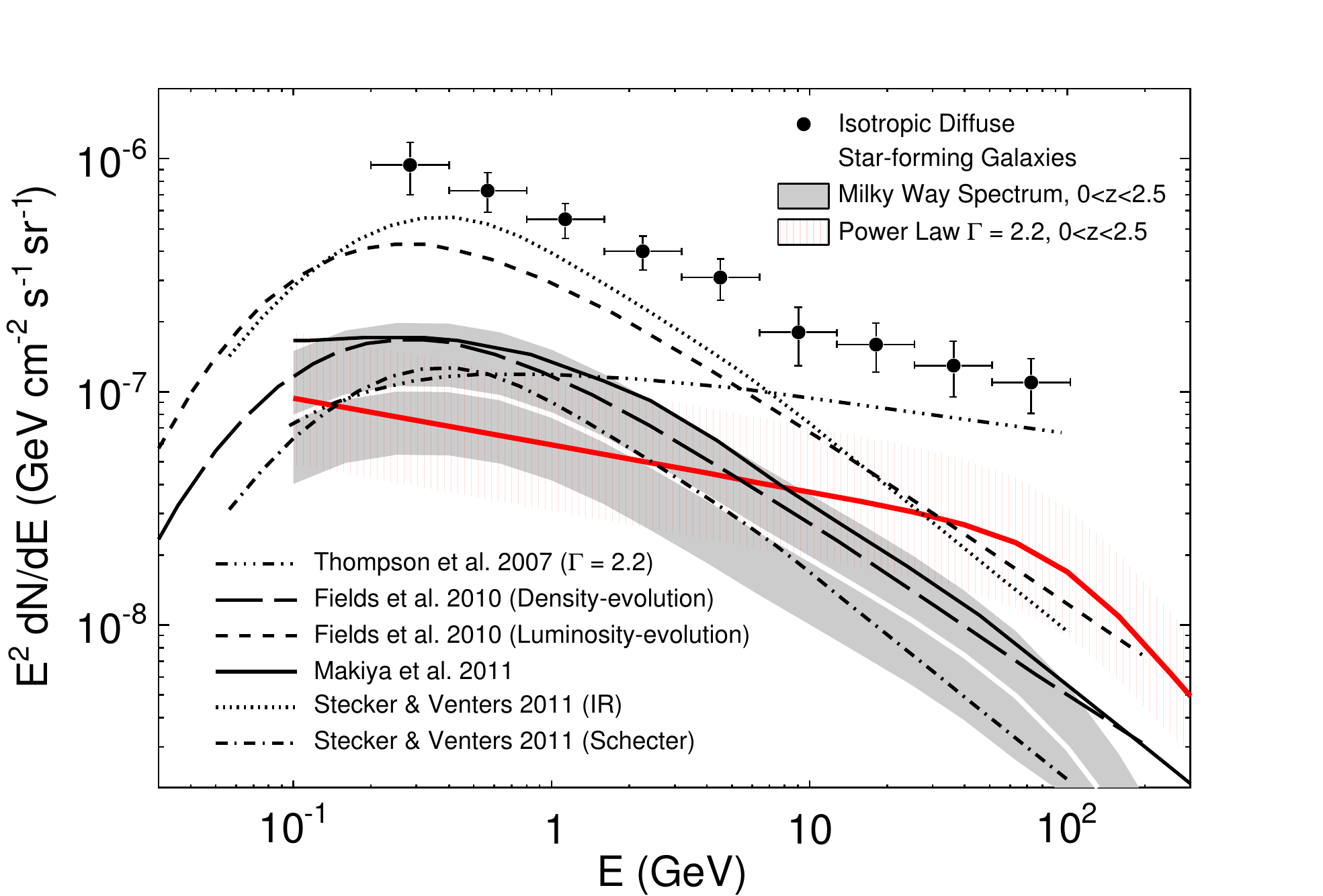}
\caption{\label{fig:Energy_spectrum_SFGs} Estimated contribution of unresolved SFGs (both quiescent and starburst) to the DGRB emission measured by the \fermi LAT (black points, taken from Ref.~\cite{Abdo:2010nz}). Two different spectral models are used: a power law with a photon index of 2.2 (red line), and a spectral shape based on a numerical model of the global gamma-ray emission of the MW \cite{Strong:2010pr} (white line, inside gray band). The shaded regions indicate combined statistical and systematic uncertainties. Several other estimates for the intensity of unresolved SFGs are shown for comparison. For the double-dotted-dashed line starburst galaxies are treated as calorimeters of CR nuclei as in Ref.~\cite{Thompson:2006qd}. Ref.~\cite{Fields:2010bw} (dashed and long-dashed line) considers the extreme cases of either pure luminosity and pure density evolutions. The solid black line shows predictions from Ref.~\cite{Makiya:2010zt}, obtained from simulated galaxies and semi-analytical models of galaxy formation. Two recent predictions from Ref.~\cite{Stecker:2010di} are plotted by dotted and dash-dotted black lines: the former assumes a scaling relation between IR and gamma-ray luminosities and the latter uses a redshift-evolving relation between the gas mass and the stellar mass of a galaxy. Taken from Ref.~\cite{Ackermann:2012vca}.}
\end{center}
\end{figure}

Ref.~\cite{Fields:2010bw} assumes the same formalism than
Ref.~\cite{Pavlidou:2002va} but it relates the gas content of a galaxy to its 
SFR, using the following relation:
\begin{equation}
M_{\rm gas}(\psi,z) = 2.8 \times 10^9 M_\odot \, (1+z)^{-0.571} 
\left( \frac{\psi}{M_\odot \mbox{yr}^{-1}} \right)^{0.714}.
\label{eqn:gas_mass}
\end{equation}
This is obtained by the so-called Kennicutt-Schmidt law \cite{Schmidt:1959pe,
Kennicutt:1997dm}, which links the surface density of star formation to that
of the gas. Then, the integration over $L_\gamma$ in 
Eq.~(\ref{eqn:unresolved_sources}) is converted into one in SFR. In turn,
the SFR is related to the IR luminosity, $L_{\rm IR}$, assuming a direct 
proportionality between the two \cite{Hopkins:2004ma}. The contribution of 
SFGs now depends on the IR LF, $\Phi_{\rm IR}$. Ref.~\cite{Fields:2010bw} 
considers the IR LF from Ref.~\cite{Nakamura:2003gn}. Results are provided for 
both a pure luminosity evolution and a pure density evolution for 
$\Phi_{\rm IR}$. The difference between the two prescriptions amounts to 
approximately a factor of 2 (see the dashed and long-dashed lines in 
Fig.~\ref{fig:Energy_spectrum_SFGs}). An intermediate scenario that mediates 
between the two evolution schemes indicates that unresolved SFGs account for 
$\sim 50\%$ of the DGRB intensity reported in Ref.~\cite{Abdo:2010nz} below 
10 GeV. Above that energy, the contribution of SFGs goes down rapidly due to 
the softness of the energy spectrum assumed, i.e. a power law with a slope of 
2.75.

Ref.~\cite{Stecker:2010di} proposes two alternatives to estimate $M_{\rm gas}$: 
in the first one, the gas mass is not obtained from Eq.~(\ref{eqn:gas_mass}) 
but taken to be some fraction of the stellar mass $M_\star$. The latter can be 
constrained by combining the findings of Refs.~\cite{Papovich:2010zx,
Elsner:2007}. In the second scenario, the mean gas mass is assumed to scale 
with the cosmic SFR divided by a parameter linking the SFR to the density of 
clouds of molecular hydrogen \cite{Leroy:2008kh,Stecker:2010di}. The two 
methods differ by approximately a factor of 10 in their predictions for the 
contribution of SFGs to the DGRB. In the second one, SFGs are able to account 
for the whole DGRB of Ref.~\cite{Abdo:2010nz} between 200 MeV and 1 GeV.

Starburst SFGs have also been considered: Ref.~\cite{Thompson:2006qd} works
under the hypothesis that these objects are in the calorimetric regime. From
Eq.~(\ref{eqn:luminosity_calorimeter}), their gamma-ray luminosity is 
proportional to the SFR which, in Ref.~\cite{Thompson:2006qd}, is related to 
the IR luminosity. Thus, the total gamma-ray emission from unresolved starburst 
SFGs can be obtained simply by rescaling the total diffuse extragalactic IR 
background taken, e.g., from Ref.~\cite{Nagamine:2006kz}. Unresolved starburst 
galaxies are found to account for $\sim 10\%$ of the DGRB detected by EGRET in 
Ref.~\cite{Strong:2004ry}, at least at the GeV scale (see the 
double-dotted-dashed line in Fig.~\ref{fig:Energy_spectrum_SFGs}). The main 
uncertainty affecting this prediction stems from the unknown starburst 
fraction, i.e. the fraction of the IR background emission associated with 
starburst galaxies in the calorimetric regime. Observations suggests that such 
a fraction is quite low at $z=0$ but that it can approximate to 1 at earlier 
epochs, when the star formation is much more efficient \cite{Yun:2001jx,
Dole:2006de}. However, see also Refs.~\cite{Stecker:2006vz,Stecker:2010di,
Hopkins:2009cw,Lacki:2012si} for different values.

Both quiescent and starburst SFGs are modeled at the same time in 
Ref.~\cite{Makiya:2010zt}. The relations between $L_\gamma$ and $\psi M_{\rm gas}$ 
and between $L_\gamma$ and $\psi$ for quiescent and starburst SFGs 
respectively (see Eqs.~(\ref{eqn:luminosity_escape}) and 
(\ref{eqn:luminosity_calorimeter})) are determined by fitting the results of 4 
SFGs detected by the \fermi LAT. On the other hand, the abundance of galaxies 
is inferred from simulated galaxy catalogs \cite{Nagashima:2004gr,
Nagashima:2005jc}. Starburst SFGs are found to be always subdominant and the 
total (quiescent {\it and} starburst) gamma-ray emission is between 5.4\% and 
9.6\% of the DGRB intensity reported by the \fermi LAT in 
Ref.~\cite{Abdo:2010nz} above 100 MeV (see the black solid line in 
Fig.~\ref{fig:Energy_spectrum_SFGs}).

In Ref.~\cite{Lacki:2012si} the authors model the emission from the MW and 
from M82, which are considered as templates of quiescent and starburst SFGs, 
respectively. The results are used to determine $L_\gamma$ for the two 
typologies of SFGs. Their total emission is assumed to evolve with redshift 
following $\dot{\rho}_\star(z)$, modulated by functions $f_i(z)$ (where $i$ 
stands for either ``starburst'' or ``quiescent'') that fix the relative 
abundance of the two sub-classes. They find that SFGs can be responsible from 
4\% to 76\% of the DGRB of Ref.~\cite{Abdo:2010nz} in the GeV range and that 
their contribution cannot reproduce the data below the GeV or above few tens 
GeV. In their fiducial model quiescent SFGs always dominate over starbursts.

Ref.~\cite{Ackermann:2012vca} follows an alternative approach to compute the
emission of unresolved SFGs. The authors consider the 8 SFGs detected by the 
\fermi LAT and 64 galaxies observed in radio and IR but for which only upper 
limits are available in the gamma-ray energy range. These are used to 
determine a correlation between $\log_{10}(L_\gamma)$ (defined between 0.1 and 
100 GeV) and the IR luminosity $\log_{10}(L_{\rm IR})$ (defined between 8 and 
1000 $\mu$m). Similarly to what done for blazars and MAGNs, this correlation 
is used to infer the properties of SFGs in the gamma-ray band from the study 
performed at IR frequencies. In particular, it is assumed that the gamma-ray 
LF can be written in terms of $\Phi_{\rm IR}$, as 
$\Phi_\gamma(z,L_\gamma) = \Phi_{\rm IR} d\log_{10}(L_{\rm IR}) / d\log_{10}(L_\gamma)$.

\begin{figure}[h]
\begin{center}
\includegraphics[width=0.95\linewidth]{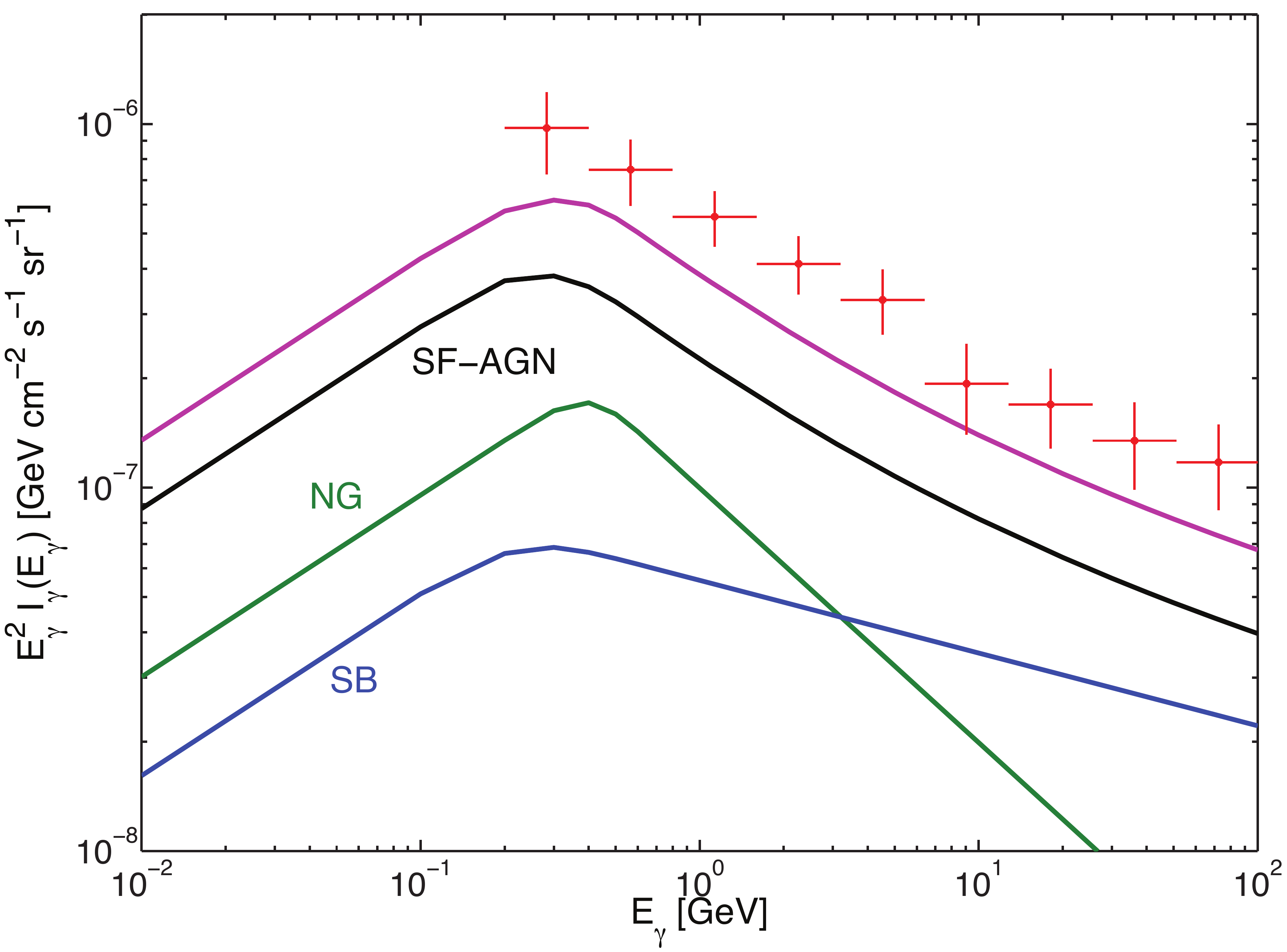}
\caption{\label{fig:Energy_spectrum_SFGs_Tamborra} Diffuse gamma-ray emission as a function of the energy (without EBL attenuation) for the three different SFG contributions to the DGRB: quiescent SFGs (green, labeled ``NG'' in the figure), starburst SFGs (blue) and SFGs hosting an AGN (black). The solid pink indicates their sum. The DGRB from Ref.~\cite{Abdo:2010nz} is plotted in red. Taken from Ref.~\cite{Tamborra:2014xia}.}
\end{center}
\end{figure}

The contribution to the DGRB, then, is computed following 
Eq.~(\ref{eqn:unresolved_sources}), with $Y=L_\gamma$.\footnote{As done for 
MAGNs, gamma-ray and IR LF are defined here per unit of $\log(L_\gamma)$ and of 
$\log(L_{\rm IR})$, respectively.} The IR LF is measured in 
Ref.~\cite{Rodighiero:2010} from data gathered by the Spitzer Space Telescope. 
In Ref.~\cite{Ackermann:2012vca} only objects below $z=2.5$ are considered, 
since $\Phi_{\rm IR}$ is not well determined for higher redshifts. The 
$L_{\rm IR}-L_\gamma$ correlation is assumed to stand valid up to that redshift, 
even if only SFGs below $z \sim 0.05$ are used in 
Ref.~\cite{Ackermann:2012vca} to derive it. Starburst galaxies are observed to 
have a harder energy spectrum than quiescent ones 
\cite{Acciari:2009wq,Abdo:2009aa}, at least in the Local Group. Assuming that 
this is a general property, the steepness of the contribution of unresolved 
SFGs to the DGRB will depend on the fraction of starburst galaxies, which is 
unknown. Ref.~\cite{Ackermann:2012vca} assumes that all SFGs share the same 
energy spectrum and it considers two extreme cases: a power-law with a slope 
of 2.2 (typical of starburst galaxies) and the spectrum of the diffuse 
foreground of the MW \cite{Strong:2010pr}, which reproduces well the behavior 
of quiescent SFGs. Results are summarized in 
Fig.~\ref{fig:Energy_spectrum_SFGs} by the red and gray bands, respectively.
The figures show that unresolved SFGs can explain between 4\% and 23\% of the 
DGRB measured by the \fermi LAT in Ref.~\cite{Abdo:2010nz} above 100 MeV.

The results of Ref.~\cite{Ackermann:2012vca} have been updated in 
Ref.~\cite{Tamborra:2014xia}, by improving the modeling of $\Phi_{\rm IR}$. 
Thanks to the recent detection of a larger number of high-$z$ sources by the 
Herschel Space Observatory \cite{Gruppioni:2013jna,Devlin:2009qn,
Bethermin:2012jd,Barger:2012st}, it is now possible to extend the LF up to
$z \simeq 4$ and to consider separately the contribution of different classes
of SFGs. In particular, Herschel classifies its sources into quiescent 
galaxies, starburst ones and SFGs hosting an obscured or low-luminosity 
AGN.\footnote{We note that, even if these SFGs host an AGN, their IR 
luminosities are expected to be dominated by the star-forming activity rather 
than the one associated with the AGN and, therefore, $L_{\rm IR}$ remains a
useful tracer of the SFR.} At $z=0$, starburst SFGs are found to be 
subdominant with respect to the other two classes, but the three families have 
different evolution and, by $z=1$, the number of AGN-hosting SFGs is 
approximately twice the number of quiescent and starburst galaxies combined. 
Ref.~\cite{Gruppioni:2013jna} provides analytical fits to the LF of the three 
classes. From these, the emission of unresolved SFGs can be computed, assuming 
a broken power law as energy spectrum.\footnote{The low-energy slope is fixed 
to -1.5 in all cases, while the high-energy one is assumed to be -2.7 for 
quiescent galaxies and -2.2 for starburst ones. The class of AGN-hosting 
objects is additionally split into ``starburst-like'' and ``quiescent-like'' 
galaxies, with slopes of -2.2 and -2.7, respectively.} The results are 
summarized in Fig.~\ref{fig:Energy_spectrum_SFGs_Tamborra} and they show that 
unresolved SFGs are responsible for $\sim 50\%$ of the \fermi LAT DGRB of 
Ref. \cite{Abdo:2010nz} in the range between 0.3 and 30 GeV. Note that the 
emission is dominated by SFGs hosting AGNs.

\subsubsection{Millisecond pulsars}
\label{sec:MSPs}
Pulsars are highly magnetized and rapidly spinning neutron stars, with a
beam of radiation that periodically intersects the Earth. Their initial spin
$P$ decreases due to magnetic dipole braking \cite{Ng:2014qja}, so that the
time derivative of the period $\dot{P}$ can be written as follows 
\cite{FaucherGiguere:2009df,SiegalGaskins:2010mp,Gregoire:2013yta,
TheFermi-LAT:2013ssa,Calore:2014oga}:
\begin{equation}
\dot{P} = 9.8 \times 10^{-26} \left( \frac{B}{\mbox{G}} \right)^2 
\left( \frac{P}{\mbox{s}} \right)^{-1},
\end{equation}
where $B$ is the surface magnetic field. The loss of kinetic energy associated 
with the slowing down of the spinning, i.e. the so-called ``spin-down 
luminosity'' $\dot{E}$, is
\begin{equation}
\dot{E} = 4\pi^2 M \frac{\dot{P}}{P^3},
\label{eqn:spin_down_luminosity}
\end{equation}
where $M$ is the moment of inertia of the neutron star. The spin-down 
luminosity is converted, with some efficiency, into radiation. MSPs were
traditionally detected in radio, while, nowadays, thanks to the \fermi LAT, an
increasingly large number of objects is observed also in the gamma-ray band.
The shape of their gamma-ray energy spectra suggests that the emission comes 
from curvature radiation. This is a mechanism similar to synchrotron 
radiation, in which gamma rays are produced by relativistic charged particles 
following the curved force lines of a magnetic field \cite{Benford:1977,
Gangadhara:2010,Wang:2012ft}. Self-synchrotron Compton possibly also 
contributes \cite{Kerr:2012xh}.

Pulsars are classified in terms of their period and sources with $P<15$ ms
are referred to as MSPs. These are generally part of a binary system and
their higher spin is the result of the large angular momentum transferred 
from the companion object \cite{Alpar:1982,Phinney:1994gf,Lorimer:2001vd,
Lorimer:2008se,Kiziltan:2009rx}. Thanks to lower surface magnetic fields, MSPs 
have smaller $\dot{P}$ and are, therefore, older than ``normal'' (i.e., young) 
pulsars. A longer life cycle may compensate the intrinsic lower birthrate so 
that MSPs are expected to be as abundant as normal pulsars \cite{Lyne:1998}. 
Moreover, having had the time to orbit many times around the Galaxy, the 
distribution of MSPs is expected to be uncorrelated with their birth locations, 
potentially extending to high Galactic latitudes 
\cite{FaucherGiguere:2009df}.

The conversion of energy loss $\dot{E}$ into $L_\gamma$ is parametrized by an 
empirical relation of the form 
\begin{equation}
L_\gamma = \eta \dot{E}^\alpha
\label{eqn:L_gamma_E}
\end{equation}
\cite{Arons:1996}, where $\eta$ is the called ``conversion efficiency''. 
The case of $\alpha=0.5$ is motivated in Refs.~\cite{Ruderman:1975ju,
Harding:1981,Arons:1996} and adopted in Refs.~\cite{FaucherGiguere:2005ny,
SiegalGaskins:2010mp,Gregoire:2013yta}, even if some of those references also 
consider the case of $\alpha=1$. Assuming that Eq.~(\ref{eqn:L_gamma_E}) is 
valid both for young pulsars and for MSPs, their luminosities are related by
\begin{equation}
\frac{L_\gamma^{\rm MSP}}{L_\gamma^{\rm young}} = 
\frac{\dot{P}_{\rm MSP}}{\dot{P}_{\rm young}} 
\left( \frac{P_{\rm MSP}}{P_{\rm young}} \right)^{-3},
\end{equation}
for $\alpha=1$. Typical values are $P_{\rm MSP}=3$ ms, $P_{\rm young}=0.5$ ms, 
$\dot{P}_{\rm MSP}=10^{-19}$ and $\dot{P}_{\rm young}=10^{-15}$. These suggest that 
MSPs can be brighter than normal pulsars, by a factor of few tens. It is, 
therefore, reasonable to speculate that these sources can contribute 
significantly to the DGRB emission \cite{FaucherGiguere:2009df,
SiegalGaskins:2010mp,Lorimer:2012hy,Gregoire:2013yta,Calore:2014oga}. On the 
other hand, the cumulative emission of normal pulsars would be quite 
anisotropic and mainly localized along the Galactic plane, thus hardly 
compatible with the DGRB \cite{McLaughlin:1999uv,FaucherGiguere:2005ny}.

\begin{figure}[h]
\begin{center}
\includegraphics[width=0.95\linewidth]{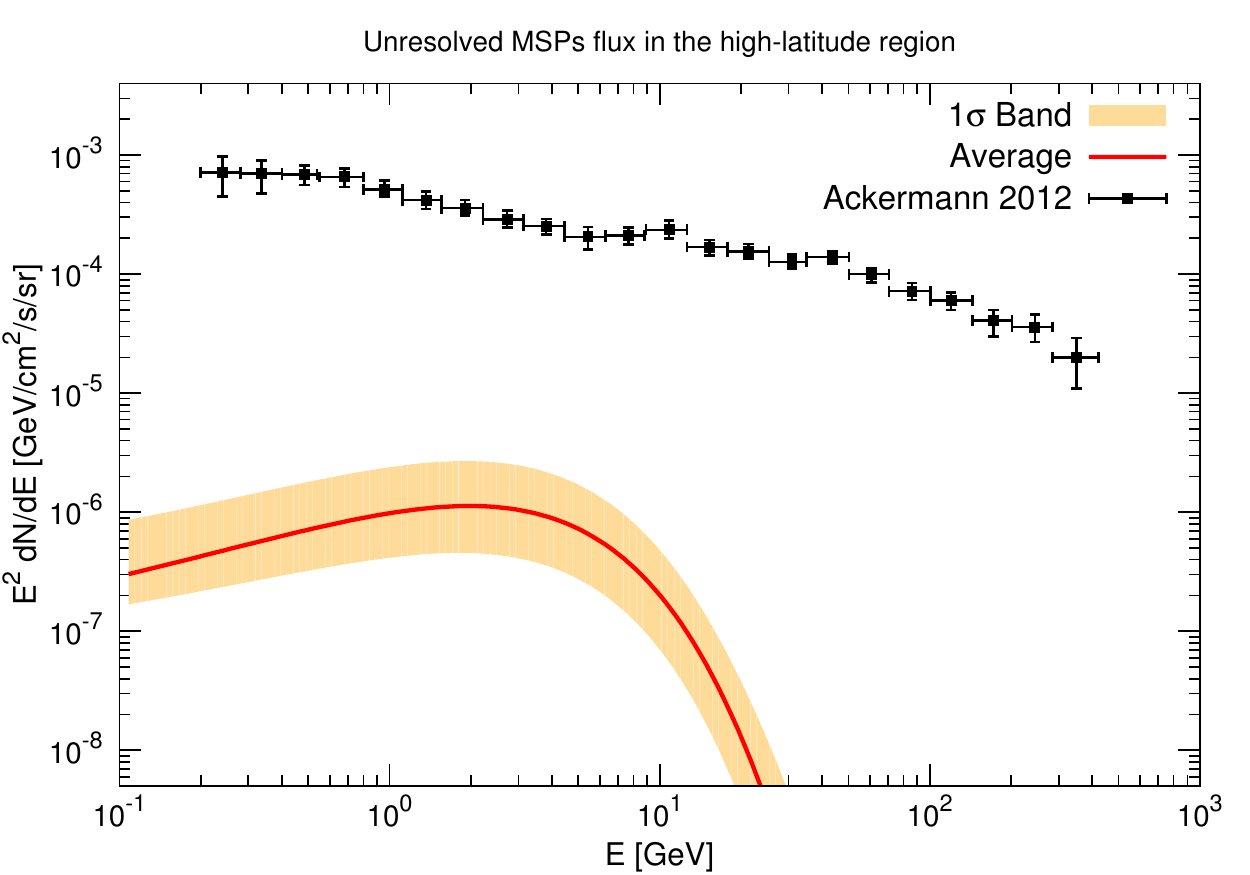}
\caption{\label{fig:Energy_spectrum_MSPs} Prediction for the contribution of unresolved MSPs to the DGRB, as derived from 1000 Monte Carlo simulations of the MSP population of the MW. The red solid line represents the mean spectrum distribution (see Ref. \cite{Calore:2014oga} for further details), while the orange band corresponds to its $1\sigma$ uncertainty band. The black points refer to the preliminary DGRB measurement anticipated in Ref.~\cite{Ackermann:2012}. Taken from Ref.~\cite{Calore:2014oga}.}
\end{center}
\end{figure}

The second \fermi LAT catalog of gamma-ray pulsars (2FPC) contains 117 
sources, 40 of which are MSPs \cite{TheFermi-LAT:2013ssa}. The number is too 
low to build a MSP LF only from \fermi LAT data. Also, given the uncertainties 
on the mechanisms of gamma-ray emission, it is not possible to postulate 
correlations with other frequencies, as done for blazars, MAGNs and SFGs. This 
is the reason why Refs.~\cite{FaucherGiguere:2009df,SiegalGaskins:2010mp,
Gregoire:2013yta,Calore:2014oga} rely on Monte Carlo simulations in order to 
properly describe the population of MSPs. Probability distributions of the 
most relevant quantities (e.g., radial and vertical position of the source, its 
period and surface magnetic field) are derived from the pulsars detected in 
radio \cite{Cordes:1997my,Lyne:1998,Story:2007xy}. At date, the largest 
catalog of pulsars is the Australia Telescope National Facility Pulsar 
catalog, containing 1509 sources, out of which 132 are MSPs 
\cite{Manchester:2004bp}. Ref.~\cite{Calore:2014oga} analyses the objects in 
the catalog and establishes that they are well described by the following 
prescriptions: $i)$ a Gaussian distribution for the radial distance, $R$, from 
the center of the MW projected on the Galactic plane, i.e. 
$dN/dR \propto \exp[- (R - \langle R \rangle) / R_0]$, $ii)$ an exponential 
distribution for the vertical distance, $z$, from the Galactic plane, i.e.
$dN/dz \propto \exp(-z/z_0)$ \cite{Cordes:1997my,FaucherGiguere:2005ny}, $iii)$ 
a log-Gaussian distribution for the period $P$ (in contrast to previous works 
\cite{FaucherGiguere:2009df,SiegalGaskins:2010mp,Gregoire:2013yta}) and $iv)$ 
a log-Gaussian distribution for the magnetic field.

Ref.~\cite{Calore:2014oga} also studies the relation between $L_\gamma$ and 
$\dot{E}$. The data, however, are affected by quite large errors which prevent
a statistically meaningful fit to the data. Thus, Ref.~\cite{Calore:2014oga} 
identifies a benchmark case with $\alpha=1$ that provides a qualitative good 
description of the data, and an uncertainty band that encompasses reasonably 
well the distribution of the data points. This completes the characterization 
of the MSP population and synthetic sources can be generated by randomly 
drawing from the assumed distributions. Each source is labeled as ``resolved'' 
or ``unresolved'' depending on whether its flux is larger or smaller than the 
sensitivity of the telescope at the position of the source in the sky. 
Simulated sources are accumulated until the number of ``resolved'' ones equals 
the amount of detected MSPs: Refs.~\cite{FaucherGiguere:2005ny,
SiegalGaskins:2010mp} consider the MSPs detected in radio in the Australia 
Telescope National Facility catalog \cite{Manchester:1996}, while 
Refs.~\cite{Gregoire:2013yta,Calore:2014oga} the sources detected by the 
\fermi LAT.\footnote{In this case the sensitivity of the \fermi LAT is taken 
from Ref.~\cite{TheFermi-LAT:2013ssa}.} This calibrates the Monte Carlo data 
and it allows the computation of the MSP contribution to the DGRB simply by 
summing over the ``unresolved'' sources. Their energy spectrum is fixed to a 
power law with an exponential cut-off, i.e a functional shape that reproduces 
well the MSPs in the 2FPC. The slopes and cut-off energies are assumed to have 
Gaussian distributions. The results of Ref.~\cite{Calore:2014oga} are shown in 
Fig.~\ref{fig:Energy_spectrum_MSPs}, from which it is evident that MSPs are 
only a subdominant component to the DGRB, responsible for less than 1\% of the 
intensity measured by the \fermi LAT in Ref.~\cite{Abdo:2010nz}. This value is 
lower than the previous estimates from Refs.~\cite{SiegalGaskins:2010mp,
Gregoire:2013yta}, in which a different modeling of the MSP population was 
adopted.

\subsubsection{Other astrophysical components}
\label{sec:other_astrophysics}
Several other potential contributors to the DGRB have been proposed in the 
past. Some correspond to unresolved astrophysical populations not considered 
in the previous sections, while others are intrinsically diffuse processes. 
In this section, we focus on astrophysical scenarios, while a potential 
DM-induced emission will be discussed in detail in Sec. \ref{sec:DM}.

\begin{itemize}
\item {\bfseries Clusters of galaxies:} it is believed that huge amounts of 
energy, of the order of $10^{61}-10^{63}$ erg, is dissipated in the shocks
associated with the assembly of clusters of galaxies \cite{Voit:2005bq,
vanWeeren:2011}. A fraction of this energy can go into the acceleration of 
CRs, even if the details of the acceleration mechanisms are still uncertain 
\cite{Pinzke:2013}. Accelerated CRs would produce gamma rays by means of $i)$
the decay of pions produced by the interaction of CR protons with the 
intracluster medium and $ii)$ IC scattering of primary CR electrons or of the 
electrons produced by the pion decays in point $i)$. Yet, no gamma-ray 
emission has been detected from galaxy clusters. On the other hand, the 
interpretation of the observed radio emission as synchrotron radiation 
confirmed the presence of accelerated electrons and magnetic fields 
\cite{Deis:1997,Brown:2011}. However, not all known clusters emit in radio 
\cite{Brunetti:2009mu,Cassano:2013nza} and it is not clear why some objects 
are radio-quiet. 

The IC-induced gamma-ray emission \cite{Loeb:2000na,Totani:2000rg} depends on 
the abundance of CR electrons, on the intensity of the magnetic field and on 
the so-called acceleration efficiency $\xi_e$, i.e. the fraction of the 
thermal energy density produced by the shocks that is transferred to CR 
electrons. A Halo Model \cite{Press:1973iz,Cooray:2002dia}, linking the 
properties of the shock to those of the DM halo hosting the cluster 
\cite{Keshet:2002sw,Pfrommer:2007}, is often used to predict the IC-induced 
emission of clusters. Results can be tested with $N$-body simulations. An 
acceleration efficiency of $\xi_e=0.05$ is typically assumed, following similar 
values measured in shocks surrounding supernova remnants. For this $\xi_e$, 
Ref.~\cite{Keshet:2002sw} finds that the cumulative IC-induced gamma-ray flux 
from unresolved clusters can explain, at most, 10\% of the EGRET DGRB in 
Ref.~\cite{Sreekumar:1997un}. Similar values are obtained by 
Refs.~\cite{Gabici:2002fg,Miniati:2002hs,Gabici:2003kr,Ando:2006mt,
Kashiyama:2014rza}. Yet, the non-detection of gamma rays from the observation 
of the Coma galaxy cluster suggests even lower efficiencies ($\xi_e < 1\%$) 
\cite{Zandanel:2013wea}, which would further decrease the contribution of this 
source class to the DGRB.

On the other hand, the gamma-ray emission expected from pion decay can be
estimated as a function of the amount of gas in the intracluster medium, of
the injected spectrum of CR protons and of the intensity of the magnetic 
fields \cite{Colafrancesco:1998us}. Ref.~\cite{Ando:2007yw} estimates that 
this hadronic gamma-ray emission can only account for less than few percents of 
the DGRB reported by EGRET in Ref.~\cite{Sreekumar:1997un}. This result was 
later reduced to less than 1\% of the \fermi LAT DGRB in 
Ref.~\cite{Abdo:2010nz}, after that mock catalogs of clusters from 
Ref.~\cite{Zandanel:2013wja} were considered in Ref.~\cite{Zandanel:2013wea}.

The most recent predictions for the emission of unresolved clusters come from 
Ref.~\cite{Zandanel:2014pva}. In this paper, a correlation between the mass of 
the cluster and its gamma-ray luminosity is assumed and it is calibrated to 
reproduce the number of clusters detected in radio during the Radio Astronomy 
Observatory Very Large Array sky survey \cite{Giovannini:1999fa}. It is also 
required that results are compatible with the non-detection of the Coma
cluster by the \fermi LAT \cite{Zandanel:2013wea} and of the Perseus cluster by 
MAGIC \cite{Aleksic:2011cp}. A second, more physically motivated, model for 
the distribution of CRs and of the intracluster medium distribution is also 
considered in Ref.~\cite{Zandanel:2014pva}. In this case, the intracluster 
medium is reconstructed from X-ray observations and the CR spatial and spectral 
distributions are based on hydrodynamic $N$-body simulations 
\cite{Zandanel:2013wja}. The two scenarios agree in finding that the 
contribution of galaxy clusters to the DGRB is negligible.

\item {\bf Interactions of ultra-high-energy cosmic rays with background
radiation:} UHECRs, with energies larger than $6 \times 10^{19}$ eV are 
attenuated due to their pion-producing interactions with the CMB, i.e. the 
so-called Greisen-Zatsepin-Kuzmin cut-off 
\cite{Greisen:1966jv,Zatsepin:1966jv}). Pions trigger electromagnetic 
cascades, effectively transferring energy from the CRs to the GeV-TeV energy 
range. This can potentially contribute to the DGRB \cite{Coppi:1996ze,
Kalashev:2007sn,Ahlers:2011sd,Cholis:2013ena}. Ref.~\cite{Kalashev:2007sn} 
estimates this emission, showing that it is subject to large uncertainties. 
Indeed, the signal strongly depends on the evolution of the UHECRs with 
redshift, on their composition and on the intensity of the magnetic fields 
encountered during the propagation of the CRs. The uncertainty on the 
intensity of the emission spans more than two orders of magnitude below 10 
GeV. The most optimistic prediction indicates that UHECRs can indeed represent 
a significant contribution to the EGRET DGRB in Refs.~\cite{Sreekumar:1997un,
Strong:2004ry}. However, above 10 GeV, the intensity of the signal reduce 
significantly so that it would not be able to explain the bulk of the sub-TeV 
DGRB detected by the \fermi LAT.  

\item {\bf Type Ia supernovae:} Type Ia supernovae are generated in the 
thermonuclear explosions of white dwarfs near the Chandrasekhar mass 
\cite{Whelan:1973,Iben:1984iz}. Gamma-ray emission results from the decay of 
the material (mainly $^{56}$Ni) produced during the detonation. 
Refs.~\cite{Ahn:2005ws,Strigari:2005hu,Horiuchi:2010kq} compute the cumulative 
emission associated to this class of supernovae as a function of their event 
rate. The latter is either measured directly or related to the cosmic SFR, 
$\dot{\rho}_\star(z)$, through a model of the delay time, i.e. the time 
required for a Type Ia supernova progenitor to become a supernova. The 
cumulative gamma-ray emission can contribute significantly to the DGRB only 
around the MeV scale (see also Ref.~\cite{Rasera:2005sa}). 

On the other hand, Ref.~\cite{Lien:2012gz} computes the emission associated 
with the decay of neutral pions produced in the interactions of CR protons with 
the interstellar medium of the galaxy hosting the supernova. This is a 
scenario very similar to the one described in Sec.~\ref{sec:SFGs} for SFGs. 
However, in Ref.~\cite{Lien:2012gz}, the authors consider CRs accelerated in 
shocks induced by the explosions of Type Ia supernovae and not of the 
core-collapse supernovae, as done in Sec.~\ref{sec:SFGs}. The predictions of
Ref.~\cite{Lien:2012gz} are affected by a significant uncertainty and they 
span a range between less than 10\% and 100\% of the DGRB reported by the 
\fermi LAT in Ref.~\cite{Abdo:2010nz} between 1-10 GeV.

\item {\bf Gamma-Ray Bursts:} gamma-ray bursts are very short and intense 
episodes of beamed gamma-ray emission with a bimodal SED \cite{Baring:1997gs,
Lithwick:2000kh,Zhang:2005fa}. As in the case of AGNs, the low- and 
high-frequency peak are associated, respectively, with synchrotron radiation 
and IC. The widely-accepted fireball internal-external shock model 
\cite{Meszaros:1999xf} allows to describe the phenomenology of the material 
inside the bursts and to predict the SED of the bursts. 
Refs.~\cite{Casanova:2008zz,Le:2009} estimate the total contribution of 
unresolved gamma-ray bursts to the DGRB by adopting the LF given in 
Ref.~\cite{Schmidt:1999iw}. Results suggest that this emission can explain 
only a small fraction of the EGRET DGRB in Ref.~\cite{Sreekumar:1997un}, 
becoming negligible at energies above $\sim 40$ GeV. Similar results have
been obtained in Ref.~\cite{Ando:2008pj} which estimates the contribution of 
unresolved gamma-ray bursts to be as large as 0.1\% of the EGRET DGRB in 
Ref.~\cite{Sreekumar:1997un} at the GeV scale.

\item {\bf Small Solar-system bodies:} our knowledge of comets populating the 
Oort Cloud is quite limited and it comes almost entirely from numerical 
simulations. We believe that more than $10^{12}$ objects, with sizes ranging 
from 1 to 50 km, may populate that region. These small Solar-system bodies 
emit gamma rays from the hadronic interactions of CRs impinging on them. Their 
abundance is the main unknown, with column densities of Solar-system bodies 
spanning over three orders of magnitude. As a consequence, a similar level of 
uncertainty affects the predictions for their cumulative gamma-ray emission. 
Their contribution to the DGRB may go from overpredicting the DGRB measured in 
Ref.~\cite{Strong:2004ry} to being responsible for only few percents of it. 
See Ref.~\cite{Moskalenko:2009tv} and references therein.

\item {\bf Radio-quiet AGNs:} AGNs at sub-Eddington luminosities are 
characterized by radiatively inefficient accretions. Since the jet is not 
beamed enough to trigger non-thermal emission, the source emits mainly between 
IR and X-ray frequencies. Gamma rays can still be produced: one possibility 
is to have proton-proton interactions producing neutral pions in the hot gas 
surrounding the supermassive Black Hole. The intensity of the signal depends 
on the spin of the Black Hole, since more rapidly rotating objects correspond 
to larger X-to-gamma flux ratio \cite{Mahadevan:1997,Oka:2003,Teng:2011fz}. 
Another possibility is a population of non-thermal electrons that can be 
accelerated in the hot corona around the AGN \cite{Inoue:2007tn,Inoue:2008pk}. 
The similarities between solar coronae and accretion disks 
\cite{Galeev:1979td} suggest that magnetic reconnection could be responsible 
for this acceleration \cite{Liu:2002}. The scenario is considered in 
Refs.~\cite{Inoue:2007tn,Inoue:2008pk} where, using a 
luminosity-dependent-density evolution, the authors determine that this 
contribution can explain the whole DGRB measured by EGRET in 
Refs.~\cite{Sreekumar:1997un,Strong:2004ry}, but only below 1 GeV. 

\begin{figure}[h]
\begin{center}
\includegraphics[width=0.95\linewidth]{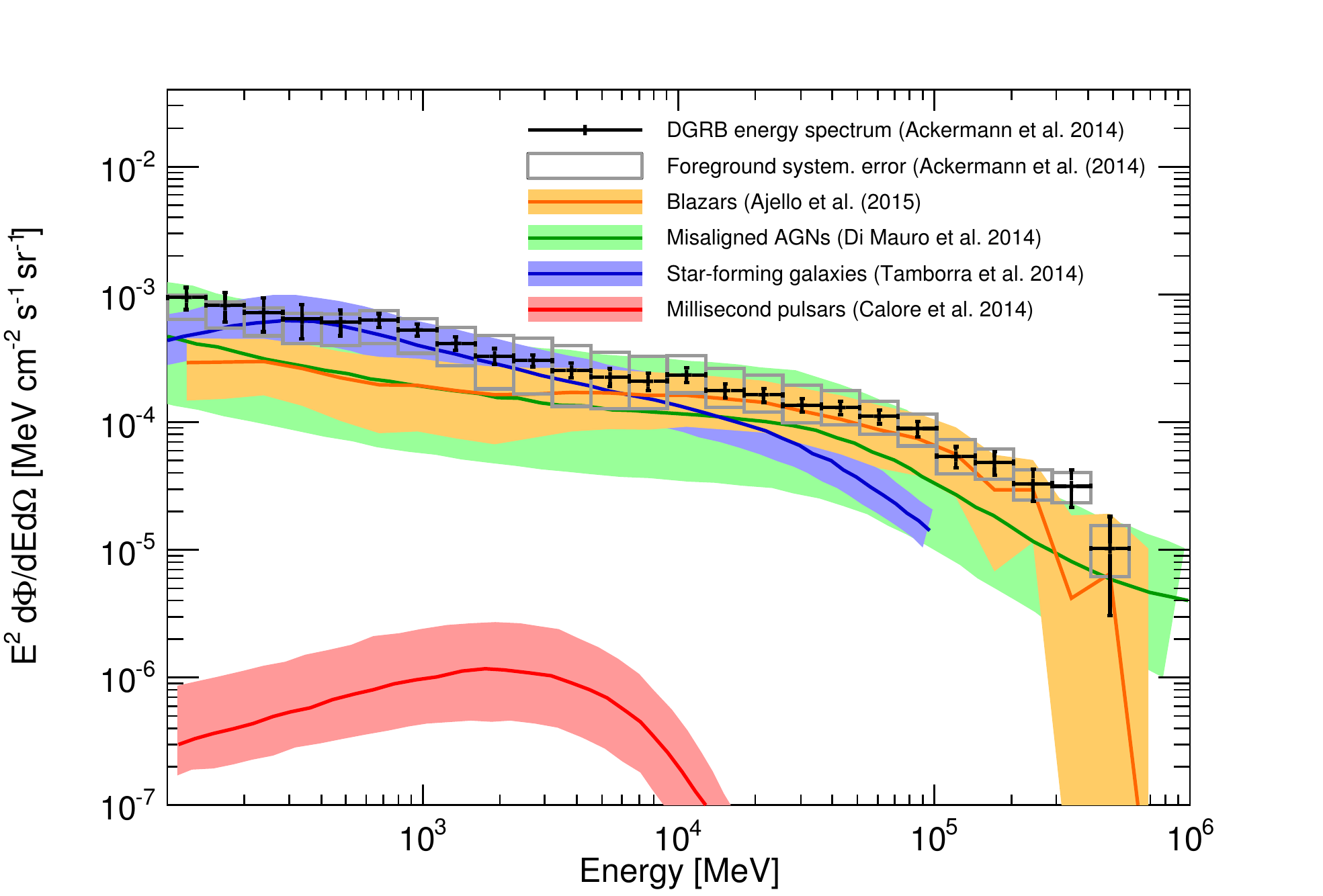}
\caption{\label{fig:summary_astrophysics} The energy spectrum of the DGRB (black points) as recently measured by the \fermi LAT \cite{Ackermann:2014usa}. Gray boxes around each data point denote the uncertainty associated with the Galactic diffuse emission. The solid color lines indicate the expected gamma-ray emission from unresolved sources, for 4 different well-established astrophysical populations: blazars (in orange), MAGNs (in green), SFGs (in blue) and MSPs (in red). Color bands represent the corresponding uncertainties on the emission of each population. Estimates are taken from Ref.~\cite{Ajello:2015mfa} (blazars), Ref.~\cite{DiMauro:2013xta} (MAGNs), Ref.~\cite{Tamborra:2014xia} (SFGs) and Ref.~\cite{Calore:2014oga} (MSPs).}
\end{center}
\end{figure}

\item {\bf Imperfect knowledge of the Galactic foreground:} the authors of 
Ref.~\cite{Keshet:2003xc} reconsidered the measurement of the DGRB by EGRET 
and noted that some modifications on the modeling of Galactic CRs could 
decrease significantly the intensity of the DGRB. They also raised some doubts 
on the approach of template fitting. In particular, the effect of an extended 
halo of electrons around the MW (with a consequent IC gamma-ray emission 
extending to high latitudes) is considered. Furthermore, 
Ref.~\cite{Feldmann:2012rx} investigates the possibility of a gas cloud with 
a mass of few $10^{10} M_\odot$, extending to hundreds of kpc from the center of 
the MW. This halo would be theoretically well motivated, as it would alleviate 
the problem of the missing baryons in spiral galaxies. A similar object around 
spiral galaxy NGC 1961 would also explain the diffuse X-ray detected in 
Ref.~\cite{Anderson:2011ih}. Hints of such large halo could be already present 
in hydrodynamical $N$-body simulations of our Galaxy \cite{Gnedin:2010iq,
Feldmann:2010pi,Feldmann:2012rx}. The gamma-ray emission associated with pion 
decay in this hypothetical gas halo would be able to explain between 3\% and 
10\% of the \fermi LAT DGRB in Ref.~\cite{Abdo:2010nz}, depending on the 
exact size of the halo. 
\end{itemize}

Other possibilities not considered in the list above include emission from 
massive black holes at $z \sim 100$ \cite{Gnedin:1992}, from the evaporation 
of primordial black holes \cite{Page:1976wx,MacGibbon:1991vc}, from the 
annihilations at the boundaries of cosmic matter and anti-matter domains 
\cite{Stecker:1971gc} and from the decays of Higgs or gauge bosons produced 
from cosmic topological defects \cite{Bhattacharjee:1997in}.

We conclude this section by discussing Fig.~\ref{fig:summary_astrophysics}.
The image gathers the most recent predictions for the ``guaranteed'' 
components to the DGRB, i.e. the emission associated with unresolved blazars, 
MAGNs, SFGs and MSPs (see sections from \ref{sec:blazars} to \ref{sec:MSPs}). 
They are taken from the results of Refs.~\cite{Ajello:2015mfa,DiMauro:2013xta,
Tamborra:2014xia,Calore:2014oga}, respectively and they are depicted in 
Fig.~\ref{fig:summary_astrophysics} by orange, green, blue and red lines,
respectively.\footnote{Ref.~\cite{Ajello:2015mfa} only provides the total 
emission from resolved {\it and} unresolved blazars. Since we are interested
in the unresolved component, the orange line in 
Fig.~\ref{fig:summary_astrophysics} is obtained by subtracting the emission
of resolved sources from Ref.~\cite{Ackermann:2014usa} from the total signal
from blazars. The width of the light orange band is, then, computed summing 
the estimated errors of the two components in quadrature. Also, the abrupt end 
of the SFG component at $\sim 100$ GeV in not physical and simply comes from 
the lack of predictions above this enegy in Ref.~\cite{Tamborra:2014xia} only 
up to that energy.} Each contribution is embedded in a band that denotes the 
level of uncertainty affecting the prediction. The largest is the one 
associated with MAGNs (light green band) spanning almost one order of 
magnitude. Black data points represent the new \fermi LAT measurement of the 
DGRB in Ref.~\cite{Ackermann:2014usa} (see 
Sec.~\ref{sec:intensity_measurement}). The gray boxes around the data points 
indicate the systematic error associated with the modeling of the Galactic 
foreground. From the figure, it is clear that MSPs are subdominant and that 
the remaining 3 astrophysical components can potentially explain the whole 
DGRB, leaving very little room for additional contributions (see also 
Refs.~\cite{Calore:2011bt,Calore:2013yia,Cholis:2013ena}). Similar results 
have been recently obtained by Ref.~\cite{DiMauro:2015tfa}. This reference 
also shows that the goodness of the fit to the \fermi LAT DGRB energy spectrum 
in terms of astrophysical sources depends significantly on the model adopted 
for the diffuse Galactic foreground and on the slope of the energy spectrum of 
unresolved SFGs. In particular, a description of SFGs with a softer energy 
spectrum (similar to that of the Galactic foreground) can provide a better fit 
to the DGRB intensity.

\subsection{The Dark Matter component of the Diffuse Gamma-Ray Background}
\label{sec:DM}
The DGRB can also be used to investigate more exotic scenarios than those 
presented in the above subsections. In particular, it has already been shown 
that the DGRB is a powerful tool to investigate the nature of DM.

Discussing the very wide range of viable DM candidates is beyond the scope
of this review (see, e.g., Ref.~\cite{Taoso:2007qk}). In the following, we 
only consider a family of candidates called Weakly Interacting Massive 
Particles (WIMPs), loosely characterized by a mass of the order of the GeV-TeV 
and by weak-scale interactions. This is a very well studied scenario since 
many extensions of the Standard Model of Particle Physics predict the 
existence of WIMPs \cite{Jungman:1995df,Martin:1997ns,Bertone:2004pz,
Hooper:2007qk,Bertone:2010at}. It is also quite natural for WIMPs to reproduce 
the DM relic density observed, e.g., by Planck \cite{Ade:2013zuv}. Yet, 
currently there is no observational confirmation of the existence of WIMPs.

WIMP DM can either annihilate or decay into Standard Model particles, including 
gamma rays. This is a general prediction of WIMP candidates and it represents 
an additional reason to focus only on WIMPs for this review. The specific 
mechanisms of gamma-ray emission (see, e.g., Ref.~\cite{Bertone:2004pz} for a 
review) depend on the DM candidate considered and include $i)$ direct 
production of monochromatic gamma rays, $ii)$ decay of neutral pions, produced 
by the hadronization of the primary annihilation/decay products, $iii)$ final 
state radiation and $iv)$ secondary emission by IC or bremsstrahlung of 
primarily produced leptons. Since no DM source has been unambiguously detected 
up to now, the entire DM-induced gamma-ray emission may be unresolved and, 
thus, it contributes to the DGRB. In Sec.~\ref{sec:DM_annihilation} we discuss 
the potential DM contribution to the DGRB in the case of self-annihilating DM 
particles, while Sec.~\ref{sec:DM_decay} is devoted to decaying DM. Note that 
some DM candidates can experience both annihilations and decays 
\cite{PalomaresRuiz:2010pn}. 

DM-induced gamma rays can be produced in the DM halo of the MW or in 
extragalactic DM structures and substructures. We refer to the two 
possibilities as the ``Galactic'' and ``cosmological'' DM components, 
respectively. The latter is isotropic by construction, while the former is 
expected to exhibit some anisotropy, due to the particular location of the 
Earth in the DM halo of the MW. We remind that, as described in 
Sec.~\ref{sec:intensity_measurement}, the intensity of the DGRB is obtained by
means of an isotropic template \cite{Ackermann:2014usa}. However, the Galactic
DM signal can exhibit a significant anisotropy and, in that case, it cannot be 
considered part of the DGRB.\footnote{Small-scale anisotropies in the DGRB,
on the other hand, are discussed in detail in Sec.~\ref{sec:angular_spectrum}.} 
In this section, we focus mainly on the contribution of the cosmological DM 
signal to the DGRB, discussing a possible Galactic DM component only towards 
the end of the section.

\subsubsection{The case of annihilating Dark Matter}
\label{sec:DM_annihilation}
In the $\Lambda$CDM cosmological framework \cite{Ade:2013zuv}, initial matter 
fluctuations in the Early Universe are the seeds of the structures that 
populate today's Universe. These fluctuations grow by accreting new matter 
and form the first protostructures, which, then, collapse and eventually 
virialize into DM {\it halos}. $\Lambda$CDM predicts that, in later epochs, 
larger halos gradually assemble by accretion and merging of smaller halos. 
Under this scenario of structure formation, a cosmological gamma-ray signal is 
expected from the annihilations of DM particles taking place in {\it all} DM 
halos at {\it all} cosmic epochs.

The cosmological gamma-ray flux $d\Phi/dE_0d\Omega$ (i.e. the number of 
photons per unit energy, time, area and solid angle) produced by DM 
annihilations at energy $E_0$ over all redshifts $z$ is given by 
\cite{Ullio:2002pj,Taylor:2002zd,Ando:2005xg}:
\begin{equation}
\frac{d\Phi}{dE_0 d\Omega} = 
\frac{c\ \langle \sigma v \rangle \, (\Omega_{\chi,0}\rho_c)^2}{8\pi m^2_\chi}
\int dz \frac{(1+z)^2}{H(z)} \zeta(z) \, e^{-\tau_{\rm EBL}(E_0(1+z),z)}
\sum_i B_i \frac{dN^i}{dE}\Big{|}_{E=E_0(1+z)}\,
\label{eqn:cosmological_flux}
\end{equation}
where $m_\chi$ is the mass of the DM particle, $\langle \sigma v \rangle$ its 
annihilation cross section\footnote{The annihilation cross section in 
Eq.~(\ref{eqn:cosmological_flux}) is assumed to be dominated by $s$-wave 
interactions. In the case of a dependence on the relative velocity of the 
annihilating DM particles, Eq.~(\ref{eqn:cosmological_flux}) has to be 
modified accordingly and the signal will, thus, depend on the velocity 
distribution of DM \cite{Campbell:2010xc,Campbell:2011kf}.} and the sum runs 
over all the possible annihilation channels, each of them corresponding to a 
specific branching ratio $B_i$ and a differential photon yield $dN^i/dE$. 
$\Omega_{\chi,0}$ is the current DM density ratio, $\rho_c$ the critical density 
of the Universe, and $H(z)$ and $c$ are the Hubble parameter and the speed of 
light, respectively. The function $\tau_{\rm EBL}(E,z)$ accounts for the 
absorption of gamma-ray photons due to interactions with the EBL. Finally, the 
quantity $\zeta(z)$ is the so-called {\it flux multiplier} and it indicates 
the variance of the fluctuations in the field of squared DM density. It is, 
therefore, a measure of the statistical clustering of DM in the Universe:
\begin{equation}
\zeta(z) = \langle \delta^2(z) \rangle = 
\frac{\langle \hat{\rho}_\chi^2(z) \rangle}{(\Omega_{\chi,0}\rho_c)^2},
\label{eqn:general_flux_multiplier}
\end{equation}
where $\hat{\rho}_\chi$ is the comoving DM density and the parentheses 
$\langle \cdot \rangle$ denote angular integration over all the possible
pointings in the sky.

Two approaches have been proposed to calculate the flux multiplier. The first 
is based on the Halo Model \cite{Press:1973iz,Cooray:2002dia} and it relies on 
the knowledge of the abundance and of the internal properties of DM halos. The 
former is described by the Halo Mass Function (HMF), $dn/dM$, i.e. the 
comoving number of halos per unit mass, while the latter is encoded in the DM 
halo density profile. More specifically, in the Halo Model framework, 
$\zeta(z)$ is proportional to the integral of the HMF times the integral of 
the DM density squared inside the halo \cite{Ullio:2002pj,Taylor:2002zd,
Ando:2005hr,Ackermann:2015gga}:
\begin{equation}
\zeta(z)= \frac{1}{\rho_c} \int_{M_{\rm min}} dM  \frac{dn}{dM} M 
\frac{\Delta(z)}{3} \langle F(M,z) \rangle,
\label{eqn:halo_model}
\end{equation}
where $M_{\rm min}$ is the minimum halo mass. Typical values for $M_{\rm min}$ 
range approximately between $10^{-12}$ and $10^{-3} M_\odot$ for supersymmetric 
DM candidates.\footnote{The value $M_{\rm min}=10^{-6} M_\odot$ has become a 
standard benchmark value in the field.} Its exact value depends on the 
position of a cut-off in the power spectrum of matter fluctuations, above which
the formation of DM structures is suppressed. This cut-off arises from the
combined effect of kinetic decoupling and baryonic acoustic oscillations
\cite{Hofmann:2001bi,Loeb:2005pm,Green:2005fa,Bringmann:2009vf}, and its 
precise location ultimately depends on the Particle Physics nature of the DM 
candidate \cite{Profumo:2004qt,Bringmann:2009vf,Cornell:2013rza}. In addition, 
the mapping between the cut-off scale and $M_{\rm min}$ is not well defined, 
depending on the assumed relation between mass and size of the small-mass halos 
\cite{Sefusatti:2014vha}. These uncertainties are responsible for the huge 
variability of $M_{\rm min}$ quoted above.

The factor $\Delta(z)$ in Eq.~(\ref{eqn:halo_model}) is the so-called halo
overdensity and it depends on the cosmology assumed and on the details of the
gravitational collapse of the halo. The radius $R_\Delta$ at which the mean 
enclosed density of a DM halo is $\Delta(z)$ times $\rho_c$ is called the 
{\it virial} radius, which can be taken as a measurement of the size of the 
halo. The DM distribution inside the halo is codified in the function 
$\langle F (M,z)\rangle$:
\begin{equation}
F(M,z) \equiv c_{\Delta}^3(M,z) 
\frac{\int_0^{c_{\Delta}} dx \, x^2 \kappa^2(x)}
{\left[ \int_0^{c_{\Delta}} dx \, x^2 \, \kappa(x) \right]^{2}} \, ,
\label{eqn:FMz}
\end{equation}
where $\kappa(x)$ is the DM density profile and $x = r/r_s$. $r_s$ is a scale 
radius, whose precise definition depends on the assumed $\kappa(x)$. The 
quantity $c_\Delta$ is the so-called {\it concentration} \cite{Bullock:1999he,
Maccio':2008xb,Prada:2011jf} and it is defined as $R_{\Delta}/r_s$. Note that 
$\langle F (M,z)\rangle$ in Eq.~(\ref{eqn:halo_model}) is the function $F(M,z)$ 
from Eq.~(\ref{eqn:FMz}) averaged over a log-normal probability distribution 
assumed for $c_{\Delta}$. Such a distribution accounts for halo-to-halo scatter 
on the value of $c_{\Delta}$ \cite{Bullock:1999he,Wechsler:2001cs,Dolag:2003ui}, 
which is a natural consequence of the stochastic process of structure 
formation in $\Lambda$CDM cosmology.

Information on the abundance and internal properties of DM halos is mainly 
extracted from $N$-body cosmological simulations (see, e.g., 
Ref.~\cite{Kuhlen:2012ft} and references therein). However, simulations do 
not resolve the whole halo hierarchy down to $M_{\rm min}$. The current 
resolution limit for MW-like DM halos is approximately at $10^{5} M_\odot$ 
at $z=0$ \cite{Springel:2008cc}, i.e. several orders of magnitude above the 
expected value for $M_{\rm min}$. Thus, extrapolations of the relevant quantities 
(e.g., the HMF and $c_\Delta(M,z)$) are needed. Since $F(M,z)$ in 
Eq.~(\ref{eqn:FMz}) depends on the third power of the halo concentration, the 
way $c_\Delta(M,z)$ is extended below the mass resolution of simulations is 
crucial in the estimation of the cosmological DM signal. Above the mass 
resolution limit, $c_\Delta(M)$ can be well described by a power law 
\cite{Bullock:1999he,Neto:2007vq,Duffy:2008pz,Maccio':2008xb,Prada:2011jf}. 
Several works assume the same behavior to be valid down to $M_{\rm min}$ 
\cite{Pieri:2007ir,Zavala:2009zr,Pinzke:2011ek,Gao:2011rf}, thus assigning 
very large concentrations to the smallest halos. This translates into a very 
high gamma-ray flux expected from DM annihilations. Nevertheless, power-law 
extrapolations for $c_\Delta(M)$ are not physically motivated and they are now 
clearly ruled out both by recent high-resolution simulations of the smallest 
DM halos \cite{Ishiyama:2014uoa,Diemand:2005vz,Anderhalden:2013wd} and by 
theoretical predictions deeply rooted in the $\Lambda$CDM cosmological 
framework \cite{Prada:2011jf,Sanchez-Conde:2013yxa,Ludlow:2013vxa}. Indeed, 
simulation-based and theoretical estimates of $c_\Delta(M,z)$ have been shown 
to agree now remarkably well over the full halo mass range, i.e. from 
Earth-like ``microhalos'' up to the scale of the heaviest galaxy clusters 
\cite{Sanchez-Conde:2013yxa}. Compared to power-law extrapolations, these 
estimates exhibit a flattening of $c_\Delta(M)$ at low masses. This leads to 
substantially less concentrated low-mass halos and, thus, to a considerably 
smaller cosmological DM signal. Overall, predictions for the cosmological 
DM-induced emission can vary by few orders of magnitude depending on the 
adopted model for $c_\Delta(M,z)$ in Eq.~(\ref{eqn:FMz}), see, e.g., 
Refs.~\cite{Zavala:2009zr,Mack:2013bja}.

Not unexpectedly, the variability of the predicted DM signal also depend on 
the particular choice of $M_{\rm min}$. For example, the cosmological signal 
increases by up to a factor of $\sim 6$ when $M_{\rm min}$ goes from 
$10^{-3} M_\odot/h$ to $10^{-12} M_\odot/h$ \cite{Fornasa:2012gu}. This refers to 
the case of a power-law extrapolation of $c_\Delta(M)$ and, by construction, 
flux multipliers that rely on power-law extrapolations are particularly 
sensitive to the choice of $M_{\rm min}$. When adopting a $c_\Delta(M)$ that 
flattens at low halo masses, the flux multiplier changes by a factor of 
$\sim 3$ over the same range of $M_{\rm min}$ \cite{Fornasa:2012gu,
Ackermann:2015gga}. 

$N$-body cosmological simulations have also been employed to understand the 
HMF and its redshift evolution \cite{Jenkins:2000bv,Springel:2005nw}. Mock 
halo catalogs have been used to test the predictions of the Press-Schechter 
formalism, according to which the HMF can be written as follows 
\cite{Press:1973iz,Bond:1990iw,Lacey:1993iv,Sheth:1999su}:
\begin{equation} 
\frac{dn}{dM}(M,z) = f(\sigma(M,z)) \frac{\rho_c\Omega_\chi(z)}{M}
\frac{d\ln\sigma^{-1}(M,z)}{dM},
\end{equation}
where $\sigma(M,z)$ is the variance of the fluctuations of the DM density
field (smoothed on a scale of mass $M$) and the exact expression for $f(x)$ 
depends on the mechanism assumed to describe the halo gravitational collapse.
An accurate fitting formula for the HMF, inspired by Ref.~\cite{Sheth:2001dp}, 
can be found in Ref.~\cite{Tinker:2008ff} for the cosmological model favored
by the first data release of the Wilkinson Microwave Anisotropy Probe (WMAP). 
More recently, Ref.~\cite{Prada:2011jf} have also adopted the functional form
proposed in Ref.~\cite{Tinker:2008ff} and derived the parameters of the HMF
compatible with the cosmological model preferred by Planck.\footnote{Note that 
the agreement of different fitting formulas with the simulations may depend on 
the algorithm used by the simulators to extract the mass of a halo from the raw 
particle data, with the caveat that different algorithms may lead to 
different results \cite{Knebe:2011rx}.} Overall, the HMF at $z=0$ can be 
qualitatively approximated by a power law with a slope between -1.9 and -2.0 
and a sharp cut-off for halos more massive than $\sim 10^{14} M_\odot$. The 
uncertainty in the calculation of the cosmological DM signal induced by 
different possible parametrizations of the HMF is only marginal when compared 
to other sources of uncertainties. For instance, assuming the HMF of 
Ref.~\cite{Sheth:2001dp} instead of the one in Ref.~\cite{Tinker:2008ff} 
changes the total gamma-ray flux by a factor of about 20\% (see also 
Ref.~\cite{Mack:2013bja}).

High-resolution $N$-body simulations have also helped to establish the 
density profiles of DM halos in great detail. Ref.~\cite{Navarro:1995iw} 
determined that DM halos exhibit a density profile that is universal, 
i.e. independent of the halo mass. A Navarro-Frenk-White (NFW) profile, 
i.e. proportional to $\kappa \propto 1/r$ in the inner region and to a steeper 
$r^{-3}$ at large radii, provides a good fit. More recently, the so-called 
Einasto profile was found to agree better with the results of $N$-body 
simulations, especially at intermediate halo radii \cite{Einasto:1965,
Graham:2005xx,Navarro:2008kc}. Even if many other parametrizations have been 
proposed over the last years \cite{Hernquist:1990be,Moore:1999gc,Dehnen:2005cu,
Navarro:2003ew,Merritt:2005xc,Stadel:2008pn}, current $N$-body simulations 
agree on a slope $\lesssim -1$ for the DM density in the inner region. However, 
these ``cuspy'' profiles are derived from simulations that only contain DM, 
without including baryons. Indeed, high-resolution observations of the 
rotation curves of DM-dominated dwarfs and low-surface-brightness galaxies 
favor DM density profiles with a flat central core \cite{Flores:1994gz,
Burkert:1995yz,deBlok:1996ns,McGaugh:2001yc,deBlok:2002tg,Gentile:2004tb,
Simon:2004sr,Salucci:2007tm,Li:2009mp}. Phenomenological cored profiles, such 
as the Burkert one \cite{Burkert:1995yz}, were proposed to accommodate such 
results. More recently, hydrodynamical simulations have been performed, which 
realistically include the complexity of baryonic physics. They begin to 
reproduce the observed properties of galaxies successfully, e.g. in 
Refs.~\cite{Guedes:2011ux,Kuhlen:2013tra}. Yet, the exact interplay between 
baryons and DM at all radial scales and for all halo masses is not fully 
understood, and the impact of complex baryonic phenomena (such as supernova 
feedback, stellar winds and baryonic adiabatic compression) on the DM density 
profile is still uncertain, particularly in the inner region 
\cite{Gustafsson:2006gr,Colin:2005rr,Tissera:2009cm,Gnedin:2011uj,
SommerLarsen:2009me,Mashchenko:2007jp,Pontzen:2011ty,Governato:2009bg,
Maccio':2011eh}. Ref.~\cite{Profumo:2009uf} finds a difference of almost one 
order of magnitude in the cosmological flux when a Burkert profile is assumed 
for all DM halos instead of a NFW one.

In addition to DM halos, a natural prediction of $\Lambda$CDM is the existence 
of a large number of subhalos, i.e. halos gravitationally bound to a larger
host halo and located within its virial radius. Since the annihilation 
luminosity of a halo is proportional to its DM density squared, the presence 
of small clumps with high DM densities has the effect of boosting the overall 
gamma-ray luminosity of the host halo \cite{Strigari:2006rd,Kuhlen:2008aw,
Lavalle:1900wn,Pieri:2007ir,Springel:2008by,Martinez:2009jh,
Kamionkowski:2010mi,Zavala:2009zr,Charbonnier:2011ft,SanchezConde:2011ap,
Pinzke:2011ek,Nezri:2012tu,Gao:2011rf,Anderhalden:2013wd,Zavala:2013lia,
Sanchez-Conde:2013yxa}. The additional contribution from substructures can 
be accounted for in the computation of the flux multiplier by adding an extra 
term in Eq.~(\ref{eqn:halo_model}). In particular, the factor 
$\langle F(M,z) \rangle$ has to be replaced by 
$\langle F(M,z) \rangle (1 + B(M,z))$, where $B(M,z)$ is called the 
``boost factor''.\footnote{This definition of the boost factor is particularly 
convenient for the computation of the flux multiplier and of the cosmological
DM signal. We warn the reader, though, that alternative definitions of $B(M,z)$ 
can be found in the literature.}

The subhalo population is characterized by a subhalo mass function:
\begin{equation}
\frac{dn_{\rm sub}}{dM}(m_{\rm sub}) \propto 
\left( \frac{m_{\rm sub}}{M_{\rm host}} \right)^{-\alpha}, 
\end{equation}
extending down to $M_{\rm min}$. The slope of the subhalo mass function has been 
measured in high-resolution $N$-body simulations, ranging between -1.9 and -2 
\cite{Diemand:2006ik,Springel:2008cc}. These values are also in line with 
theoretical expectations from the Press-Schechter theory of structure 
formation \cite{vandenBosch:2004zs,Giocoli:2007uv,Blanchet:2012vq}. It has 
been noted, though, that several effects may prevent the subhalo mass function 
to reach the lowest subhalo masses, since processes such as tidal disruption, 
accretion or merging may be particularly efficient and deplete the low-mass 
tail of the mass function. It is difficult to estimate the survival probability 
of these small subhalos, while it remains computationally very expensive to 
simulate and keep track of such processes with the resolution needed. Although
the properties of low-mass DM subhalos are expected to follow those of the 
more massive counterparts, the abundance and distribution of DM substructures 
below the resolution of current simulations remain uncertain. The internal 
properties of subhalos are also subject to debate. We still lack a 
comprehensive understanding of the subhalo concentration, even if subhalos 
have been shown to exhibit larger $c_\Delta(M)$ than halos of the same mass 
\cite{Diemand:2007qr,Springel:2008cc,Diemand:2008in,Klypin:2010qw}. Similarly 
to the case of main halos, the assumptions made on $c_\Delta(M)$ and on the 
abundance of low-mass subhalos have a large impact on the amplitude of the 
subhalo boost factor. Overall, under the most extreme scenarios (corresponding 
to blind power-law extrapolations of $c_\Delta(M)$), the presence of subhalos 
can increase the total cosmological annihilation signal by more than one 
order of magnitude \cite{Ullio:2002pj,Fornasa:2012gu,Cholis:2013ena,
Ando:2013ff}.

As an alternative to the Halo Model described above, the Power Spectrum 
approach has been recently introduced to compute the flux multiplier 
\cite{Serpico:2011in,Sefusatti:2014vha}. In this new framework, $\zeta(z)$ can 
be calculated by means of the non-linear matter power spectrum $P_{NL}$, i.e.
the Fourier transform of the two-point correlation function of the matter 
density field, as follows:
\begin{equation} 
\zeta(z) \equiv \langle \delta^2 (z) \rangle = 
\int ^{k_{\rm max}} \frac{d\,k}{k} \frac{k^3 P_{NL}(k,z)}{2\pi^2},
\label{eqn:PS}
\end{equation} 
where $k_{\rm max}(z)$ is a maximal scale that can be related to $M_{\rm min}$ in 
the Halo Model formalism.\footnote{The exact relation between $k_{\rm max}$ and
$M_{\rm min}$ is not trivial, as discussed in detail in 
Refs.~\cite{Sefusatti:2014vha,Ackermann:2015gga}.} The main benefit of the 
Power Spectrum approach relies on the fact that only one single quantity is 
needed for the calculation of $\zeta(z)$, i.e. $P_{NL}$ in Eq.~(\ref{eqn:PS}). 
$P_{NL}$ can be measured directly in $N$-body cosmological simulations using 
only a matter density map and it does not rely on the concept of DM 
halos. It is, thus, also independent on the complex issue of halo finding and 
on the uncertainties associated with it \cite{vanDaalen:2015msa}.\footnote{In 
Refs.~\cite{Zavala:2013bha,Zavala:2013lia}, the authors introduce the concept 
of particle phase space average density, an estimate of the coarse-grained 
phase-space density of DM structures. Interestingly, they show in 
Ref.~\cite{Zavala:2013lia} how this observable can be used to determine the 
DM-induced emission of MW-like halos. The particle phase space average density 
can be directly computed directly from $N$-body simulations' raw data and, 
therefore, the formalism shares some similarities with the Power Spectrum 
approach, introduced here to estimate the cosmological DM signal.} This is, 
indeed, what the authors of Ref.~\cite{Sefusatti:2014vha} did, guided by the 
results of the Millennium-I and Millennium-II simulations 
\cite{Springel:2005nw,BoylanKolchin:2009nc}. Yet, as in the Halo Model 
approach, current $N$-body simulations do not reach the highest $k$ values 
needed in Eq.~(\ref{eqn:PS}) and, thus, extrapolations beyond the simulation 
resolution are again required.\footnote{At redshift zero, for instance, the 
maximum $k$ values resolved in current large-scale structure simulations are 
typically of the order of few hundreds, while $k_{\rm max}$ may take values up to
$10^6$.} Note, however, that only $P_{NL}$ is extrapolated in this case, as 
opposed to the Halo Model approach that relies on the knowledge of several 
quantities in the low-mass regime. This is indeed the reason why the authors 
in Ref.~\cite{Ackermann:2015gga} adopted the Power Spectrum approach to obtain 
a realistic estimate of the uncertainty on $\zeta(z)$, even if their fiducial 
flux multiplier is computed within the Halo Model framework. Furthermore, 
within the PS approach, it is possible to motivate in a realistic way both the 
choice of $k_{\rm max}$ and the way the extrapolation is performed. Note also 
that, by construction, the Power Spectrum approach naturally includes the 
contribution of substructures down to length scales $\sim \pi/k_{\rm max}$, while
in the Halo Model the description of substructures requires additional 
knowledge, often leading to further debatable extrapolations and 
uncertainties.\footnote{We remind the reader that Eq.~(\ref{eqn:PS}) only
determines the cosmological DM signal. Any gamma-ray emission associated with
the halo of the MW and its subhalos is not accounted for by the Power Spectrum
model and, thus, needs to be added by hand (see further discussion at the end 
of this section).}

\begin{figure}
\begin{center}
\includegraphics[width=0.49\linewidth]{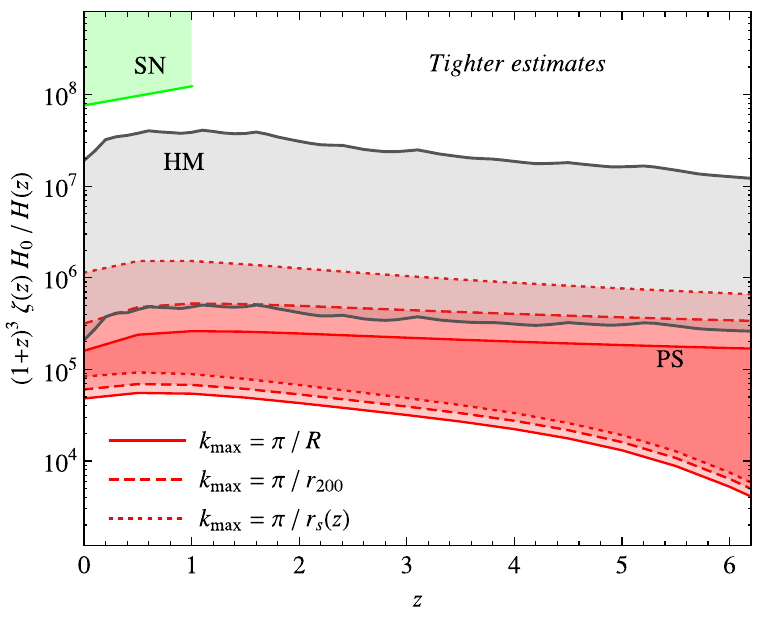}
\includegraphics[width=0.49\linewidth]{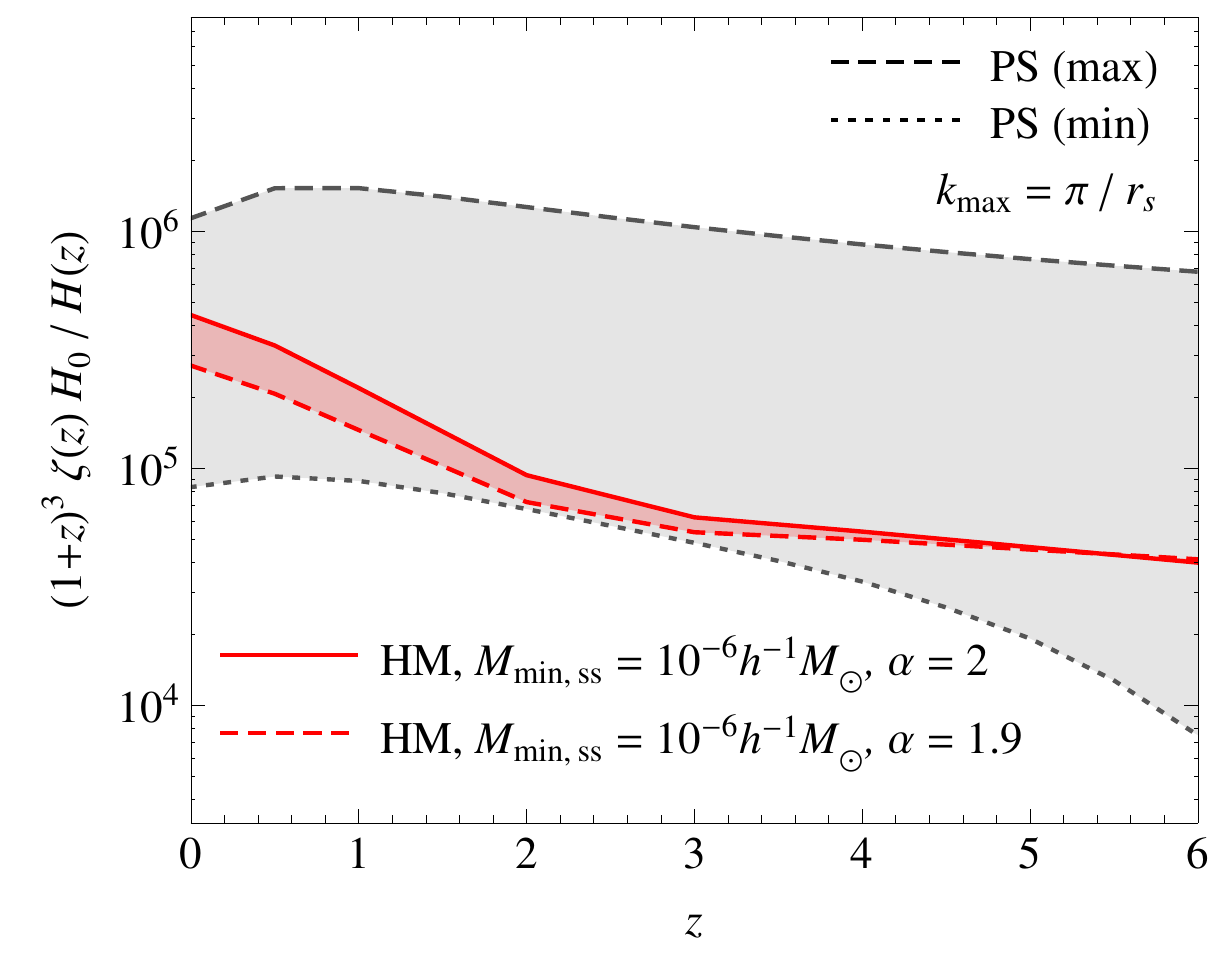}
\caption{\label{fig:HMvsPS} Comparison of the Halo Model and Power Spectrum approaches for the calculation of the flux multiplier. The Halo Model makes use of Eq.~(\ref{eqn:FMz}) while the Power Spectrum formalisms adopts Eq.~(\ref{eqn:PS}). {\it Left:} Power Spectrum predictions are shown in red. The different bands correspond to different prescriptions to derive the cut-off scale $k_{\rm max}$ from the size of the DM halo with mass $M_{\rm min}$. The width of the bands accounts for different extrapolation schemes up to $k_{\rm max}$. The Halo Model predictions (gray band) are obtained following Refs.~\cite{Zavala:2009zr} and \cite{Abdo:2010dk}. Figure taken from Ref.~\cite{Sefusatti:2014vha}. {\it Right:} Predictions according to the Power Spectrum approach (gray) and the Halo Model (red). The Halo Model formalism follows what done in Ref.~\cite{Ackermann:2015gga} for two possible subhalo mass functions. The gray band corresponds to a particular choice of $k_{\rm max}$ in the Power Spectrum approach, for different ways (labeled as ``PS (max)'' and ``PS (min)'' in the figure) to extend $P_{NL}$ beyond the resolution of the simulations. For both panels, a value of $M_{\rm min} = 10^{-6}M_\odot/h$ is adopted in the Halo Model formalism. Figure taken from Ref.~\cite{Ackermann:2015gga}.}
\end{center}
\end{figure}

A comparison between the Power Spectrum and Halo Model approaches is performed 
in Refs.~\cite{Sefusatti:2014vha,Ng:2013xha,Ackermann:2015gga}. Results from
Refs.~\cite{Sefusatti:2014vha,Ackermann:2015gga} are reproduced in 
Fig.~\ref{fig:HMvsPS}. In the left panel, the red bands indicate the 
predictions for $\zeta(z)$ (multiplied by factors depending on redshift) 
obtained by means of the Power Spectrum method for different choices of 
$k_{\rm max}$. For each value of $k_{\rm max}$ adopted, the corresponding band 
accounts for different ways of extrapolating $P_{NL}$ beyond the resolution of 
the Millennium simulations. The predictions for the Halo Model (gray band) 
refer to a $M_{\rm min}=10^{-6} M_\odot/h$ and have been obtained following 
Refs.~\cite{Zavala:2009zr,Abdo:2010dk}. In the right panel of 
Fig.~\ref{fig:HMvsPS}, the predictions of the Power Spectrum approach (for a 
specific value of $k_{\rm max}$) are given by the gray band. The width of the 
band corresponds to two different prescriptions to perform the extrapolation 
(labeled ``PS (min)'' and ``PS (max)'' in the figure). On the other hand, the 
red band reproduces the results of the Halo Model in 
Ref.~\cite{Ackermann:2015gga} for $M_{\rm min}=10^{-6} M_\odot/h$ and two values 
of the slope for the subhalo mass function. Remarkably, the two methods agree 
well in their predictions for $\zeta(z)$, within uncertainties. This is so 
despite the caveats mentioned above and the intrinsic differences of the two
formalisms.

From Sec.~\ref{sec:components} we know that gamma-ray emission of astrophysical
origin is able to explain a significant fraction, if not all, of the DGRB, 
therefore leaving very little room for a potential DM contribution (see also 
Fig.~\ref{fig:summary_astrophysics}). Moreover, the gamma-ray energy spectrum 
expected from DM annihilations is generally slightly curved, with a cut-off at 
the mass of the DM particle and the possibility of spectral features like 
bumps or lines. On the other hand, the DGRB detected by the \fermi LAT 
exhibits a power-law spectrum at low energies and a cut-off compatible with 
EBL attenuation at higher energies (see Sec.~\ref{sec:intensity_measurement}). 
This suggests, again, that DM annihilations cannot provide a dominant 
contribution to the DGRB. Thus, the DGRB can be used to set limits on the 
intensity of the DM-induced emission. These are usually translated into upper 
limits in the $(m_\chi,\langle \sigma v \rangle$) plane, identifying the region 
that, given a model for the abundance and properties of DM halos and subhalos, 
is excluded as it overproduces the measured DGRB.

\begin{figure}
\begin{center}
\includegraphics[width=0.95\linewidth]{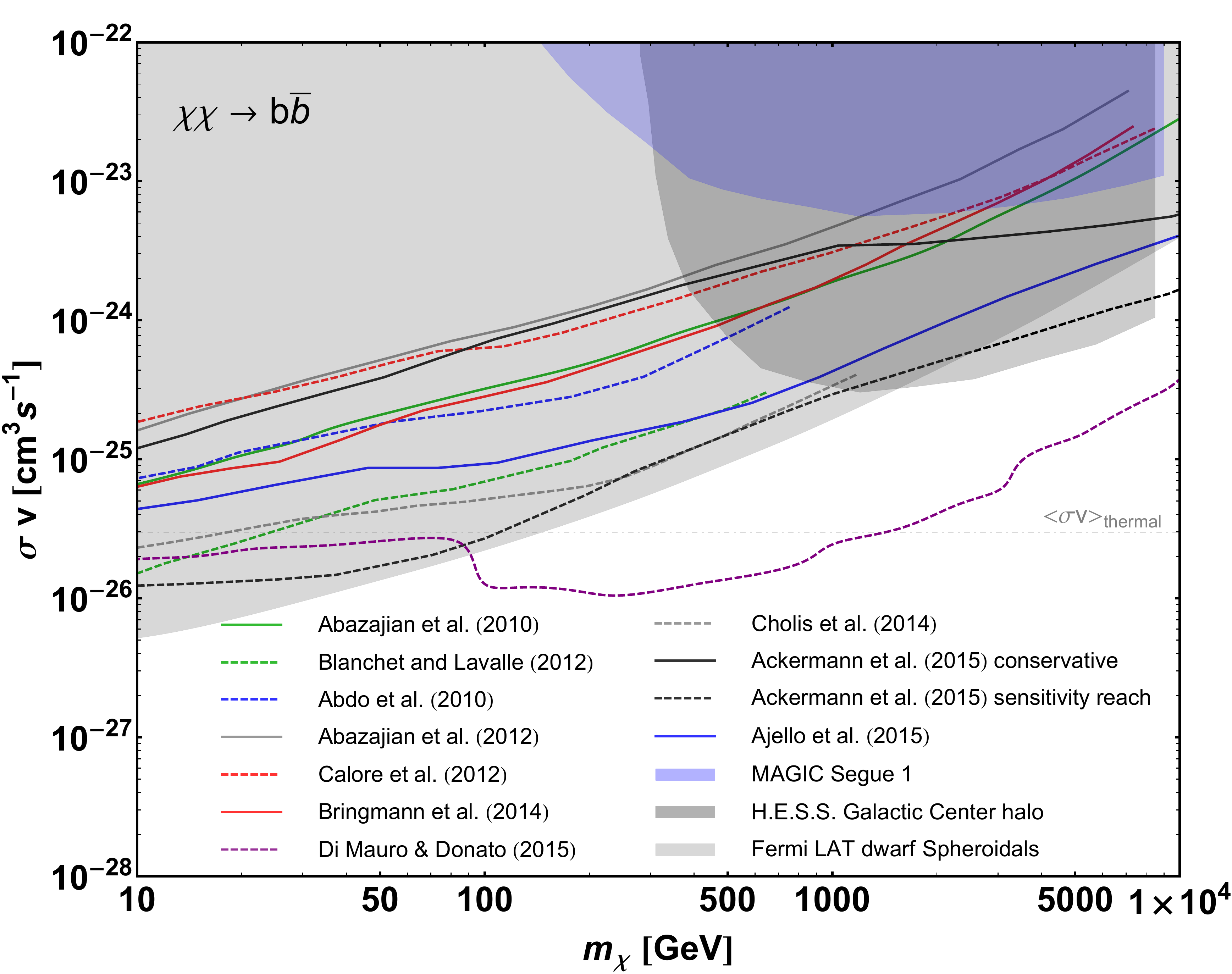}
\caption{\label{fig:DM_annihilation_limits} Upper limits obtained by considering the DGRB energy spectrum measured in Refs.~\cite{Abdo:2010nz,Ackermann:2014usa}. Annihilations into $b$ quarks are assumed. The regions above the colored lines are excluded because the cumulative DM-induced emission would overproduce the DGRB. Different lines correspond to different assumptions for the properties of DM halos (especially for low halo masses) and different methods to compute the upper limits. The solid green line is taken from Fig.~2 of Ref.~\cite{Abazajian:2010sq} while the dashed green line is from Fig.~8 of Ref.~\cite{Blanchet:2012vq} (lower bound of the band relative to $\alpha_m=2$ for the emission from the Galactic Poles). The dashed blue line is taken from Fig.~5 of Ref.~\cite{Abdo:2010dk} (conservative limits for model MSII-Sub1) and the solid gray one from Fig.~2 of Ref.~\cite{Abazajian:2010zb}. The solid and dashed red lines are taken from Fig.~3 of Ref.~\cite{Calore:2011bt} (limits labeled ``Fermi EGB'') and from Fig.~5 of Ref.~\cite{Calore:2013yia} (panel labeled ``best-fit background''), respectively. The dashed gray line is from Fig.~15 of Ref.~\cite{Cholis:2013ena} (default substructures' model). The black lines correspond to the predictions obtained in Ref.~\cite{Ackermann:2015gga} by means of the Halo Model (reference scenario). The solid blue line is taken from Fig.~4 of Ref.~\cite{Ajello:2015mfa}, while the dashed purple one is from Fig.~4 of Ref.~\cite{DiMauro:2015tfa}. The blue region indicates the portion of the ($m_\chi,\langle \sigma v \rangle$) plane already excluded by the observation of the Segue 1 dwarf Spheroidal galaxy performed by the MAGIC telescopes (see Fig.~6 of Ref.~\cite{Aleksic:2013xea}). The dark gray region is excluded by the analysis performed by the H.E.S.S. telescopes in Ref.~\cite{Abramowski:2011hc} from the so-called ``Galactic Center halo'' (see their Fig.~4 for an Einasto DM density profile). Finally, the light gray region indicates the DM candidates not compatible with the combined analysis of 15 dwarf Spheroidal galaxies by the \fermi LAT \cite{Ackermann:2015zua}. A comparison between the \fermi LAT DGRB and the DM-induced signal can also be found in Refs.~\cite{Cuoco:2010jb,Hutsi:2010ai,Zavala:2011tt,Cavadini:2011ig,Fornasa:2012gu,Ando:2013ff}. The dash-dotted horizontal line marks the value of the thermally averaged annihilation cross section.}
\end{center}
\end{figure}

Most of the works following this idea employ the Halo Model approach to 
predict the cosmological DM annihilation signal. Given the large number of 
parameters involved in the Halo Model framework, as well as the large 
uncertainties associated with some of them, it is very hard to perform a 
detailed, one-to-one comparison among the different DM limits available in the 
literature. In particular, predictions obtained by different groups differ 
mainly due to different assumptions on $c_\Delta(M,z)$ and $M_{\rm min}$. In
addition, some works also consider the Galactic DM signal. Differences can be 
further amplified by the various statistical prescriptions employed to compute 
the upper limits. Yet, we believe that a comparison among the limits in the 
literature is useful, as it showcases the potential of searching for DM in the 
DGRB compared to other indirect DM probes. 
Fig.~\ref{fig:DM_annihilation_limits} summarizes some of the upper limits 
available.\footnote{We only consider limits derived from the two measurements 
of the DGRB performed by the \fermi LAT \cite{Abdo:2010nz,Ackermann:2014usa}. 
Older works based on earlier DGRB measurements can be found, e.g., in 
Refs.~\cite{Stecker:1978du,Gao:1991rz,Bergstrom:2001jj,Ullio:2002pj,
Taylor:2002zd,Elsaesser:2004ap,Elsaesser:2004ck,Ahn:2004yd,Ando:2005hr,
Oda:2005nv,Ando:2005xg,Ando:2006cr,Horiuchi:2006de,deBoer:2006tv,Ahn:2007ty,
Fornasa:2007nr,Cuoco:2007sh,Pieri:2007ir,deBoer:2007zc,Baltz:2008wd,
Fornasa:2009qh,SiegalGaskins:2009ux,Profumo:2009uf,Belikov:2009cx,
Kawasaki:2009nr,CyrRacine:2009yn,Huetsi:2009ex,Zavala:2009zr,
Dodelson:2009ih}.} They all refer to annihilations entirely into $b$ quarks.
Note that the limits obtained by assuming a power-law $c_\Delta(M)$ below 
the mass resolution of $N$-body simulations are among the most constraining in 
Fig.~\ref{fig:DM_annihilation_limits} (see, e.g., the gray dashed line from
Ref.~\cite{Cholis:2013ena}). However, as argued above, power-law $c_\Delta(M)$ 
are not well motivated, putting the corresponding limits into question.

Two approaches are possible when deriving DM limits from the DGRB measurement. 
In the first one it is required that the DM signal does not overshoot the 
measured DGRB at a particular confidence level, typically 2 or 3$\sigma$. See,
e.g., Refs.~\cite{Abazajian:2010sq,Hutsi:2010ai,Cuoco:2010jb,Zavala:2011tt,
Fornasa:2012gu,Blanchet:2012vq,Ando:2013ff}. This leads to {\it conservative} 
and robust upper limits on the DM annihilation cross section, shown by the 
green lines in Fig.~\ref{fig:DM_annihilation_limits}. Alternatively, one or 
more astrophysical contributions to the DGRB can be modeled and included in 
the analysis. The DM component is, then, required not to overshoot the 
fraction of the DGRB not already accounted for by astrophysics. These limits 
are more constraining than the previous ones \cite{Abdo:2010dk,
Abazajian:2010zb,Calore:2011bt,Cavadini:2011ig,Calore:2013yia,Cholis:2013ena}. 
Nevertheless, since the exact contribution from astrophysical sources is not 
known, it must be noted that they are subject to larger uncertainties.

The most recent constraints on annihilating DM comes from 
Ref.~\cite{Ackermann:2015gga} and are shown by black lines in 
Fig.~\ref{fig:DM_annihilation_limits}. The ``conservative limit'' (solid 
black line) is obtained without including any modeling of astrophysical 
contributors to the DGRB. These limits exclude annihilation cross sections a 
factor of $\sim 3$ lower than the thermal value of 
$3 \times 10^{-26} \mbox{cm}^3 \mbox{s}^{-1}$ for a mass of 10 GeV, while, for 
$m_\chi \geq 1$ TeV, the upper limit is around 
$5 \times 10^{-24} \mbox{cm}^3 \mbox{s}^{-1}$. Ref.~\cite{Ackermann:2015gga} 
also shows that an improvement of one order of magnitude (a factor $\sim 2$) 
for DM masses of the GeV scale (around 30 TeV) is possible when $i)$ the 
cumulative astrophysical contribution to the DGRB is modeled as a power-law in
energy with an exponential cut-off and $ii)$ slope and position of the cut-off
are fixed to the best-fit values to the DGRB data in 
Ref.~\cite{Ackermann:2014usa}. Note that this approach (referred to as 
``sensitivity reach'' and denoted by the dashed black line in 
Fig.~\ref{fig:DM_annihilation_limits}) does not account for any uncertainty in 
the description of the astrophysical emission. A more realistic scenario is 
the one of Ref.~\cite{Ajello:2015mfa}, where the authors estimate the 
unresolved astrophysical emitters using the most up-to-date information 
from resolved sources or from other frequencies (see 
Sec.~\ref{sec:components}). A renormalization factor $\mathcal{A}$ is included
in front of the total astrophysical contribution to account for possible 
fluctuations in its intensity. The limit of $\langle \sigma v \rangle$ (solid 
blue line in Fig.~\ref{fig:DM_annihilation_limits}) is then obtained by 
profiling over $\mathcal{A}$. This results in an improvement of a factor of 
$\sim 3$ for $m_\chi=10$ GeV, with respect to the conservative limits of 
Ref.~\cite{Ackermann:2015gga}. A negligible improvement is expected at masses 
larger than 10 TeV, where the limits are determined by the Galactic DM signal.

Similar results have recently been obtained in Ref.~\cite{DiMauro:2015tfa}, 
which also employs a model for the astrophysical components to the DGRB. The 
different statistical analysis performed in Ref.~\cite{DiMauro:2015tfa} 
suggests that a description of the DGRB of Ref.~\cite{Ackermann:2014usa} in 
terms of unresolved astrophysical sources {\it and} gamma rays from DM 
annihilations in the MW halo provides a better fit, compared to a purely 
astrophysical interpretation with no DM. Such a hint of a DM signal in the 
DGRB suggests a DM particle with a mass of $\sim 10-20$ GeV (for annihilation 
into $b$ quarks), depending on the model adopted for the diffuse Galactic 
foreground.\footnote{A similar indication of a DM component to the DGRB was 
also reported in Refs.~\cite{Elsaesser:2004ap,deBoer:2007zc} based on the 
EGRET measurement of the DGRB in Ref.~\cite{Strong:2004ry}.}

Indeed, some of the limits shown in Fig.~\ref{fig:DM_annihilation_limits} have
been obtained assuming that DM annihilations in the MW halo also contribute to 
the DGRB \cite{Abdo:2010dk,Abazajian:2010sq,Abazajian:2010zb,Hutsi:2010ai,
Cuoco:2010jb,Calore:2011bt,Cavadini:2011ig,Fornasa:2012gu,Blanchet:2012vq,
Ando:2013ff,Calore:2013yia,Cholis:2013ena}. Note that, as mentioned in 
Sec.~\ref{sec:intensity_measurement}, the DGRB is obtained from the 
normalization of the isotropic template in the multi-component fit to the 
gamma-ray data at high Galactic latitudes. Thus, when a Galactic DM signal is 
included, it is implicitly assumed that the DM signal is sufficiently 
isotropic in this particular region of the sky.

Two distinct components contribute to the Galactic DM signal. The first 
accounts for the {\it smooth} DM distribution of the host DM halo of the MW, 
while the second one comes from the population of Galactic subhalos. The 
former depends on the DM density profile assumed for the MW main halo. This is 
uncertain in the innermost parts of the Galaxy, but for $|b|>20^\circ$, i.e. 
more than 3 kpc from the Galactic Center, $N$-body simulations roughly agree.
The signal from this smooth component peaks towards the Galactic Center and, 
between $b=20^\circ$ and $90^\circ$, typical variations are of a factor of 
$\sim 16$ \cite{Ackermann:2015gga}. Thus, even outside the Galactic plane the 
signal is largely anisotropic and, therefore, this component cannot be 
described by an isotropic template as the DGRB. Indeed, the presence of an 
emission with such a well-defined morphology may impact the procedure used in 
Ref.~\cite{Ackermann:2014usa} to measure the DGRB energy spectrum. This was 
tested in Ref.~\cite{Ackermann:2015gga}, where the authors re-derived the DGRB 
including an additional template for the smooth Galactic DM signal. They found 
that this signal can be degenerate with other diffuse Galactic emissions, 
especially the one from IC. They also checked the impact that the new template 
would have on the upper limits on $\langle \sigma v \rangle$. The result 
suggests that, at least for DM candidates not excluded by the conservative 
limits in Ref.~\cite{Ackermann:2015gga}, this additional Galactic DM template 
has only a moderate effect. 

The second contribution to the Galactic DM signal comes from the subhalos of 
the MW: the brightest or closest of them may potentially give rise to bright 
spots in the gamma-ray sky. However, none of the unassociated sources in the 
2FGL catalog has been robustly interpreted as a DM subhalo 
\cite{Ackermann:2012nb,Zechlin:2012by}. The overall subhalo population is 
expected to give rise to a diffuse smooth emission \cite{Pieri:2007ir,
SiegalGaskins:2008ge,Fornasa:2009qh,SiegalGaskins:2009ux}. Its morphology 
depends on the abundance and distribution of subhalos in the Galaxy. A general 
prediction is that the cumulative emission of subhalos is more extended (thus 
more isotropic) than that of the main halo. More specifically, factors between 
0.1 and 2 are quoted in Ref.~\cite{Ackermann:2015gga} for the variation of the 
signal of Galactic substructures between $b=20^\circ$ and $90^\circ$. These 
small variations motivate the authors of Ref.~\cite{Ackermann:2015gga} to 
assume this component as isotropic and, thus, to include it when setting their 
DM limits. Its impact can be quite significant since Galactic subhalos can
boost the Galactic DM signal by a factor of 3 to 15, depending on the slope
of the subhalo mass function ~\cite{Sanchez-Conde:2013yxa}.

We end this section by comparing the upper limits on the annihilation cross
section derived in Ref.~\cite{Ackermann:2015gga} to the results of other
indirect searches for DM. In particular, in 
Fig.~\ref{fig:DM_annihilation_limits}, the shaded regions indicate which
portion of the parameter space has been already excluded by these other probes. 
The blue region is derived from the observation of the Segue 1 dwarf 
Spheroidal galaxy by the MAGIC telescopes \cite{Aleksic:2013xea}, while the 
dark gray region indicates the portion of the 
$(m_\chi,\langle \sigma v \rangle)$ space not compatible with the analysis of 
H.E.S.S. data from the so-called ``Galactic Center halo'' 
\cite{Abramowski:2011hc}. Finally, the light gray area is excluded by the 
non-detection of gamma rays from the observation of 15 dwarf Spheroidal 
galaxies with the \fermi LAT \cite{Ackermann:2015zua}. Note that the 
conservative upper limits derived in Ref.~\cite{Ackermann:2015gga} (solid 
black line) are always inside the area already excluded, while the most 
stringent {\it sensitivity reach} (dashed black line) provides the strongest 
constraints on $\langle \sigma v \rangle$ for DM masses up to 1 TeV, above 
which the limit from H.E.S.S. becomes more stringent.

\subsubsection{The case of decaying Dark Matter}
\label{sec:DM_decay}
Decaying particles can be a viable DM candidates if their decay lifetime is
larger than the age of the Universe \cite{Bolz:2000fu,Choi:2009ng,
Arvanitaki:2009yb}. As in the previous section, we will not discuss here the
models that can accommodate such particles or the mechanisms that guarantee 
long lifetimes. We simply consider generic WIMPs that decay emitting gamma-ray
photons. 

In contrast to the case of annihilating DM, the gamma-ray signal expected 
from decaying DM particles is linearly proportional to the DM density. For 
instance, the contribution of the MW halo (at a certain energy $E_0$ and 
towards the direction $\psi$) can be written as follows:
\begin{equation}
\frac{d\phi}{dE_0 d\Omega} = \frac{1}{4\pi} \frac{1}{\tau m_\chi} 
\int ds \, \rho_{\rm MW}[r(s,\psi)] \sum_i B_i \frac{dN^i}{dE}\Big{|}_{E=E_0}\,
\label{eqn:MW_decay}   
\end{equation}
where $\tau$ is the DM particle lifetime, $\rho_{\rm MW}$ the DM density profile 
of the MW halo and $s$ is the line-of-sight variable pointing towards the 
direction of observation. In the case of prompt emission, the $s$-wave 
annihilation of 2 non-relativistic DM particles with a mass $m_\chi$ is 
equivalent to the decay of a $s$-wave state with mass 2$m_\chi$ and integer 
spin \cite{Cirelli:2008pk}. Under those assumptions, the photon yield $dN^i/dE$ 
in Eq.~(\ref{eqn:MW_decay}) can be determined from the same quantity in 
Eq.~(\ref{eqn:cosmological_flux}). A decaying DM candidate can also produce 
gamma rays through semileptonic channels (provided that it is characterized 
by a semi-integer spin) or through a three-body decays 
\cite{PalomaresRuiz:2010pn}. Mono-chromatic lines are also a typical signature 
of a vast class of decaying DM candidates \cite{Bertone:2007aw,Ibarra:2007wg,
Buchmuller:2007ui,Choi:2009ng,Huang:2011xr}, while the so-called 
``box-shaped'' spectra have been recently introduced in 
Ref.~\cite{Ibarra:2012dw}. Finally, secondary gamma rays can also be produced 
when primary leptons from DM decay interact via bremsstrahlung with the 
interstellar medium of the Galaxy or via IC off its radiation fields 
\cite{Zhang:2009ut}.

Given the linear dependence on the DM density in Eq.~(\ref{eqn:MW_decay}), the
gamma-ray signal expected from the smooth MW DM halo has a different 
morphology, compared to the case of annihilating DM. For a decaying particle, 
the emission is now more isotropic and, thus, there exists a better motivation
to include it among the contributors to the DGRB. Note also that, on the case
of decaying DM, the emission of a DM halo is proportional to the its total 
mass. This means that DM substructures are not expected to boost significantly
the predicted DM signal. Therefore, for a specific DM candidate, the main 
uncertainty affecting the emission in Eq.~(\ref{eqn:MW_decay}) comes from the 
unknown DM density profile of the MW halo.\footnote{When considering secondary 
emission, one should also add the uncertainty associated with our imperfect 
knowledge of the propagation of charged particles in the MW 
\cite{Zhang:2009ut}.} Remarkably, this translates into a considerably smaller 
uncertainty on the intensity of the DM-induced emission compared to the case 
of annihilating DM.

\begin{figure}
\begin{center}
\includegraphics[width=0.9\linewidth]{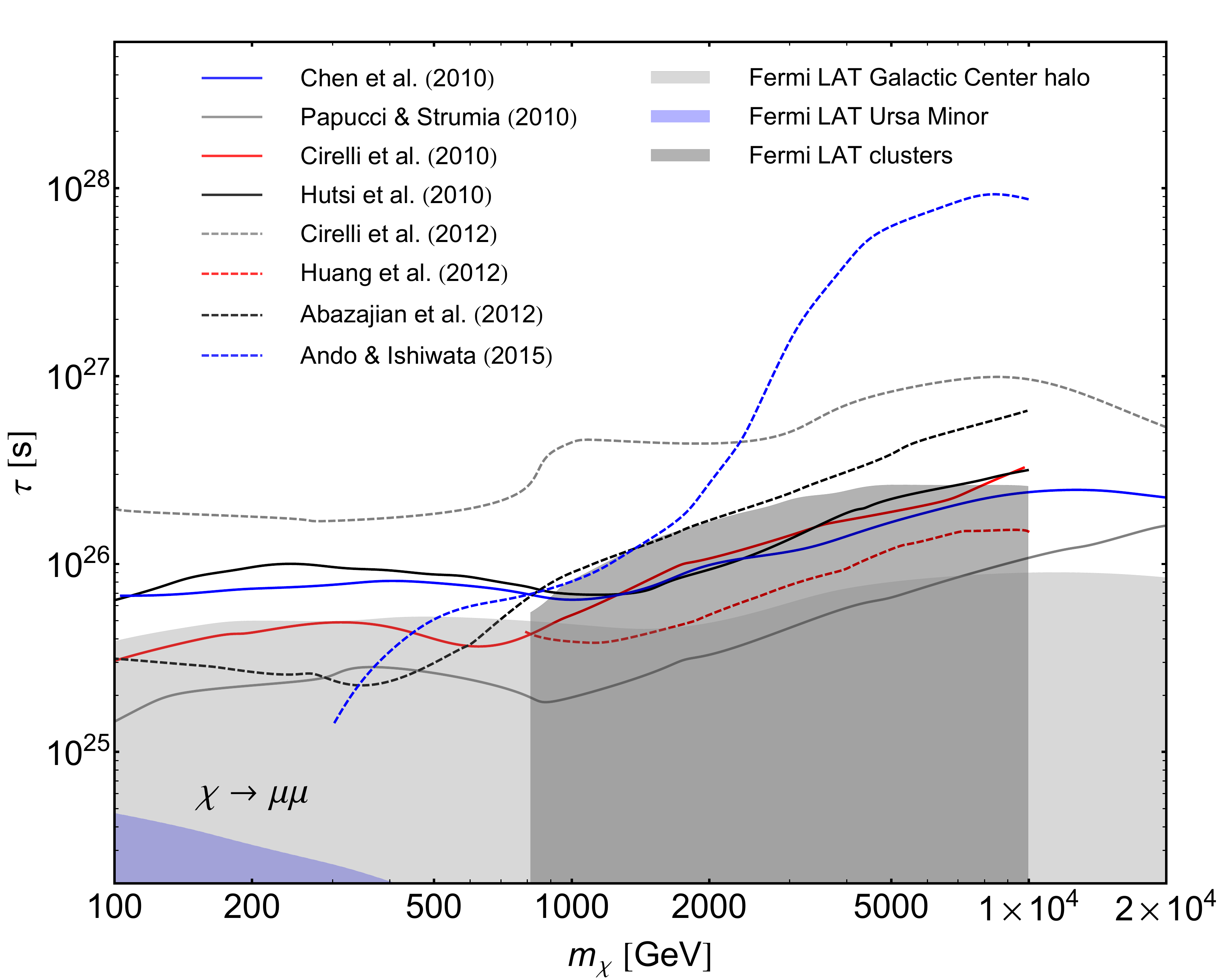}
\caption{\label{fig:summary_decay} Lower limits obtained by considering the DGRB energy spectrum measured in Ref.~\cite{Abdo:2010nz}. Decays into $\mu^+\mu^-$ are assumed. The regions below the colored lines are excluded because the cumulative DM-induced emission would overproduce the DGRB. Different lines correspond to different assumptions for the properties of DM halos and different methods to compute the lower limits. The solid blue line is taken from Fig.~3 of Ref.~\cite{Chen:2009uq}, while the solid gray one is from Fig.~8 of Ref. \cite{Papucci:2009gd} (NFW DM density profile). The solid red and solid black lines are taken from Fig.~4 of Ref.~\cite{Cirelli:2009dv} and from Fig.~6 of Ref.~\cite{Hutsi:2010ai} (NFW DM density profile), respectively. The dashed gray line is from Fig.~2 of Ref.~\cite{Cirelli:2012ut} (DM signal and power-law background) and the dashed red one from Fig.~4 of Ref.~\cite{Huang:2011xr}. The dashed black line is taken from Fig.~4 of Ref.~\cite{Abazajian:2010zb}. The dashed blue line comes from Fig.~3 of Ref.~\cite{Ando:2015qda} (for model A of the Galactic foreground). The blue region indicates the portion of the $(m_\chi,\tau)$ plane already excluded by the observation of the Ursa Minor dwarf Spheroidal galaxy performed by the \fermi LAT (see Fig.~3 of Ref.~\cite{Dugger:2010ys}, for the case of no IC emission). The dark gray region is excluded by the analysis performed in Ref.~\cite{Huang:2011xr} combining \fermi LAT data from the position of 8 galaxy clusters (see their Fig.~4). Finally, the light gray region indicates the DM candidates not compatible with the observation of the so-called ``Galactic Center halo'' performed by the \fermi LAT in Ref.~\cite{Ackermann:2012rg} (see their Fig.~5 for a NFW DM density profile).}
\end{center}
\end{figure}

The cosmological signal for a decaying DM candidate is written as follows:
\begin{equation}
\frac{d\phi}{dE_0 d\Omega} = \frac{c}{4\pi} 
\frac{ \Omega_{\rm DM} \, \rho_c }{\tau m_{DM}} 
\int dz \frac{e^{-\tau_{\rm EBL}(E_0,z)}}{H(z)} \frac{dN}{dE}\Big{|}_{E=E_0(1+z)}.
\label{eqn:extragalactic_decay}
\end{equation}
There is no need of including the flux multiplier since the average emission
depends on the total amount of DM, accounted for by $\Omega_{\rm DM}$. This 
eliminates any dependence on $M_{\rm min}$ or on the shape of $c_\Delta(M,z)$ and 
it translates into more robust predictions.\footnote{Even if one rewrites 
Eq.~(\ref{eqn:extragalactic_decay}) in terms of the Halo Model formalism, 
the result of integrating the HMF times the number of gamma rays expected from
a DM halo of mass $M$ will be quite insensitive to the value adopted for 
$M_{\rm min}$ \cite{Fornasa:2012gu}.}

Many works have used the DGRB to derive constraints on the nature of decaying 
DM candidates. Refs.~\cite{Asaka:1998ju,Takayama:2000uz,Overduin:2004sz,
Bertone:2007aw,Ibarra:2007wg,Ishiwata:2009dk,Ibarra:2009dr,Choi:2009ng} 
consider the DGRB measured by EGRET in Refs.~\cite{Sreekumar:1997un,
Strong:2004ry}, while Refs.~\cite{Cirelli:2012ut,Cirelli:2009dv,Chen:2009uq,
Huang:2011xr,Hutsi:2010ai,Abazajian:2010zb,Ando:2015qda} rely on the DGRB 
measurements by the \fermi LAT. Results are summarized in 
Fig.~\ref{fig:summary_decay}, which collects lower limits on $\tau$ as a 
function of $m_\chi$. The regions below the lines in 
Fig.~\ref{fig:summary_decay} are excluded since the corresponding DM particle 
(for a specific decay channel) produces a gamma-ray emission which is not 
compatible with the DGRB data. The scatter among the different lines in the 
figure is due both to the different statistical techniques used to derive the 
limits and to the different modeling adopted for the DM-induced emission. It 
should be noted that all lines are obtained by assuming that the DM-induced 
emission is the only component of the DGRB. The only exceptions are 
the dashed gray line from Refs.~\cite{Cirelli:2012ut} and the dashed blue one
from Ref.~\cite{Ando:2015qda}, in which the authors also model the 
astrophysical component of the DGRB. Both the Galactic and cosmological 
signals are considered when deriving all lower limits in 
Fig.~\ref{fig:summary_decay}, but the ones given by the solid gray and dashed
blue lines, which correspond to Ref.~\cite{Papucci:2009gd} and 
Ref.~\cite{Ando:2015qda}, respectively. In these two cases, limits refer to 
the cosmological DM component only. Other predictions for the contribution of 
decaying DM to the DGRB can be found in Refs.~\cite{Arina:2009uq,
Ishiwata:2010am,Fornasa:2012gu,Ibarra:2013cra}.

The most constraining lower limit in Fig.~\ref{fig:summary_decay} comes from
Ref.~\cite{Cirelli:2012ut} (dashed gray line) below $\sim 2$ TeV, and from
Ref.~\cite{Ando:2015qda} (dashed blue line) for larger DM masses. Decay 
lifetimes as large as $2 \times 10^{26}$ s are excluded for DM masses below 
$\sim 500$ GeV, while the limit goes up to $10^{28}$ s for $m_\chi \sim 10$ TeV. 
Note that the results of Ref.~\cite{Ando:2015qda} are obtained from the most
recent \fermi LAT measurement of the DGRB reported in 
Ref.~\cite{Ackermann:2014usa}. The authors also shows how the lower limit 
changes depending on the model employed to describe the diffuse Galactic 
foreground. Furthermore, the lower limit at large DM masses is found to 
vary by up to a factor of a few for different ways of parametrizing the 
astrophysical component of the DGRB and its uncertainty.

Compared to the limits derived from other DM targets, the dashed gray and 
dashed blue lines in Fig.~\ref{fig:summary_decay} represent the most 
constraining information available on $\tau$. In particular, in 
Fig.~\ref{fig:summary_decay} we show the regions excluded by three different 
analyses performed with \fermi LAT data, namely the observation of $i)$ the 
Segue 1 dwarf Spheroidal galaxy \cite{Dugger:2010ys} (blue region), $ii)$ 8 
galaxy clusters \cite{Huang:2011xr} (in dark gray) and $iii)$ the ``Galactic 
Center halo'' \cite{Ackermann:2012rg} (in light gray).

\section{The angular power spectrum of anisotropies}
\label{sec:angular_spectrum}
In this section we move away from the study of the all-sky average of the
DGRB, focusing on what can be learnt from its spatial fluctuations. Note that,
following the procedure outlined in Sec.~\ref{sec:intensity_measurement} and 
used in Ref.~\cite{Ackermann:2014usa}, the DGRB should be isotropic by 
construction. Yet, the template fitting is not sensitive to moderate 
small-scale anisotropies in the emission.\footnote{Moreover, as we will see in 
Sec.~\ref{sec:measurement_APS}, the measurement of the DGRB anisotropies 
performed in Ref.~\cite{Ackermann:2012uf} does not rely on the template 
fitting used in Ref.~\cite{Ackermann:2014usa}.} A well-established strategy 
to quantify the amount of spatial fluctuations is the anisotropy APS. 
Traditionally employed for the study of the CMB \cite{Ade:2013zuv}, the 
technique consists in decomposing a 2-dimensional map $I(\mathbf{n})$ in 
spherical harmonics $Y_{\ell,m}(\mathbf{n})$: 
$I(\mathbf{n}) = \sum_{\ell,m} a_{\ell,m} Y_{\ell,m}(\mathbf{n})$. The APS $C_\ell$
is, then, computed as follows:
\begin{equation}
C_{\ell} = \frac{\sum_{|m| \leq \ell} |a_{\ell,m}|^2}{2\ell+1}.
\label{eqn:intensity_APS}
\end{equation}

A measurement of the APS of the DGRB provides information on its composition 
that is complementary to the study of its intensity energy spectrum. In 
Sec.~\ref{sec:measurement_APS} we summarize the measurement of the DGRB 
anisotropies performed by the \fermi LAT in Ref.~\cite{Ackermann:2012uf}. 
Then, in Sec.~\ref{sec:implications_APS}, we describe how such a measurement 
can be used to constrain the DGRB contributors.\footnote{Other observables 
have been used to quantify the anisotropies in the DGRB: 
Refs.~\cite{Ando:2006mt,Ave:2009id,Xia:2011ax} consider the 2-point 
correlation function in real space, while Ref.~\cite{Soltan:2011nz} relies on 
the nearest-neighbor statistics. Ref.~\cite{Slatyer:2009zi} compares the 
number of ``isolated'' gamma-ray events with the ``empty regions'' in the sky. 
In this section, we focus only on the APS since it is the most commonly used 
technique.}

\subsection{The \fermi LAT measurement of gamma-ray anisotropies}
\label{sec:measurement_APS}
The measurement performed by the \fermi LAT Collaboration in 
Ref. \cite{Ackermann:2012uf} is currently the only observational data 
available on the APS of the DGRB. 22 months of data are analyzed between 1 and 
50 GeV, divided in 4 energy bins. Gamma rays are binned into a HEALPix 
map\footnote{http://healpix.jpl.nasa.gov/index.shtml} \cite{Gorski:2004by} 
with $N_{\rm side}=512$. The count maps are divided by the exposure of the 
instrument in order to obtain gamma-ray intensity maps. This is required in 
order to eliminate any spurious spatial fluctuations due to the non-uniform 
exposure of \fermi LAT.

Two definitions of the APS are used in Ref.~\cite{Ackermann:2012uf}. The
so-called {\it intensity} APS $C_\ell$ is obtained from 
Eq.~(\ref{eqn:intensity_APS}), while the {\it fluctuation} APS 
$C_\ell^{\rm fluct}$ comes from the decomposition of the {\it relative} 
fluctuations $I(\mathbf{n})/\langle I \rangle$, where $\langle I \rangle$ is
the all-sky average intensity. The two definitions are related by 
$C_\ell^{\rm fluct} = C_\ell / \langle I \rangle^2$. Note that the fluctuation APS 
is a dimensionless quantity, while $C_\ell$ inherits the units of the intensity 
map (squared).

Contrary to what is done in Ref.~\cite{Ackermann:2014usa} (see 
Sec.~\ref{sec:intensity_measurement}), no template fitting is employed in 
Ref.~\cite{Ackermann:2012uf} to isolate the DGRB. Nevertheless, a mask is 
applied, screening the regions in the sky where the emission is dominated by 
the diffuse Galactic foreground or by the resolved point sources. The mask 
covers the strip with $|b|<30^\circ$ around the Galactic plane and a 
$2^\circ$-radius circle around each source in the 1FGL catalog. The 
contamination of the Galactic foreground is not completely removed by the use 
of the mask, but the residual Galactic emission only induces large-scale 
features that contribute to the APS at small multipoles. That is why only 
multipoles larger than 105 are considered in Ref.~\cite{Ackermann:2012uf}.

The mask may alter the shape and normalization of the APS. Following 
Ref.~\cite{Komatsu:2001wu}, the APS is corrected for the effect of the mask
simply by dividing $C_\ell$ by the fraction of unmasked sky, $f_{\rm sky}$. 
Therefore, the APS estimator considered in Ref.~\cite{Ackermann:2012uf} is 
\begin{equation}
C_\ell = \frac{C_\ell^{\rm raw}/f_{\rm sky} - C_{\rm N}}{(W_\ell^{\rm beam})^2},
\label{sec:APS_signal}
\end{equation}
where $C_\ell^{\rm raw}$ is the intensity APS computed directly from the masked 
intensity maps. $C_{\rm N}$ is the so-called {\it photon noise}, i.e. 
$C_{\rm N} = \langle I \rangle^2 4 \pi f_{\rm sky} / N_\gamma$, where $N_\gamma$ is 
the total number of events detected outside the mask. Finally, the factor 
$W_\ell^{\rm beam}$ corrects for the smearing induced by the Point Spread Function 
(PSF) of the telescope (see Ref.~\cite{Ackermann:2012uf} for a definition of
$W_\ell^{\rm beam}$). The effect of the PSF becomes too extreme for angular 
scales beyond $\ell=504$ so that, in Ref.~\cite{Ackermann:2012uf}, no data 
are considered for multipoles larger than $\ell=504$.

\begin{figure}
\begin{center}
\includegraphics[width=0.49\linewidth]{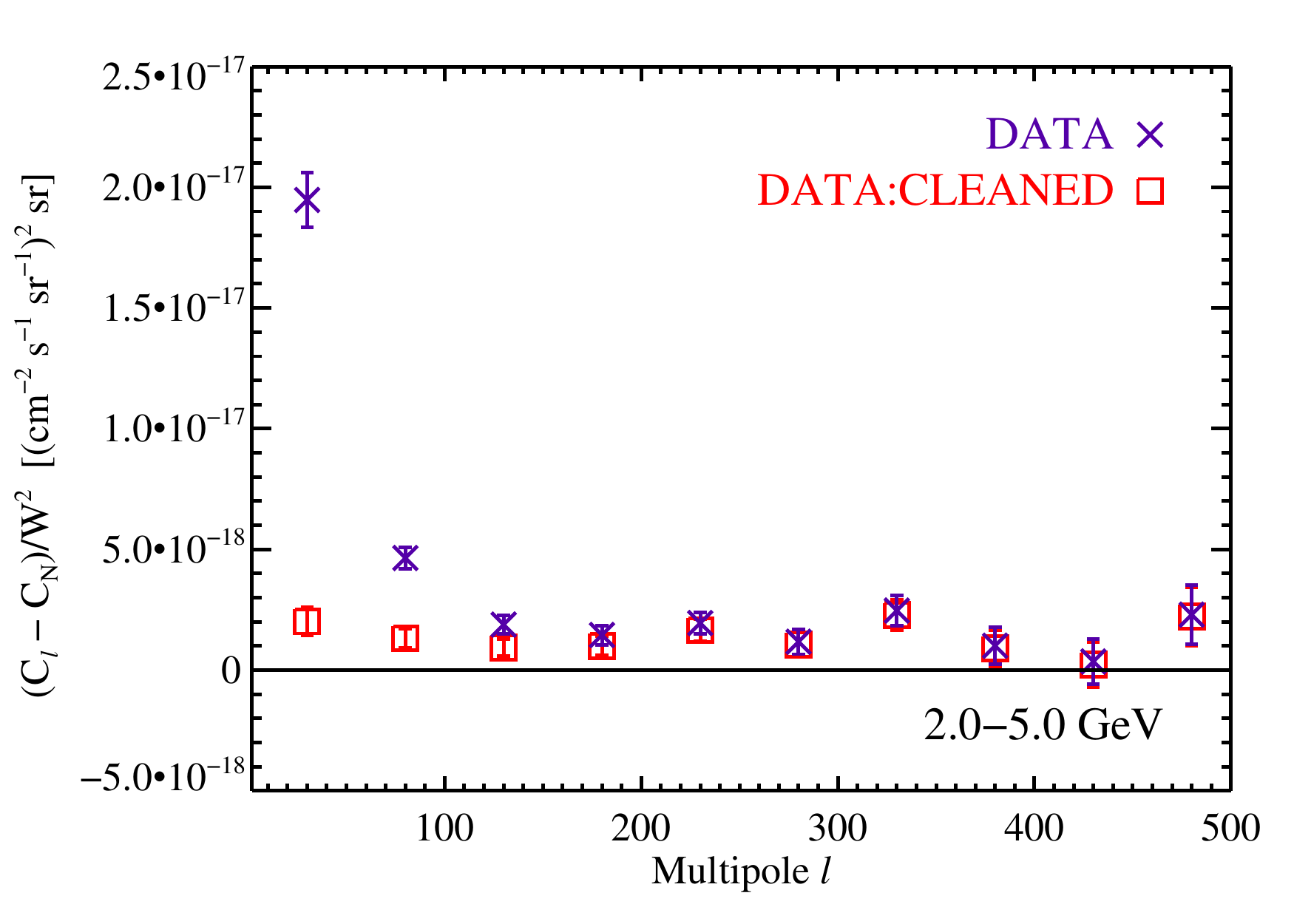}
\includegraphics[width=0.49\linewidth]{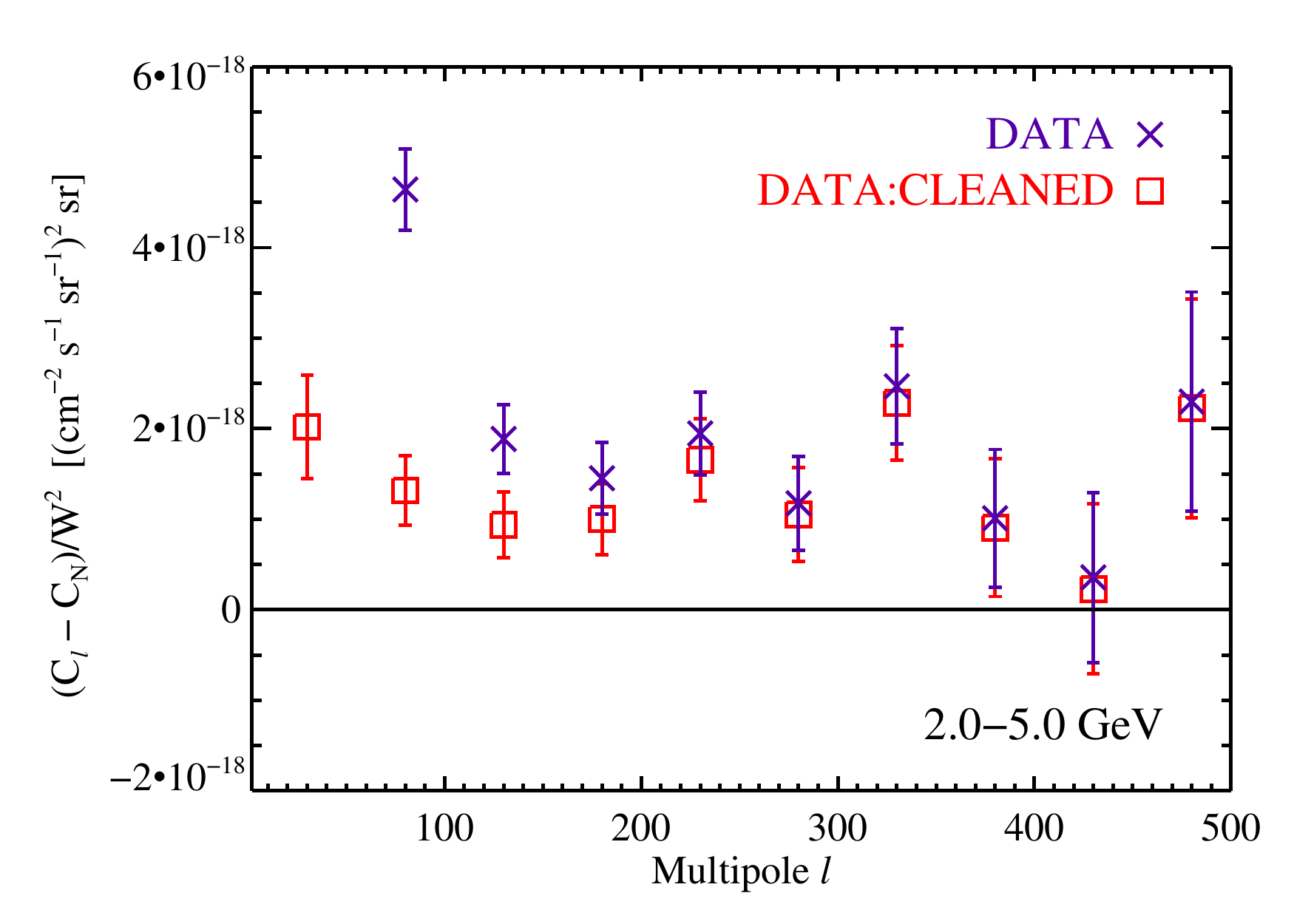}
\caption{\label{fig:APS_second_bin} Intensity APS of the data before (purple crosses) and after (red boxes) the removal of the diffuse Galactic foreground. The signal region is defined between $\ell=155$ and 504. The large increase in the intensity APS with no foreground subtraction (purple crosses) at $\ell<155$ is likely attributable to contaminations from the Galactic foreground emission. The right panel is an expanded version of the left panel and it focuses on the high-multiple angular power. Taken from Ref.~\cite{Ackermann:2012uf}.}
\end{center}
\end{figure}

Fig.~\ref{fig:APS_second_bin} shows the APS for gamma rays between 1.99 and 5.0 
GeV. The data points show the weighted average of the APS inside bins in 
multipoles with $\Delta\ell=50$. The purple crosses correspond to the APS 
estimator $C_\ell$ of Eq.~(\ref{sec:APS_signal}) derived from the intensity 
maps. On the other hand, the empty red boxes indicate the APS measured from 
the residual maps obtained after the subtraction of a model for the diffuse 
Galactic foreground. The error bars are computed following 
Ref.~\cite{Knox:1995dq}.\footnote{More recently, Ref.~\cite{Campbell:2014mpa} 
presented an alternative method to estimate the error on $C_\ell$, developed 
specifically for scenarios like the DGRB, where the emission is affected by 
limited statistics.}

The measured APS is different from zero in the signal region, i.e. 
$155 \leq \ell \leq 504$. The (unbinned) APS is fitted with a power law 
$(\ell/155)^n$ in order to determine any dependence of the APS on $\ell$. It 
is found that the $n=0$ case is compatible with the best fit at 95\% 
confidence level for all energy bins. This means that the measured APS is 
compatible with being Poissonian, i.e. independent of $\ell$. The significance 
of the detection is 6.5 (7.2) between 1.04 and 1.99 GeV (between 1.99 and 5.0 
GeV), while it decreases to 4.1 and 2.7 in the two remaining bins at higher 
energies.

The fluctuation APS of a population of sources depends on the energy only if 
their spatial clustering is itself energy-dependent or if they are 
characterized by significantly different energy spectra. On the other hand, 
their intensity APS scales with the energy as $\langle I \rangle^2$. When 
computing the APS of an emission that is the sum of different components (as 
in the case of the DGRB), it may be interesting to study how the APS of the 
total emission $C_{\ell}^{\rm total}$ depends on the spectra of the 
individual components $C_{\ell}^i$. By construction, the intensity APS is 
an addictive quantity:\footnote{In 
Eqs.~(\ref{eqn:intensity_APS_summation_rule}) and 
(\ref{eqn:fluctuation_APS_summation_rule}) we are neglecting, for simplicity, 
possible cross-correlation terms between different components. Under the 
hypothesis that gamma-ray sources trace the same LSS of the Universe, these 
cross-correlation terms can contribute significantly to the total APS 
$C_\ell^{\rm total}$ and, thus, should be taken into account \cite{Ando:2006cr,
Ando:2009nk,Ando:2013ff,Ando:2013xwa,Fornengo:2013rga,Ando:2014aoa}.}
\begin{equation}
C_\ell^{\rm total} = \sum C_{\ell}^i,
\label{eqn:intensity_APS_summation_rule}
\end{equation}
while the fluctuation APS follows the following summation rule:
\begin{equation}
C_\ell^{\rm fluct,total} = \sum f_i^2 \, C_\ell^{{\rm fluct,}i}
\label{eqn:fluctuation_APS_summation_rule}
\end{equation}
where $f_i$ is the fraction of the emission associated with the $i$-th 
contribution, with respect to the total, i.e. 
$f_i=\langle I_i \rangle / \langle I^{\rm total}\rangle$.

Assuming that the components have an energy-independent $C_\ell^{{\rm fluct},i}$,
any energy dependence in $C_\ell^{\rm fluct,total}$ must arise from the 
$f_i$-factors in Eq.~(\ref{eqn:fluctuation_APS_summation_rule}). Indeed, 
Ref.~\cite{SiegalGaskins:2009ux} proves that detecting an energy modulation in 
the fluctuation APS of the DGRB may indicate that the emission is the sum of 
more components (see also Refs.~\cite{Hensley:2009gh,Hensley:2012xj}). In 
that case, the behavior of the {\it intensity} APS as a function of energy 
would follow the energy spectrum of the dominant component. Thus, studying if 
(and how) fluctuation and intensity APS depend on the energy may be crucial to 
unravel the composition of the DGRB. 

\begin{figure}
\begin{center}
\includegraphics[width=0.45\linewidth]{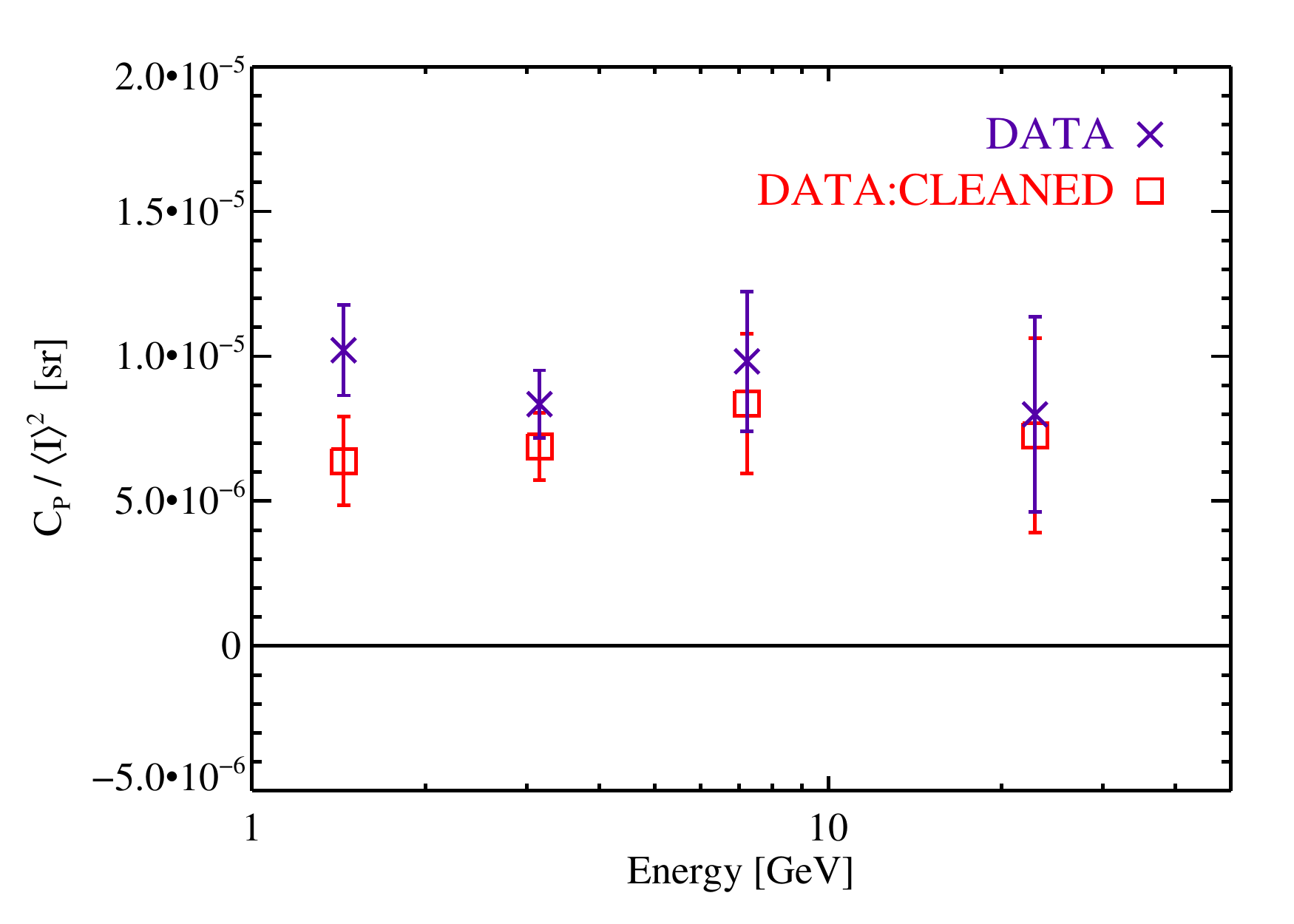}
\includegraphics[width=0.45\linewidth]{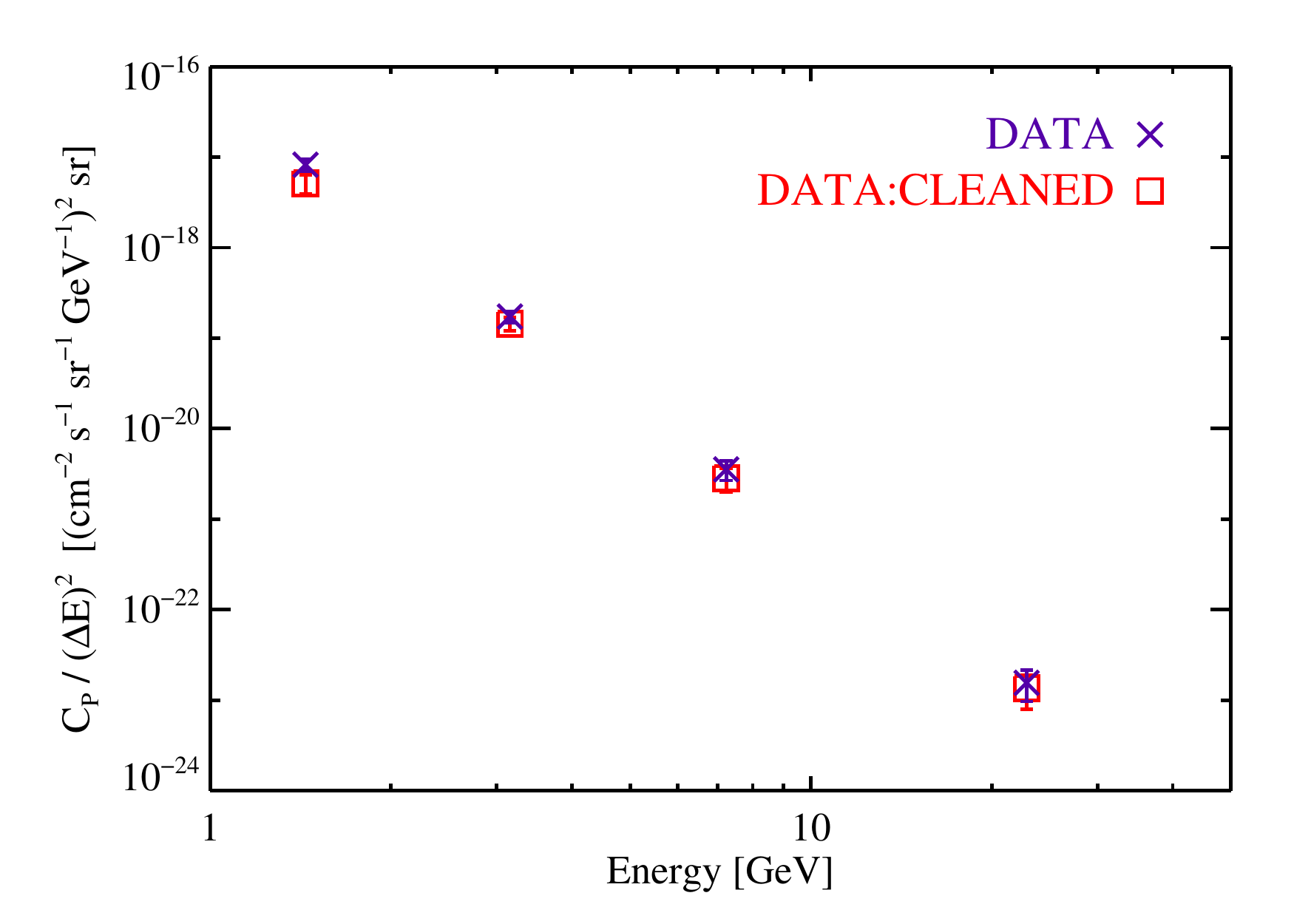}
\caption{\label{fig:APS_energy} Anisotropy energy spectra of the data (purple crosses) and of the Galactic-foreground-cleaned data (red boxes). {\it Left:} Fluctuation anisotropy energy spectrum. {\it Right:} Differential intensity anisotropy energy spectrum. Taken from Ref. \cite{Ackermann:2012uf}.}
\end{center}
\end{figure}

Fig. \ref{fig:APS_energy} shows the Poissonian $C_{\rm P}$ measured in 
Ref.~\cite{Ackermann:2012uf} in the 4 energy bins\footnote{Since, as commented
before, the APS is compatible with being constant in multipole, the whole APS
can be completely characterized by just one number, that we refer to as 
$C_{\rm P}$.}. The left panel proves that the fluctuation APS does not depend 
on the energy.\footnote{However, the large error bars and the fact that only 4 
energy bins are available do not allow to exclude large-scale modulations or 
very localized peaks.} On the other hand, the right panel shows that $C_{\rm P}$ 
decreases with energy as $C_{\rm P} \propto E^{-(4.79 \pm 0.13)}$. This result 
suggests that the \fermi LAT APS is produced by one single population of 
unclustered sources with an energy spectrum proportional to $\propto E^{-2.40}$. 
In the next section we will see that unresolved blazars fits this description.

After the publication of the \fermi LAT APS measurement, 
Ref.~\cite{Chang:2013ada} pointed out that the \fermi LAT Collaboration used 
22 months of data to measure the APS, but masked only the contribution of the 
point-like sources in the 1FGL (relative to an exposure of only 11 months). 
Thus, the emission considered in Ref.~\cite{Ackermann:2012uf} will probably 
be contaminated by the contribution of sources that, being not bright enough 
to be included in the 1FGL, could have been detected in the larger dataset 
of Ref.~\cite{Ackermann:2012uf}.

\subsection[Deducing the nature of the DGRB from its anisotropies]{Deducing the nature of the Diffuse Gamma-Ray Background from its anisotropies}
\label{sec:implications_APS}
We start this section by summarizing the technique used to estimate the
APS of a generic class of sources. The discussion will, then, focus on the
gamma-ray emitters considered in Secs.~\ref{sec:components} and \ref{sec:DM}. 
We will finally show how the comparison of the model predictions to the APS 
signal detected by the \fermi LAT in Ref.~\cite{Ackermann:2012uf} can be used 
to constrain the contribution of different populations to the DGRB.

As done at the beginning of Sec.~\ref{sec:components}, the formalism proposed 
here assumes that the sources are characterized by a generic parameter $Y$. 
However, Eq.~(\ref{eqn:unresolved_sources}) only determines the all-sky average 
gamma-ray emission associated with population $X$, referred to here as 
$\langle I_X \rangle$ to underline that it is obtained by integrating over all
the possible pointings in the sky. The emission $I_X(\mathbf{n})$ from 
direction $\mathbf{n}$ can be written as follows \cite{Ando:2005xg,Ando:2006mt,
Fornengo:2013rga}:
\begin{equation}
I_X(\mathbf{n}) = \int d\chi \, g_X(\chi,\mathbf{n}) W_X(\chi),
\label{eqn:average_emission_X}
\end{equation}
where $\chi=\chi(z)$ is the comoving distance relative to redshift $z$. 
$W_X(\chi)$ is the so-called {\it window function} and it gathers all the 
quantities that, in the definition of $I_X(\mathbf{n})$, do not depend on the 
direction of observation. In particular, $W_X(\chi)$ may depend on the energy 
at which $I_X$ is computed. The factor $g_X(\chi,\mathbf{n})$ is called the 
``source field'' and it describes how $I_X$ changes from point to point in the 
sky. It encodes the dependence on the abundance and distribution of the 
sources. The averaged source field, $\langle g_X \rangle$, can be used to write 
$\langle I_X \rangle$ as $\int d\chi \langle g_X(\chi) \rangle W_X(\chi)$. The 
average source field depends only on the abundance of sources and it can be 
written as:\footnote{Compared to Eq.~(\ref{eqn:unresolved_sources}), we are 
neglecting the dependence on the spectral shape $\Gamma$.}
\begin{equation}
\langle g_X(\chi) \rangle = \int_{Y_{\rm min}}^{Y_{\rm max}} dY 
\frac{dN}{dVdY}(\chi,Y).
\label{eqn:average_source_field}
\end{equation}

The $a_{\ell,m}^{{\rm fluct},X}$ coefficients of the fluctuation APS of population
$X$ can be computed decomposing the {\it relative} intensity fluctuations 
$I_X(\mathbf{n}) / \langle I_X \rangle$ in spherical harmonics, 
as follows:
\begin{equation}
a_{\ell,m}^{{\rm fluct},X} = \frac{1}{\langle I_X \rangle} \int d\Omega_{\mathbf{n}} \, 
I_X(\mathbf{n}) Y^\star_{\ell,m}(\mathbf{n}),
\label{eqn:alm_fluctuation_APS}
\end{equation}
where $d\Omega_{\mathbf{n}}$ indicates the angular integration. Now, from 
Eq.~(\ref{eqn:average_emission_X})
\begin{eqnarray}
a_{\ell,m}^{{\rm fluct},X} & = & \frac{1}{\langle I_X \rangle} \int d\Omega_{\mathbf{n}} 
\, \int d\chi g_X(\chi,\mathbf{n}) \, W_X(\chi) Y^\star_{\ell,m}(\mathbf{n}) \\
& = & \frac{1}{\langle I_X \rangle} \int d\Omega_{\mathbf{n}} \int d\chi \, 
f_{X}(\chi,\mathbf{n}) \langle g_X(\chi) \rangle W_X(\chi), \nonumber
\end{eqnarray}
where $f_{X}(\chi,\mathbf{n}) = g_X(\chi,\mathbf{n}) / \langle g_X \rangle$. 
The fluctuation APS is obtained from the average of $|a_{\ell,m}^{{\rm fluct},X}|^2$
with the same multipole $\ell$, see Eq.~(\ref{eqn:intensity_APS}). It can be 
shown \cite{Ando:2005xg,Fornengo:2013rga} that it is equivalent to:
\begin{equation}
C_{\ell}^{{\rm fluct},X} = 
\frac{1}{\langle I_X \rangle^2} \int \frac{d\chi}{\chi^2} \, 
\langle g_X(\chi) \rangle^2 {W}_X^2(\chi) \, 
P_{X} \left( k = \frac{\ell}{\chi},\chi \right),
\label{eqn:theory_APS}
\end{equation}
where $P_{X}(k,\chi)$ is the 3-dimensional power spectrum of the field 
$f_{X}(\chi,\mathbf{n})$:
\begin{equation}
\langle \tilde{f}_X(\chi,\mathbf{k}) \tilde{f}_X(\chi,\mathbf{k^\prime}) 
\rangle = (2\pi)^3 \delta^3 (\mathbf{k} + \mathbf{k^\prime}) P_{X}(k,\chi),
\end{equation}
and $\tilde{f}_X$ indicates the Fourier transform of the field $f_X$.
Eq.~(\ref{eqn:theory_APS}) makes use of the so-called Limber approximation
\cite{Limber:1954zz,Kaiser:1987qv}, valid for $\ell \gg 1$. This is indeed the
regime of interest here since the \fermi LAT APS measurement is robust only 
for $\ell \geq 155$.

The hypothesis of working with a collection of unresolved sources allows the 
decomposition of the 3-dimensional power spectrum $P_X$ into a 1-halo term and
a 2-halo term, $P_{1h}$ and $P_{2h}$, respectively. This helps in the 
interpretation of $P_X$ and it allows to express it easily in the context of 
the Halo Model. The 1-halo term accounts for the correlation between two 
points located within the same source, while $P_{2h}$ describes the correlation 
between points that reside in different objects and, thus, it depends on the 
spatial clustering of the sources. Within the Halo Model formalism, 1-halo and 
2-halo terms can be written, respectively, as follows 
\cite{Ando:2005xg,Ando:2006cr,Fornengo:2013rga}:
\begin{equation}
P_{1h}(k,z) = \int_{Y_{\rm min}(z)}^{Y_{\rm max}(z)} dY \, \frac{dN}{dVdY}(Y,z) \, 
\left( \frac{\tilde{u}(k|Y,z)}{\langle g_{X} \rangle} \right)^2
\label{eqn:1_halo}
\end{equation}
and
\begin{equation}
P_{2h}(k,z) = \left[ \int_{Y_{\rm min}(z)}^{Y_{\rm max}(z)} dY \, \frac{dN}{dVdY}(Y,z) \, 
b_{X}(Y,z) \, \frac{\tilde{u}(k|Y,z)}{\langle g_{X} \rangle} \right]^2 \,
P^{\rm lin}(k,z).
\label{eqn:2_halo}
\end{equation}
The factor $\tilde{u}(k|Y,z)$ is the Fourier transform of the radial brightness
profile of a source characterized by parameter $Y$ at a distance $z$. 
Astrophysical sources are normally considered to be point-like, i.e. 
intrinsically smaller than the PSF of the telescope at any energy. This 
implies that $\tilde{u}$ is proportional to the source gamma-ray flux $S$, 
without any dependence on $k$. In that case, the 1-halo power spectrum becomes 
Poissonian and it depends only on the abundance of sources. In fact, taking 
$Y=S$, Eq.~(\ref{eqn:1_halo}) can be re-written as follows:
\begin{equation}
P_{1h}^{\rm Poissonian} = \int_{S_{\rm min}(z)}^{S_{\rm max}(z)} dS \, \frac{dN}{dVdS}(S,z) \, 
\left( \frac{S}{\langle g_X \rangle} \right)^2.
\label{eqn:Poissonian}
\end{equation}

On the other hand, the 2-halo term in Eq.~(\ref{eqn:2_halo}) is obtained under
the hypothesis that the fluctuations in the source distribution traces the 
fluctuations in the matter field, except for a bias factor $b_X(Y,z)$. That is
the reason to include the linear power spectrum of matter fluctuations 
$P^{\rm lin}$ in Eq.~(\ref{eqn:2_halo}). Thus, even for point-like sources, the 
2-halo term inherits a dependence on $k$.

The balance between the 1-halo and 2-halo terms determines the shape of 
$P_X(k,\chi)$ and of $C_\ell$ through Eq.~(\ref{eqn:theory_APS}). For a 
population of point-like sources that are very bright but scarce in number,
$P_{1h}^{\rm Poissonian}$ usually dominates and it is difficult (if not impossible) 
to use the APS to extract information on their clustering. On the other hand, 
if the point-like emitters are very abundant, $P_{1h}^{\rm Poissonian}$ is smaller, 
allowing for some sensitivity to the 2-halo term. On the other hand, if the
sources appear extended, the 1-halo is no longer Poissonian and it is 
suppressed above a certain scale associated with the typical size of the 
sources \cite{Ando:2005xg,Ando:2006mt,Fornengo:2013rga}.

Following the formalism presented above, Refs.~\cite{Ando:2006cr,Ando:2006mt} 
compute the APS of unresolved blazars. Blazars are characterized in 
Refs.~\cite{Ando:2006cr,Ando:2006mt} by their gamma-ray luminosity 
($Y \equiv L_\gamma$) which is assumed to correlate with the X-ray luminosity 
$L_X$, see Sec.~\ref{sec:blazars}. The free parameters in the model are fitted 
to reproduce the abundance of sources detected by EGRET. 
Ref.~\cite{Ando:2006mt} finds that the APS of unresolved blazars is within 
reach of the \fermi LAT and that it is dominated by the Poissonian 1-halo term 
for multipoles larger than a few. Similar results are obtained in 
Ref.~\cite{Ando:2006cr}.

Ref.~\cite{Cuoco:2012yf} revises the predictions for the APS of unresolved 
blazars by modeling the differential source count distribution $dN/dS$ as 
done in Ref.~\cite{Collaboration:2010gqa}. Eq.~(\ref{eqn:Poissonian}) is used
to derive the {\it intensity} APS from $dN/dS$:
\begin{equation}
C_{\rm P} = \int_{0}^{S_{\rm thr}} \frac{dN}{dS} S^2 \, dS,
\label{eqn:APS_Poissonian}
\end{equation}
where $S_{\rm thr}$ is the flux sensitivity threshold for point-like sources.
$dN/dS$ is assumed to be a broken power law, tuned to reproduce the measured 
abundance of blazars. Interestingly, the best-fit $dN/dS$ corresponds to a 
model in which unresolved blazars exhibit an APS that is consistent with the 
value measured by the \fermi LAT in Ref.~\cite{Ackermann:2012uf}. This 
suggests that blazars {\it alone} are able to explain the whole APS signal. 
The same best-fit model predicts that unresolved blazars can account only for 
$\sim 30\%$ of the DGRB intensity reported by the \fermi LAT in 
Ref.~\cite{Abdo:2010nz} between 1 and 10 GeV.

The authors of Ref.~\cite{Harding:2012gk} employ a more detailed model of the 
blazar population, based on the correlation between $L_\gamma$ and $L_X$ and on 
a parametrization of their SED (see Ref.~\cite{Abazajian:2010pc} and 
Sec.~\ref{sec:blazars}). They confirm the existence of a scenario in which 
blazars fit {\it at the same time} the \fermi LAT $dN/dS$ and the measured 
APS. In this case, unresolved AGNs account for at most 4.3\% of the DGRB 
intensity in Ref.~\cite{Abdo:2010nz} (in any energy bin). The constraints 
obtained in Refs.~\cite{Cuoco:2012yf,Harding:2012gk} on the contribution of 
unresolved blazars to the DGRB intensity showcase how informative and 
complementary the study of gamma-ray anisotropies can be for the 
reconstruction of the composition of the DGRB.

\begin{figure}
\begin{center}
\includegraphics[width=0.9\linewidth]{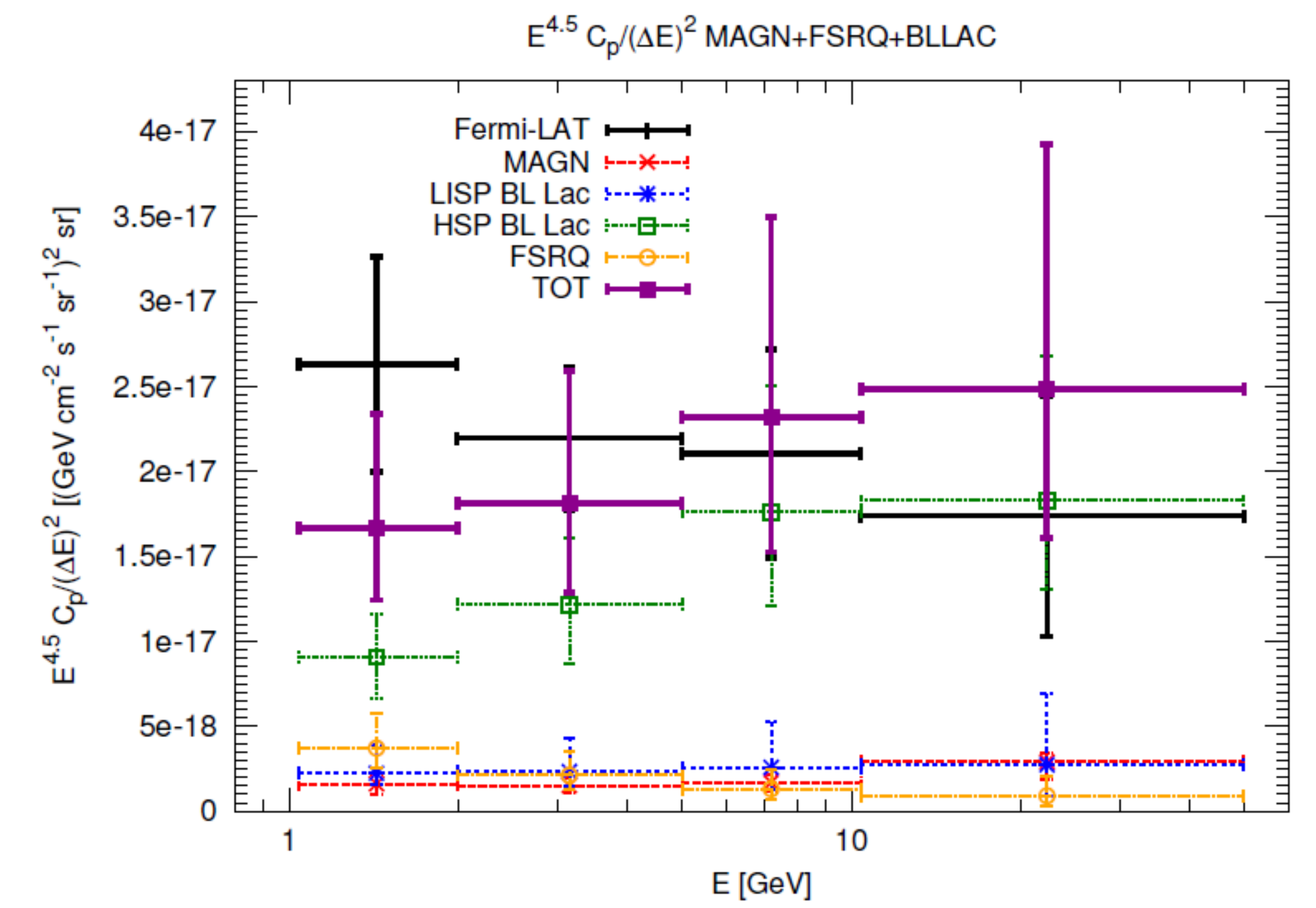}
\caption{\label{fig:APS_astrophysics} The angular power $C_{\rm P}(E)$ in units of $E^{4.5}C_{\rm P}(\Delta E)^{-2}$ for MAGNs (red long-dashed points), a class of sources combining BL Lacs LSPs and BL Lacs ISPs (blue short-dashed), HSP BL Lacs (green dotted), FSRQs (yellow dot-dashed) and the total (violet solid) of all radio-loud AGNs. The APS measurement by the \fermi LAT Collaboration is also shown (black solid points). Taken from Ref.~\cite{DiMauro:2014wha}.}
\end{center}
\end{figure}

The most recent estimate of the APS of blazars can be found in 
Ref.~\cite{DiMauro:2014wha}, where the authors consider three distinct AGN
subclasses: $i)$ FSRQs, $ii)$ HSP BL Lacs and $iii)$ a combination of LSP and 
ISP BL Lacs (see Sec.~\ref{sec:blazars}). Their Poissonian intensity APS is 
computed separately, according to Eq.~(\ref{eqn:APS_Poissonian}). The $dN/dS$ 
is taken from Ref.~\cite{Ajello:2011zi} for FSRQs and from 
Ref.~\cite{DiMauro:2013zfa} for the two classes of BL Lacs. They conclude that,
as found previously, unresolved blazars can explain the whole APS signal,
see Fig.~\ref{fig:APS_astrophysics}. HSP BL Lacs are responsible for the 
largest fraction of the measured intensity APS: between 1 and 2 GeV, they 
account for $34.5^{+9.5}_{-9.4}\%$ of the total APS signal and the fraction 
increases to $105^{+49}_{-30}\%$ above 10 GeV.

Ref. \cite{DiMauro:2014wha} also provides the first estimate of APS 
associated to MAGNs. It is computed from Eq.~(\ref{eqn:APS_Poissonian}), 
assuming that the 2-halo term can be neglected. The properties of MAGNs are 
inferred from a modeling of the sources at radio frequencies, via the 
$L_\gamma-L_{\rm r,core}$ relation discussed in Sec.~\ref{sec:MAGNs}. Unresolved 
MAGNs are found to contribute to approximately 10\% of the \fermi LAT APS 
(6.1\% between 1 and 2 GeV and 16.7\% above 10 GeV).

The APS of unresolved SFGs is computed in Ref.~\cite{Ando:2009nk}. SFGs are 
described assuming that their luminosity is proportional to the product of 
their SFR and of gas mass. The model is tuned to reproduce the properties of 
the MW. A power law with an index of 2.7 is assumed in Ref.~\cite{Ando:2009nk} 
as an universal energy spectrum (see Sec.~\ref{sec:SFGs}). The APS is computed 
from Eqs.~(\ref{eqn:1_halo}) and (\ref{eqn:2_halo}) assuming a bias factor of 
1.11, independent of redshift and luminosity \cite{Afshordi:2003xu} (see also 
Ref.~\cite{Xia:2011ax}). Contrary to the case of blazars or MAGNs, the APS of 
SFGs is not dominated by the Poissonian 1-halo term, at least below multipoles 
of few hundreds. As commented before, this is expected from a population of 
dim but very abundant sources. Thus, the signal may be used to constrain the 
clustering of SFGs. Unfortunately, the signal is overall too faint to 
contribute significantly to the APS detected by the \fermi LAT.

Refs.~\cite{SiegalGaskins:2010mp} and \cite{Calore:2014oga} performed Monte
Carlo simulations of the APS signal expected from unresolved MSPs. Sources are 
modeled combining radio and gamma-ray data (see Sec.~\ref{sec:MSPs}). 
The results of the two references differ from each other, due to the different 
models employed for the description of the MSPs: Ref.~\cite{Calore:2014oga} 
assumes that the APS of MSPs is Poissonian and it finds that MSPs are 
responsible for no more than 1\% of the APS measured by the \fermi LAT, while 
a larger fraction is allowed in the reference model considered by 
Ref.~\cite{SiegalGaskins:2010mp}.

Among the other astrophysical components considered in 
Sec.~\ref{sec:other_astrophysics}, we mention that the clustering of Type Ia
supernovae is considered in Ref.~\cite{Zhang:2004tj}: the authors find that
unresolved supernovae can exhibit a moderately large APS but the emission 
peaks in the MeV energy range. Refs.~\cite{Berezinsky:1996wx,
Colafrancesco:1998us,Waxman:2000pf,Keshet:2002sw,Keshet:2004dr,Ando:2006mt,
Zandanel:2014pva} study the case of galaxy clusters. In particular, the model 
in Ref.~\cite{Zandanel:2014pva} is tested against radio data and it proves 
that the intensity APS associated to galaxy clusters is expected to be two 
orders of magnitude lower than the APS measurement by the \fermi LAT.

\begin{figure}[h]
\begin{center}
\includegraphics[width=0.8\linewidth]{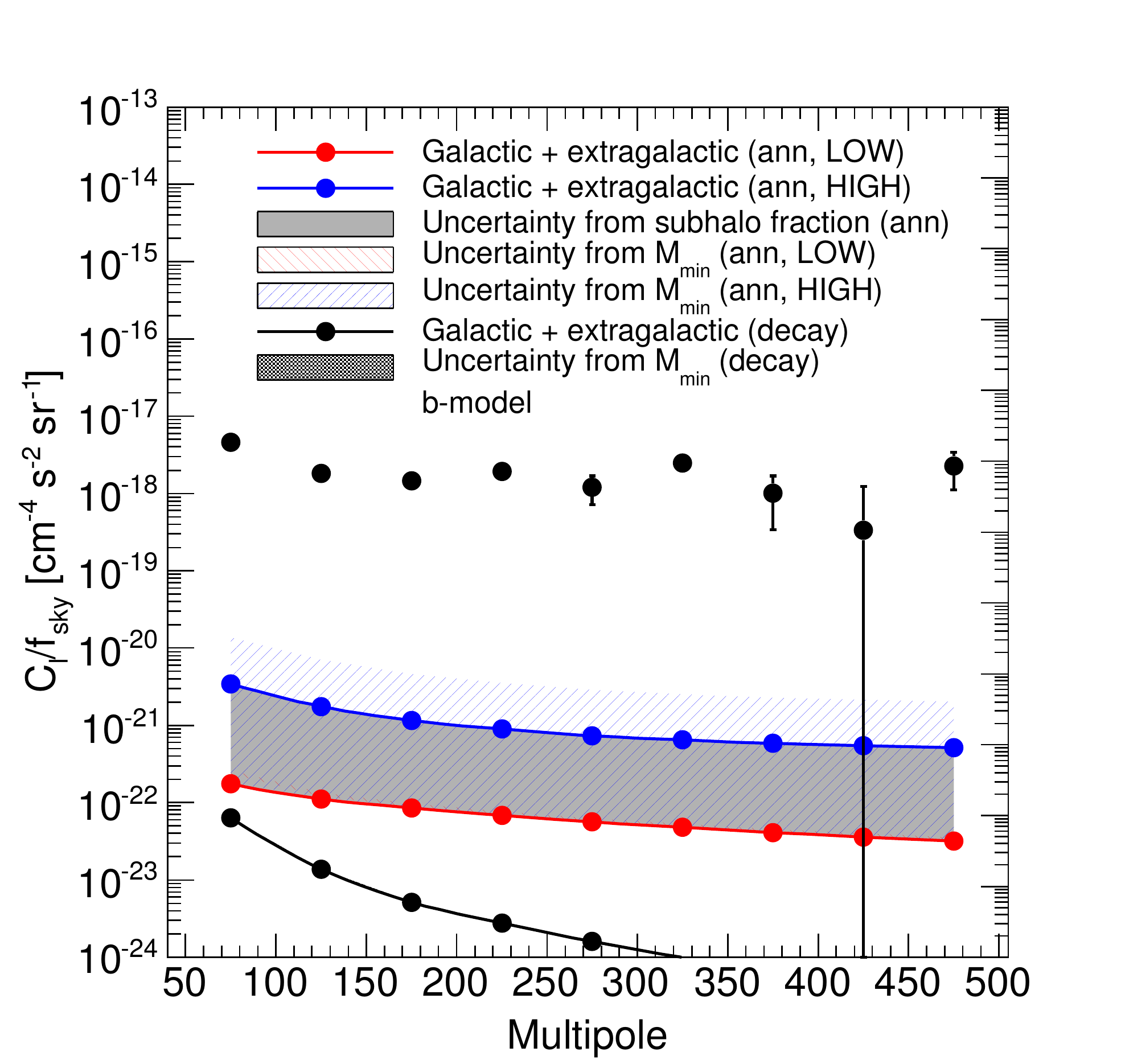}
\caption{\label{fig:APS_Fornasa_et_al} Total intensity APS of the gamma-ray emission from DM annihilation (color lines) or decay (black line) in extragalactic and galactic (sub)halos. The blue and red lines correspond to the LOW and HIGH subhalo boosts, respectively, so that the filled gray area between them corresponds to the uncertainty due to the subhalo boost, for a fixed value of $M_{\rm min}$. The red (blue) shaded area around the red (blue) solid line indicates the uncertainty in changing the value of $M_{\rm min}$ from $10^{−12}$ to 1 $M_\odot/h$, for the LOW (HIGH) case. The red-shaded area is very thin and difficult to see. The solid black line shows the prediction for a decaying DM candidate. The observational data points with error bars refer to the measurement of the APS between 2 and 5 GeV as given in Ref.~\cite{Ackermann:2012uf}. The predictions refer to a DM particle with a mass of 200 GeV, an annihilation cross section of $3 \times 10^{-26} \mbox{cm}^{3} \mbox{s}^{-1}$ and annihilations entirely into $b$ quarks (for annihilating DM) and to a  mass of 2 TeV, a decay lifetime of $2 \times 10^{27} \mbox{s}$ and decaying entirely into $b$ quarks (for decaying DM). Taken from Ref. \cite{Fornasa:2012gu}.}
\end{center}
\end{figure}

Turning to the case of gamma-ray emission induced by DM, the study of its 
angular anisotropies has also been traditionally considered as a very powerful
strategy to single out the DM component of the DGRB \cite{Campbell:2013rua}. 
Refs.~\cite{Ando:2005xg,Ando:2006cr} adopt the Halo Model approach to determine 
the 1-halo and 2-halo terms from Eqs.~(\ref{eqn:1_halo}) and 
(\ref{eqn:2_halo}), modeling the fluctuations in the gamma-ray emission 
produced by DM in extragalactic halos and subhalos. In Ref.~\cite{Ando:2005xg} 
the authors neglect the contribution of subhalos, which is included in 
Ref.~\cite{Ando:2006cr} assuming that the number of subhalos hosted by a halo 
with mass $M$ scales with $M$. Their results suggest that, depending on the 
value of $M_{\rm min}$, the DM-induced {\it fluctuation} APS can be within reach 
of the \fermi LAT and that it can be larger than the APS expected for 
unresolved blazars.\footnote{The formalism was extended to include $p$-wave 
annihilations in Ref.~\cite{Campbell:2011kf}.} Similar results are obtained 
also in Ref.~\cite{Cuoco:2007sh}, where only the 2-halo term is considered
and modeled from $N$-body simulation data.

The analytic formalism of Refs.~\cite{Ando:2005xg,Ando:2006cr} is extended
to the case of Galactic DM in Ref.~\cite{Ando:2009fp}. The DM halo of the MW 
is responsible only for large-scale anisotropies. On the other hand, the 
emission induced by Galactic subhalos can exhibit an APS that is even larger 
than the one from extragalactic DM structures. Results are tested in 
Ref.~\cite{Ando:2009fp} against different parametrizations of the Galactic 
subhalo population.\footnote{Ref.~\cite{Ando:2009fp} and the majority of the 
references mentioned in this section do not consider the effect of baryons on 
the clustering of DM and, thus, on its anisotropy pattern. One exception is 
Ref.~\cite{Taoso:2008qz}, which computes the APS expected by the so-called DM 
``mini-spikes'', i.e. DM overdensities induced by the presence of 
Intermediate-Mass Black Holes. The authors find that this scenario can lead to 
a significant increase in the amplitude of the DM-induced APS.}

\begin{figure}[h]
\begin{center}
\includegraphics[width=0.9\linewidth]{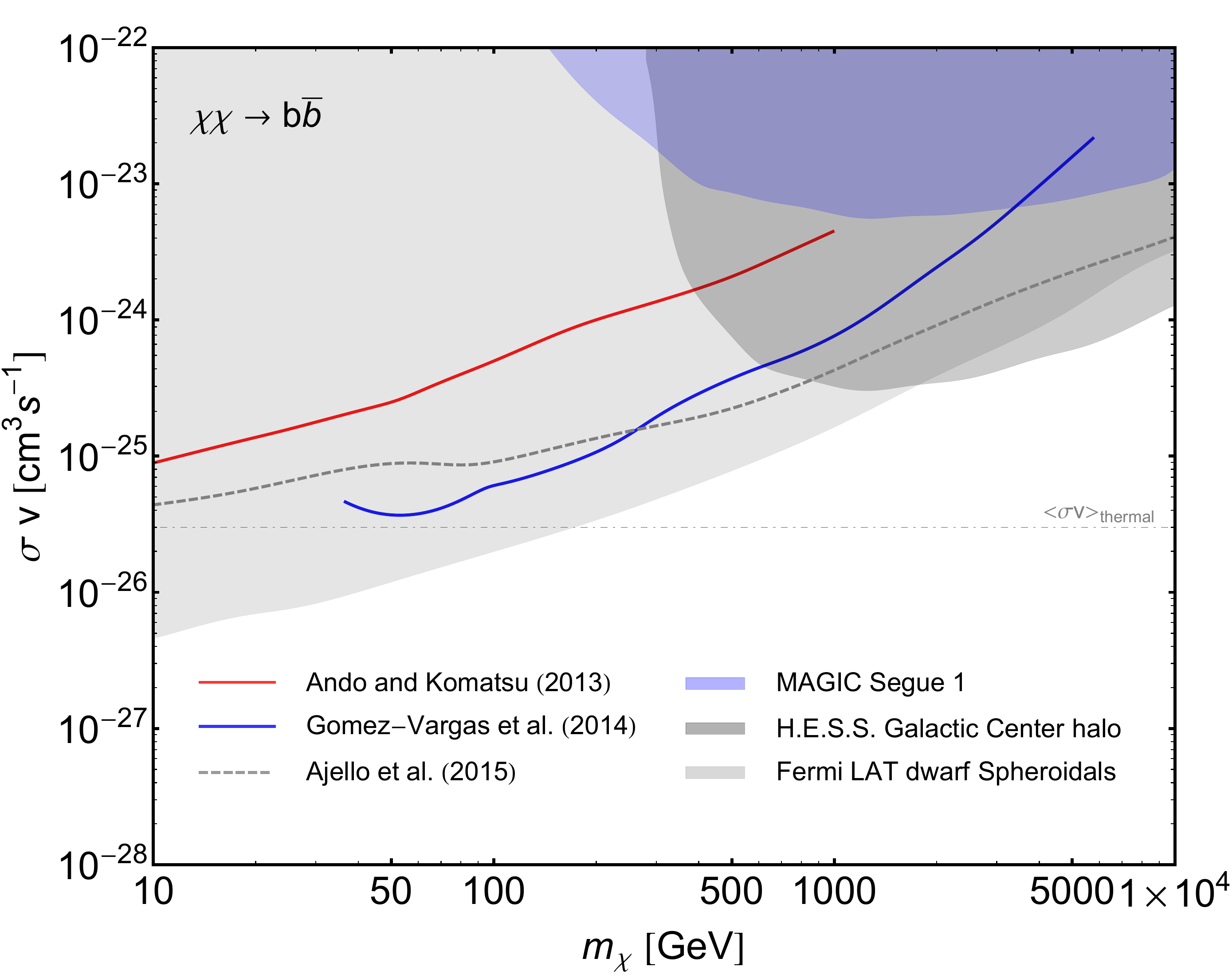}
\caption{\label{fig:APS_DM} Upper limits obtained by considering the DGRB APS measured in Ref.~\cite{Ackermann:2012uf}. The regions above the solid red and blue lines are excluded because the cumulative DM-induced emission would overproduce the APS of the DGRB. The solid red line is taken from Ref.~\cite{Ando:2013ff} while the solid blue one is from Ref.~\cite{Gomez-Vargas:2014yla}. The dashed gray line show the upper limits on $\langle \sigma v \rangle$ obtained by requiring that the DM-induced emission does not overproduce the DGRB emission measured in Ref.~\cite{Ackermann:2014usa}. The limit is obtained in Ref.~\cite{Ajello:2015mfa} by modeling the astrophysical DGRB contributors. The blue region indicates the region already excluded by the observation of the Segue 1 dwarf Spheroidal galaxy performed by the MAGIC telescope ~\cite{Aleksic:2013xea}. The dark gray region is excluded by the analysis performed by the H.E.S.S. telescope in Ref.~\cite{Abramowski:2011hc} on the so-called ``Galactic Center halo'' region (assuming an Einasto DM density profile). The light gray region indicates the DM candidates not compatible with the combined analysis of \fermi LAT data from 15 dwarf Spheroidal galaxies \cite{Ackermann:2015zua}. The dash-dotted line marks the thermal annihilation cross section.}
\end{center}
\end{figure}

Galactic and extragalactic DM signals are considered at the same time in 
Ref.~\cite{Ando:2013ff}. For multipoles larger than few tens, the APS of the 
extragalactic component is dominated by the 1-halo term, which depends mainly
on the amount of subhalos. In the fiducial model of Ref.~\cite{Ando:2013ff},
their luminosity is described by a power-law extrapolation of the 
$c_\Delta(z,M)$ relation. As discussed in Sec.~\ref{sec:DM_annihilation}, this 
probably leads to an overestimation of the DM signal, which is, anyway, much 
smaller than the measured intensity APS. The latter can be used to derive 
upper limits on the annihilation cross section, excluding the region in the 
$(m_\chi,\langle \sigma v \rangle)$ plane associated with too large 
anisotropies.\footnote{We note that, in general, the intensity APS associated
with DM scales quadratically on $\langle \sigma v \rangle$. The study of 
anisotropies is, therefore, quite sensitive to a DM signal. Whether the upper
limits on $\langle \sigma v \rangle$ derived from the \fermi LAT APS are more 
constraining than those from the DGRB intensity energy spectrum, will depend 
on the amplitude of the intrinsic fluctuations in the DM-induced gamma-ray 
emission. See also Fig.~\ref{fig:APS_DM}.} The upper limits obtained in 
Ref.~\cite{Ando:2013ff} exclude cross sections larger than 
$10^{-25} \mbox{cm}^3 \mbox{s}^{-1}$ ($5 \times 10^{-24} \mbox{cm}^3 \mbox{s}^{-1}$)
for a DM mass of 10 GeV (1 TeV), see the solid red line in 
Fig.~\ref{fig:APS_DM}. Similar results are obtained in 
Ref.~\cite{Camera:2014rja}.

Refs.~\cite{SiegalGaskins:2008ge,Fornasa:2009qh,SiegalGaskins:2009ux} follows
an alternative approach for the study of the anisotropies induced by Galactic 
DM subhalos. Instead of computing the APS analytically like in 
Refs.~\cite{Ando:2005xg,Ando:2006cr,Ando:2009fp,Cuoco:2007sh,Ando:2013ff}, 
Refs.~\cite{SiegalGaskins:2008ge,SiegalGaskins:2009ux} simulate sky-maps of 
the gamma-ray emission expected from mock realizations of DM subhalos in the 
MW. The APS is, then, computed by means of the HEALPix package (see also 
Ref.~\cite{Calore:2014hna}). In Ref.~\cite{Cuoco:2010jb} the results of 
Ref.~\cite{SiegalGaskins:2008ge} on the APS of Galactic subhalos are combined 
with the predictions in Ref.~\cite{Cuoco:2007sh} for the anisotropies of 
extragalactic DM to determine the sensitivity of \fermi LAT to the detection 
of a DM component in the DGRB through its APS. 

Similarly, Ref.~\cite{Fornasa:2009qh} employs a {\it hybrid approach} 
(inspired by Ref.~\cite{Pieri:2007ir}) in which Galactic DM subhalos are 
simulated only inside a sphere of radius $r_{\rm max}$ centered on the observer. 
The value of $r_{\rm max}$ is related to the distance beyond which subhalos 
become point-like. It is assumed that subhalos located further than $r_{\rm max}$ 
cumulatively generate a smooth emission. This is equivalent to assume that APS 
generated by subhalos beyond $r_{\rm max}$ is dominated by the 2-halo term.

Semi-analytic hybrid methods are used also in Refs.~\cite{Zavala:2009zr,
Fornasa:2012gu} to compute the anisotropies in the emission of extragalactic
DM structures. In this case, the distribution and properties of extragalactic
(sub)halos with a mass larger than the mass resolution of the Millennium-II 
$N$-body simulation \cite{BoylanKolchin:2009nc} are taken directly from the 
halo catalogs of the simulation. Mock sky-maps are generated by replicating 
the Millennium-II simulation box until it covers the region within 
$z \sim 2$.\footnote{Beyond this distance, at the energies considered, the 
gamma-ray flux is almost entirely attenuated by the interaction with the EBL.} 
DM halos less massive than the resolution of Millennium-II are included 
assuming that they share the same clustering of those immediately above the 
mass resolution \cite{Zavala:2009zr,BoylanKolchin:2009nc}. Subhalos of 
extragalactic DM clumps are also accounted for considering multiple scenarios 
for their abundance and internal properties. Ref.~\cite{Fornasa:2012gu} 
completes the prediction by modeling also the Galactic signal. The all-sky DM 
maps produced in Ref.~\cite{Fornasa:2012gu} represent complete and accurate 
templates of the total DM-induced gamma-ray emission.

As in Sec.~\ref{sec:DM_annihilation}, the main sources of uncertainties in the 
DM-induced APS are the value of $M_{\rm min}$ and the subhalo boost factor. 
Extreme scenarios are identified in Ref.~\cite{Fornasa:2012gu} for both 
$M_{\rm min}$ and the subhalo boost factor, bracketing their theoretical 
uncertainties. The effect of their variability on the intensity APS sums up 
to approximately two orders of magnitude, as it can be seen in 
Fig.~\ref{fig:APS_Fornasa_et_al}. The red and blue lines show the intensity 
APS for two different prescriptions of the subhalo boost, labeled ``LOW'' 
and ``HIGH''. The blue line is obtained following the prescription of 
Ref.~\cite{Gao:2011rf}, which employs power-law extrapolations for low-mass
halos. On the contrary, the red line makes use of the model by 
Ref.~\cite{Kamionkowski:2010mi}, extended in Ref.~\cite{SanchezConde:2011ap}. 
The blue-shaded region shows the effect of changing $M_{\rm min}$ on the HIGH 
scenario.\footnote{Changing $M_{\rm min}$ has no effect on the red line (LOW 
case), see Ref.~\cite{Fornasa:2012gu} for details.} The black data points 
indicate the \fermi LAT APS measurement between 2 and 5 GeV. As in 
Ref.~\cite{Ando:2013ff}, the DM-induced emission is found to contribute only 
marginally to the measured APS.

In Ref.~\cite{Gomez-Vargas:2014yla} the predictions of the HIGH case (blue 
line) are, then, used to derive upper limits on the DM annihilation cross 
section, shown as a solid blue line in Fig.~\ref{fig:APS_DM}. The upper limit 
reaches values as low as $3 \times 10^{-26} \mbox{cm}^3 \mbox{s}^{-1}$ for 
$m_\chi=50$ GeV, while going to $10^{-23} \mbox{cm}^3 \mbox{s}^{-1}$ if the mass 
approaches the TeV scale. Fig.~\ref{fig:APS_DM} also summarizes other 
constraints on $\langle \sigma v \rangle$ available in the literature. In 
particular, we plot the limit from the analysis of Ref.~\cite{Ajello:2015mfa}
of the DGRB intensity energy spectrum (dashed gray lines). The shaded regions 
indicate the areas already excluded by other indirect searches of DM, i.e. the 
observation of Segue 1 with MAGIC (blue), of the Galactic Center halo region 
with H.E.S.S. (dark gray) and the \fermi LAT combined analysis of dwarf 
Spheroidals (light gray). Fig.~\ref{fig:APS_DM} proves that the study of the 
anisotropies of the DGRB can produce competitive upper limits on 
$\langle \sigma v \rangle$, but probably not as strong as those induced by 
the DGRB energy spectrum.

Ref.~\cite{Fornasa:2012gu} also estimates the APS associated with a decaying
DM candidate. In this case, as found also by Ref.~\cite{Ibarra:2009nw}, the
predictions are subject to less theoretical uncertainties than for an 
annihilating DM candidate. In fact, the signal is less affected by the value 
of $M_{\rm min}$ and there is no subhalo boost (see also 
Sec.~\ref{sec:DM_decay}). Yet, in decaying DM scenarios, DM halos yield a more
extended emission. This is particularly true for Galactic subhalos which are 
still close enough not to be point-like. Thus, the APS is expected to 
decreases rapidly at high multipoles being, therefore, hard to detect. See, 
e.g., the black line in Fig.~\ref{fig:APS_Fornasa_et_al}.

\section{The photon count distribution}
\label{sec:1-point}
Another powerful statistic tool to constrain the nature of the DGRB is 
provided by the photon count Probability Distribution Function (PDF). This 
technique can be used when the emission is represent by a pixelated sky-map. 
The photon count PDF is, then, built from the number of pixels $n_k$ in which 
$k$ photons are detected. The study of the photon count PDF is commonly used 
in radio and X-ray astronomy for the analysis of diffuse emissions, in 
particular when trying to estimate the contribution of faint unresolved 
sources. It can also be used at gamma-ray frequencies to single out different 
components in the DGRB, even if they are subdominant. Indeed, different PDFs 
are expected for different populations of gamma-ray emitters: bright but rare 
sources generate pixels with a large (or moderately large) number of photons, 
i.e. $n_k$ will be significantly different from zero even at large $k$. On the 
other hand, a population of faint but numerous sources is normally associated 
with a Poissonian PDF. Intrinsically diffuse emissions also correspond to 
Poissonian PDFs. 

\begin{figure}[h]
\begin{center}
\includegraphics[width=0.9\linewidth]{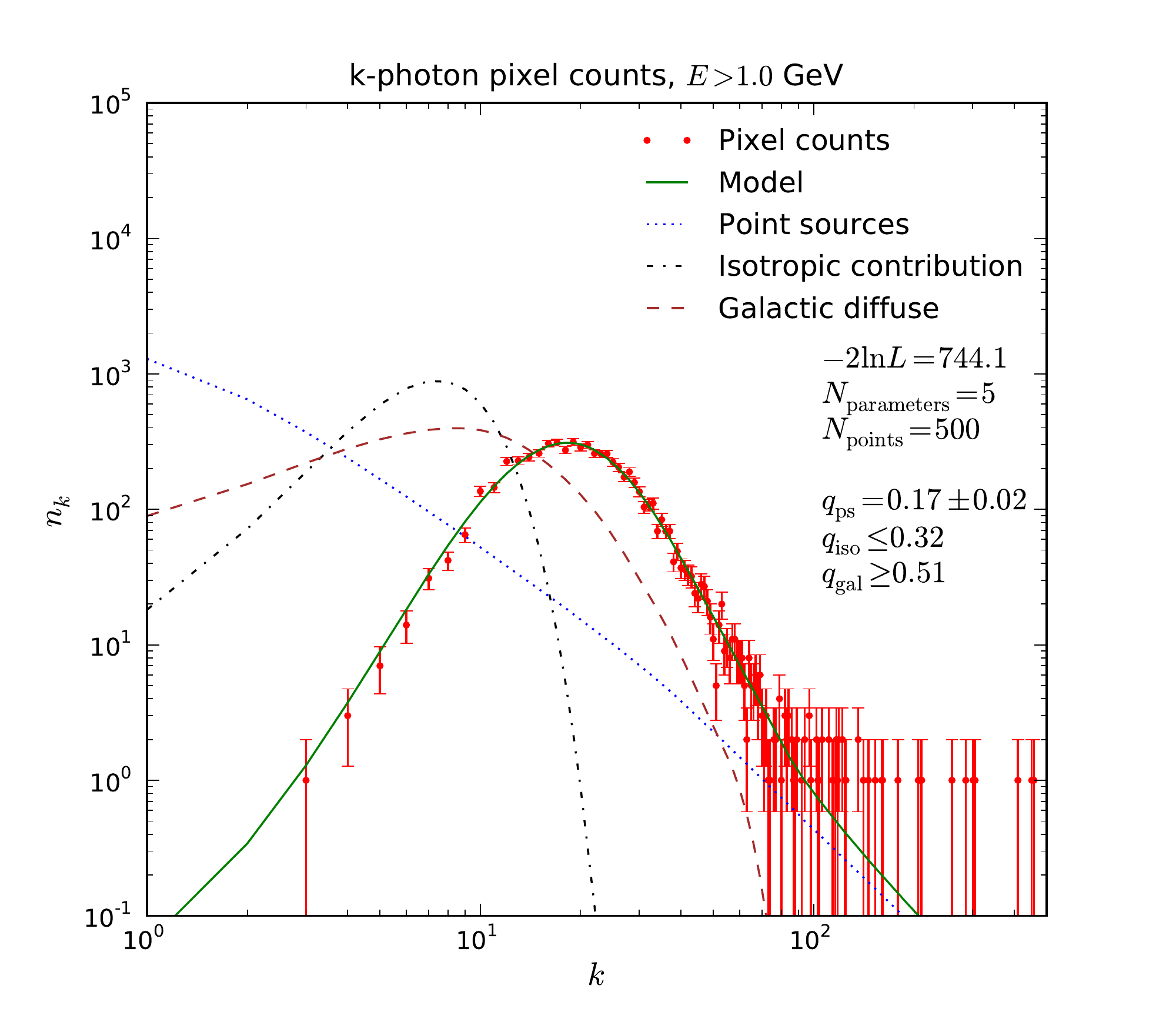}
\caption{\label{fig:PDF_Malyshev_et_al} $n_k$ is the number of pixels with $k$ photons. The red dots correspond to the pixel counts derived from the data of the \fermi LAT, the error bars are equal to $\sqrt{n_k}$. The model devised to interpret the data considers three classes of sources: $i)$ AGN-like point-like objects (blue dotted line), $ii)$ isotropic Poisson contribution (brown dashed line) and $iii)$ anisotropic Galactic diffuse emission (black dash-dotted line). Note that the photon count PDF for the total (solid black line) on this plot is not the sum of the components, but the corresponding generating function of the PDF is the product of the generating functions of the three contributions. $N_{\rm parameters}$ is the number of parameters used in the fit. $q_{\rm ps}$, $q_{\rm iso}$ and $q_{\rm gal}$ are the relative contributions of the point sources, of the isotropic and of the Galactic foreground, respectively. $N_{\rm points}$ is the number of points in the $x$-axis employed in the fit and the log-likelihood of the best-fit is indicated in the legend. Taken from Ref.~\cite{Malyshev:2011zi}.}
\end{center}
\end{figure}

Ref.~\cite{Malyshev:2011zi} measures the PDF directly from 11 months of \fermi
LAT data between 1 and 300 GeV. The data are binned into a HEALPix map with 
$N_{\rm side}=32$, corresponding to a binsize of approximately $0.4^\circ$. A 
region of $30^\circ$ around the Galactic plane is masked, while point sources 
are not masked. The observed PDF is represented as red data points in 
Fig.~\ref{fig:PDF_Malyshev_et_al}. A model based on the so-called 
{\it generating functions} is developed to interpret the data. If $p_k$ is the 
probability of finding $k$ photons in a certain pixel (i.e. 
$p_k=n_k/N_{\rm pixels}$, where $N_{\rm pixels}$ is the total number of pixels
considered in the analysis), the corresponding generating function $P(t)$
is defined as 
\begin{equation}
P(t)=\sum_{k=0}^{\infty} p_k t^k.
\end{equation}
Viceversa, the probabilities $p_k$ can be derived from $P(t)$ as the 
coefficients in front of each term in the power-law expansion of $P(t)$. The 
generating function of an emission composed by multiple components is the 
product of the generating functions of the single components \cite{Hoel:1971}. 
In particular, the model considered in Ref.~\cite{Malyshev:2011zi} consists of 
3 terms:
\begin{itemize}
\item point sources, both resolved and unresolved. The corresponding 
generating function is $P(t)=\exp [\sum_{m=1}^\infty (x_mt^m-x_m)]$, where $x_m$ 
is the average number of sources emitting exactly $m$ photons in a pixel. The
set of $\{x_m\}$ can be derived from the differential source count 
distribution $dN/dS$. In Ref.~\cite{Malyshev:2011zi}, $dN/dS$ is taken to be a 
broken power law \cite{Collaboration:2010gqa}, leaving its 4 parameters free.
\item the diffuse Galactic foreground, whose emission, in each pixel, is 
interpreted as a collection of sources that emit exactly 1 photon. The number
of those sources is proportional to the intensity of the foreground in that 
pixel. Its generating function $P(t,i)$ depends on the particular pixel $i$
considered and it can be written as 
$P(t,i)=\exp(x_{\rm diff}^i t - x_{\rm diff}^i)$, where $x_{\rm diff}^i$ is the
number of photons of foreground emission in the $i$-th pixel. In 
Ref.~\cite{Malyshev:2011zi}, the intensity and morphology of the foreground 
are derived from the \fermi LAT data itself.
\item an isotropic component, included to represent the emission associated 
to faint abundant sources. As for the Galactic foreground, this term is 
modeled as a collection of $x_{\rm iso}$ sources emitting exactly 1 photon and 
its generating function is $P(t)=\exp(x_{\rm iso}t - x_{\rm iso})$, for all the 
pixels. The quantity $x_{\rm iso}$ is left free in the fit.
\end{itemize}

The model is fitted to the photon count PDF in 
Fig.~\ref{fig:PDF_Malyshev_et_al} in order to determine the free parameters, 
i.e. those characterizing $dN/dS$ and the normalization of the isotropic 
component. The best-fit $dN/dS$ is compatible with what found in 
Ref.~\cite{Collaboration:2010gqa}. The analysis of the PDF is different and 
complementary to Ref.~\cite{Collaboration:2010gqa} but it succeeds in 
reconstructing the properties of a population of non-Poissonian blazar-like 
sources. By integrating the best-fit $dN/dS$ obtained in 
Ref.~\cite{Malyshev:2011zi} below the sensitivity of the \fermi LAT, 
unresolved blazars are found to account for $\sim 23\%$ of the DGRB reported 
in Ref.~\cite{Abdo:2010nz} above 1 GeV. We remark that this result is 
consistent with the independent estimation obtained in 
Ref.~\cite{Collaboration:2010gqa}. Similar results are also obtained in 
Ref.~\cite{Dodelson:2009ih} using 5 years of simulated \fermi LAT data and a 
pixel size of $0.25^\circ$. 

Predictions for the PDF are available not only for blazars: 
Ref.~\cite{FaucherGiguere:2009df} computed the PDF expected from unresolved
MSPs finding that their PDF is highly non-Poissonian.

In the case of DM-induced emission, the shape of the PDF is expected to depend 
on which DM structures are considered. Ref.~\cite{Lee:2008fm} computes the 
probability $P(F)$ of detecting a certain flux $F$ due to Galactic DM subhalos 
in a generic pixel. The probability depends on the modeling of the subhalo 
population and, in particular, on the value of $M_{\rm min}$. It can be written 
as follows (see the Appendix of Ref.~\cite{Lee:2008fm} for a detailed 
derivation):
\begin{equation}
P(F) = \mathcal{F}^{-1} \{ \exp [ \mu ( \mathcal{F}\{P_1(F)\} - 1 )] \},
\end{equation}
where $\mathcal{F}\{f(x)\}$ indicates the Fourier transform of $f(x)$ and
$\mathcal{F}^{-1}$ its inverse. The quantity $\mu$ is the average number of 
sources expected inside one pixel and $P_1(F)$ is the probability of having
exactly one source emitting the flux $F$. The DGRB model in 
Ref.~\cite{Lee:2008fm} also includes a background component with a Poissonian 
PDF. The authors prove that there exists a region in their parameter space in
which Galactic DM subhalos are faint enough to escape detection as individual
sources after 5 years of \fermi LAT data, but bright enough to be detected 
by studying the photon count PDF. 

Ref.~\cite{Baxter:2010fr} also studies the possibility of separating the
emission associated with Galactic DM subhalos from the diffuse Galactic 
foreground and from an isotropic background (both characterized by a Poissonian 
PDF). The authors choose some benchmark models for the description of the 
DM-induced emission and of the Poissonian backgrounds. After building the PDFs 
of the different components, they use them to generate mock sky-maps of the 
expected gamma-ray emission. This mimics a real observation and it allows to 
estimate the sensitivity of a PDF analysis to the DM signal. The mock photon 
count PDF is compared to model predictions in order to determine the free 
parameters in the model, as done in Ref.~\cite{Malyshev:2011zi}. This relies 
on a {\it a priori} knowledge of the shape of the PDF for the different 
components in the model. Ref.~\cite{Baxter:2010fr} proves that, in an 
idealized case without background, the reconstructed intensity of the 
DM-induced emission depends significantly on the assumed shape of the PDF. 
Employing the wrong PDF would lead to biased reconstruction of the intensity
of the DM component. However, when the Poissonian backgrounds are included, 
Ref.~\cite{Baxter:2010fr} finds that an unbiased reconstruction (within 
statistical uncertainties) can be obtained independently of the shape of the 
PDF.

Ref.~\cite{Feyereisen:2015cea} considered the gamma-ray emission of 
extragalactic DM halos and subhalos, instead. The authors derive $P_1(F)$ 
analytically, relying on the halo model, and they show how its shape changes
when using three different prescriptions for the subhalo boost. Then, they
compute $P(F)$ by means of the central limit theorem below a reference flux 
$F_\star$, and through MC simulations above that. The resulting flux 
distribution $P(E)$ exhibits two regimes: below $~5 \times 10^{-12} 
\mbox{cm}^{-2}\mbox{s}^{-1}\mbox{GeV}^{-1}\mbox{sr}^{-1}$ it follows a Gaussian 
distribution, while it is a power law for larger fluxes.\footnote{The 
gamma-ray flux is computed at 1 GeV.} The latter is the case when a few 
bright DM structures dominate the flux distribution. 
Ref.~\cite{Feyereisen:2015cea} also developed a model for the $P(F)$ of the
astrophysical components of the DGRB: for a fiducial subhalo boost model 
inspired by Ref.~\cite{Sanchez-Conde:2013yxa}, the authors of 
Ref.~\cite{Feyereisen:2015cea} show that a measurement of $\{p_k\}$ after 5 
years of \fermi LAT data can lead to a detection of a DM signal, provided 
that the annihilation cross section is, at least, twice the thermal value, for 
a DM mass of 85 GeV and annihilations into $b$ quarks.

\section{The cross-correlation with independent probes}
\label{sec:cross_correlation}
The most recent development in our understanding of the composition of the 
DGRB has focused on the study of its cross-correlation with other observables.
For instance, the fraction of the DGRB that originates from extragalactic
objects (whether astrophysical sources or DM halos) traces the LSS of the 
Universe, up to a maximal redshift that depends on the EBL attenuation. Thus, 
a certain level of cross-correlation is expected with any LSS tracer, e.g. the 
distribution of resolved galaxies \cite{Xia:2011ax,Ando:2013xwa,
Ando:2014aoa,Xia:2015wka} or the gravitational lensing effect of cosmic shear 
\cite{Camera:2012cj,Shirasaki:2014noa,Camera:2014rja}. These are new and 
independent observables that can complement the information inferred from the 
DGRB energy spectrum (see Sec.~\ref{sec:energy_spectrum}) or from its 
auto-correlation APS (see Sec.~\ref{sec:angular_spectrum}). Also, note that 
the gamma-ray sky-maps analyzed in Ref.~\cite{Ackermann:2012uf} to measure the 
auto-correlation APS are noise dominated. Thus, even if it was still possible 
to report a significant auto-correlation signal by subtracting the photon 
noise, one may expect the cross-correlation with other signal-dominated 
quantities to be, in principle, very informative.

In the following sections, we present the formalism proposed to predict the 
cross-correlation of the DGRB with LSS tracers. We also summarize the data 
currently available. Sec.~\ref{sec:correlation_galaxies} focuses on the 
cross-correlation with galaxy catalogs, while in 
Sec.~\ref{sec:correlation_shear} we discuss the case of the cosmic shear. 
Sec.~\ref{sec:correlation_others} presents the results of the 
cross-correlation of the DGRB with other observables.

\subsection{The cross-correlation with galaxy catalogs}
\label{sec:correlation_galaxies}
Ref.~\cite{Xia:2011ax} was the first work that measured the 2-point 
correlation function of the \fermi LAT DGRB with 4 galaxy surveys, namely: 
$i)$ the Sloan Digital Sky Survey (SDSS) Data Release 6 of optically-selected 
quasars from Ref.~\cite{Richards:2008eq}, $ii)$ the IR galaxies of the 2 
Micron All-Sky Survey (2MASS) Extended Source Catalog from 
Ref.~\cite{Jarrett:2000qt}, $iii)$ the radio sources from the NRAO VLA Sky 
Survey (NVSS) \cite{Condon:1998iy} and $iv)$ the luminous red galaxies in SDSS 
Data Release 8 from Ref.~\cite{Abdalla:2008ze}. The authors of 
Ref.~\cite{Xia:2011ax} analyzed 21 months of \fermi LAT data but no 
significant cross-correlation was observed. More recently, 
Ref.~\cite{Xia:2015wka} have updated the analysis using 60 months of data and 
exploring also the cross-correlation with the main galaxy sample of SDSS Data 
Release 8 from Ref.~\cite{Aihara:2011sj}. The region with $|b|<30^\circ$ around 
the Galactic plane was masked out, as well as the 1-degree region around the 
point sources in the 3FGL catalog. Both the Large and Small Magellanic Clouds 
were also left out of the analysis, together with the Fermi Bubbles and of the 
so-called Loop-I. A model for the diffuse Galactic foreground was subtracted 
from the sky maps. The residuals of the gamma-ray maps and the distribution of 
galaxies in the catalogs were re-binned into HEALPix maps with 
$N_{\rm side}=512$. 

The authors computed both the cross-correlation APS and the 2-point 
correlation functions\footnote{We note that the two are related by a Fourier 
transform and they contain the same information. However, they probe different 
scales with different efficiency and, thus, it is interesting to consider 
both.} using the PolSpice 
package\footnote{http://www2.iap.fr/users/hivon/software/PolSpice/}. The 
signal region was defined between 0.1 and 100 degrees for the 2-point 
correlation function and between multipoles of 10 and 1000 for the 
cross-correlation APS. Three different energy thresholds were considered for 
the gamma rays, including all events above $i)$ 500 MeV, $ii)$ 1 GeV or $iii)$ 
10 GeV. The authors in Ref.~\cite{Xia:2015wka} also validated their results 
against changes in the mask and in the model adopted for the diffuse Galactic 
foreground. They also tested their analysis pipeline on a simulated sky map 
with no signal.

\begin{figure}
\begin{center}
\includegraphics[width=0.46\linewidth]{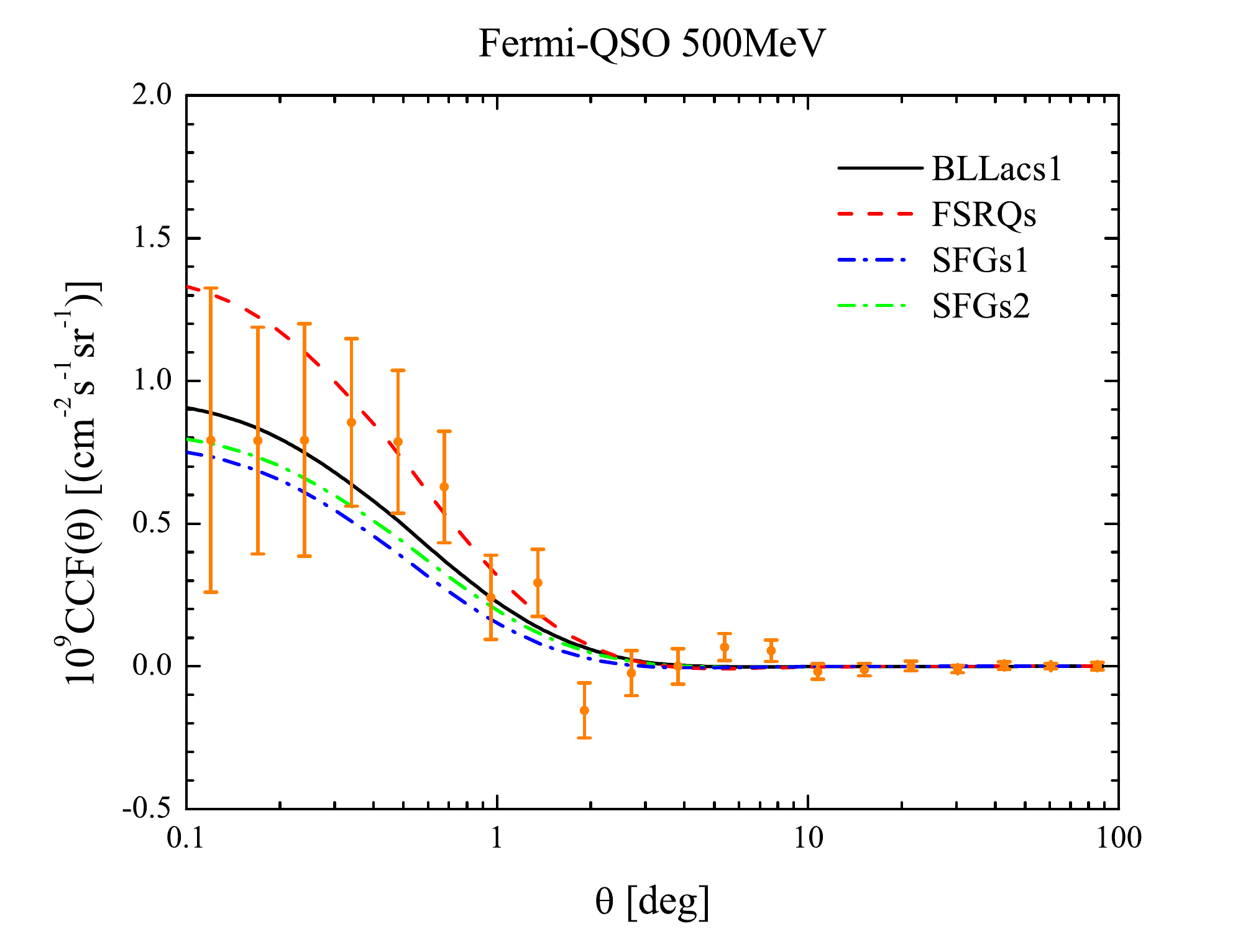}
\includegraphics[width=0.46\linewidth]{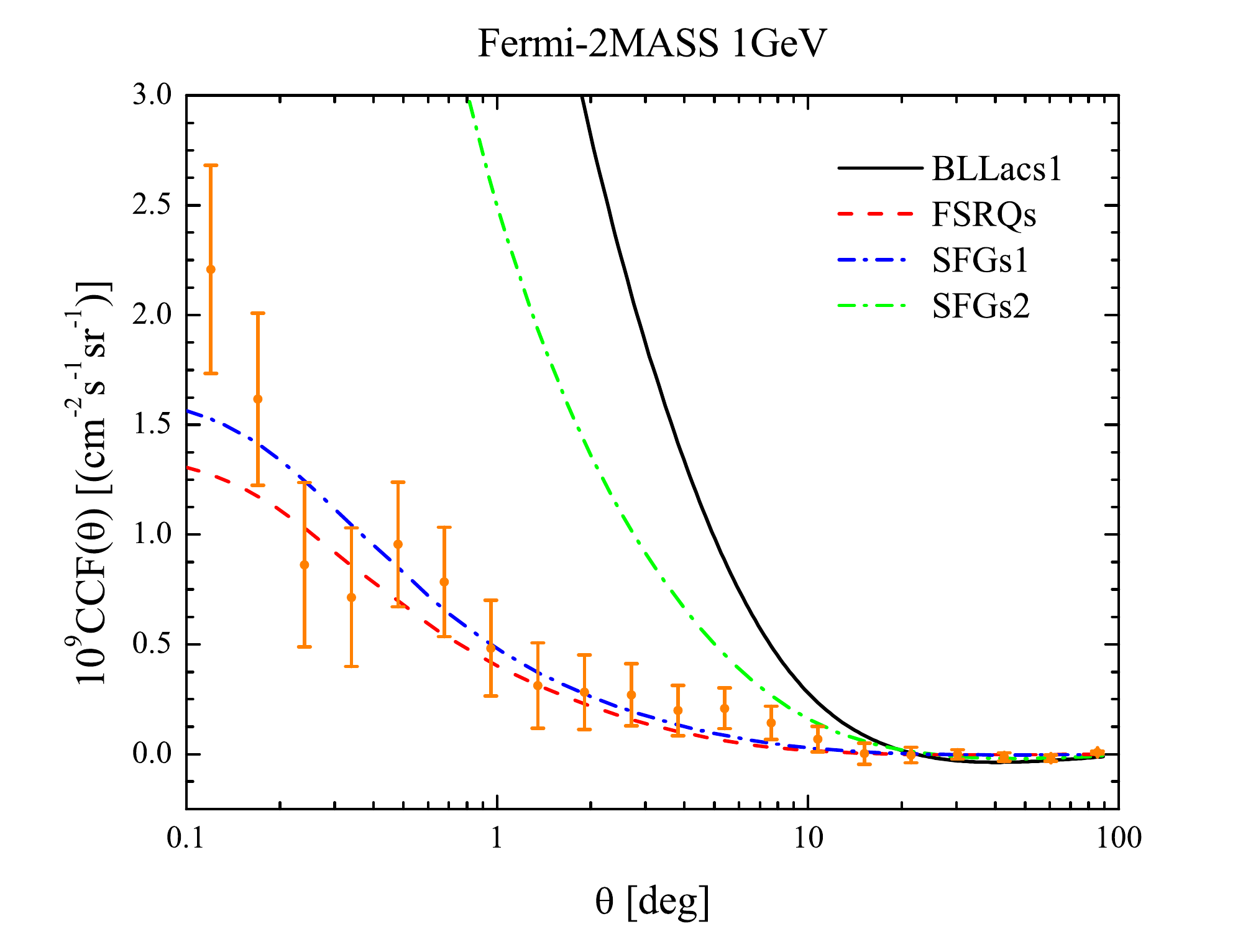}
\includegraphics[width=0.46\linewidth]{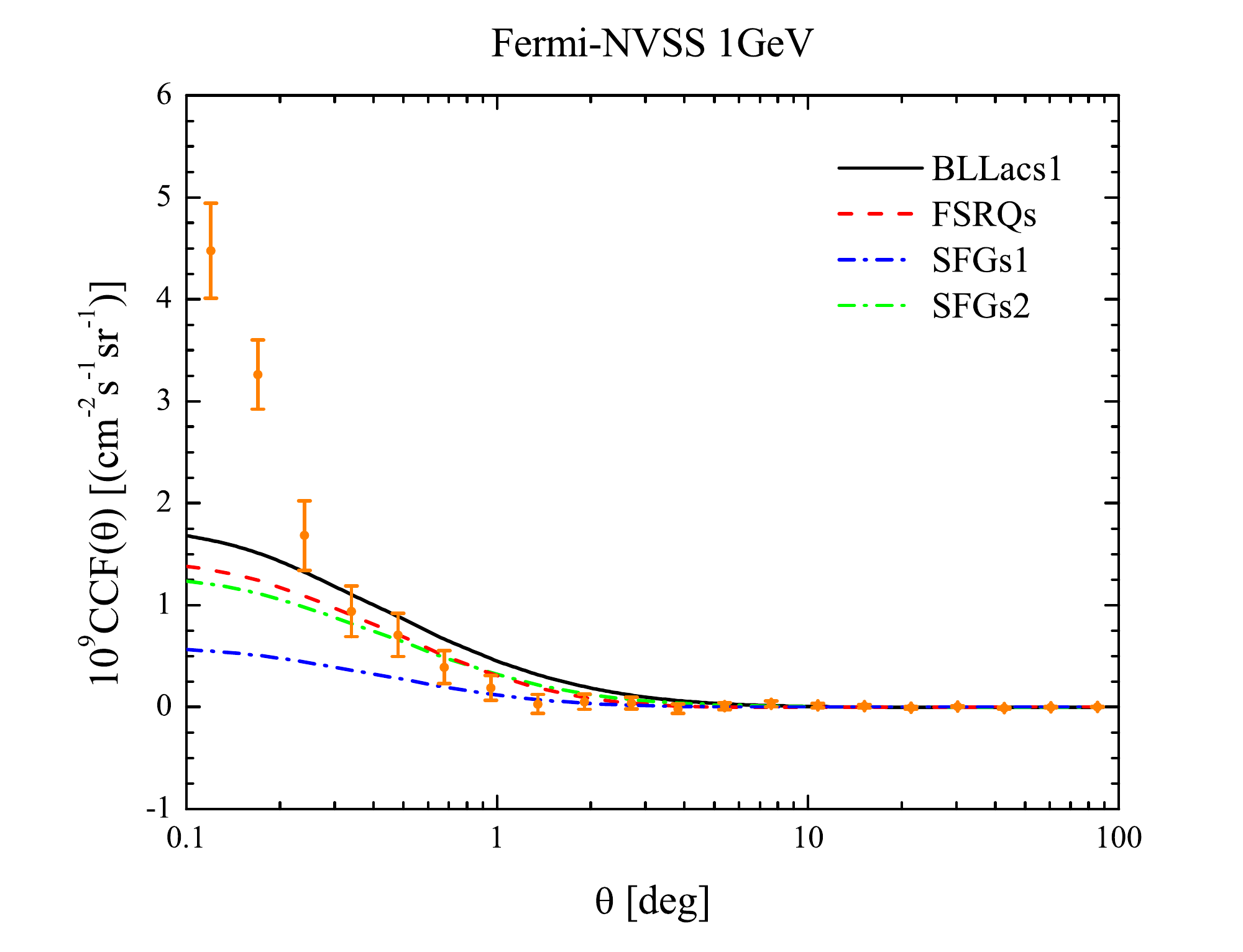}
\includegraphics[width=0.46\linewidth]{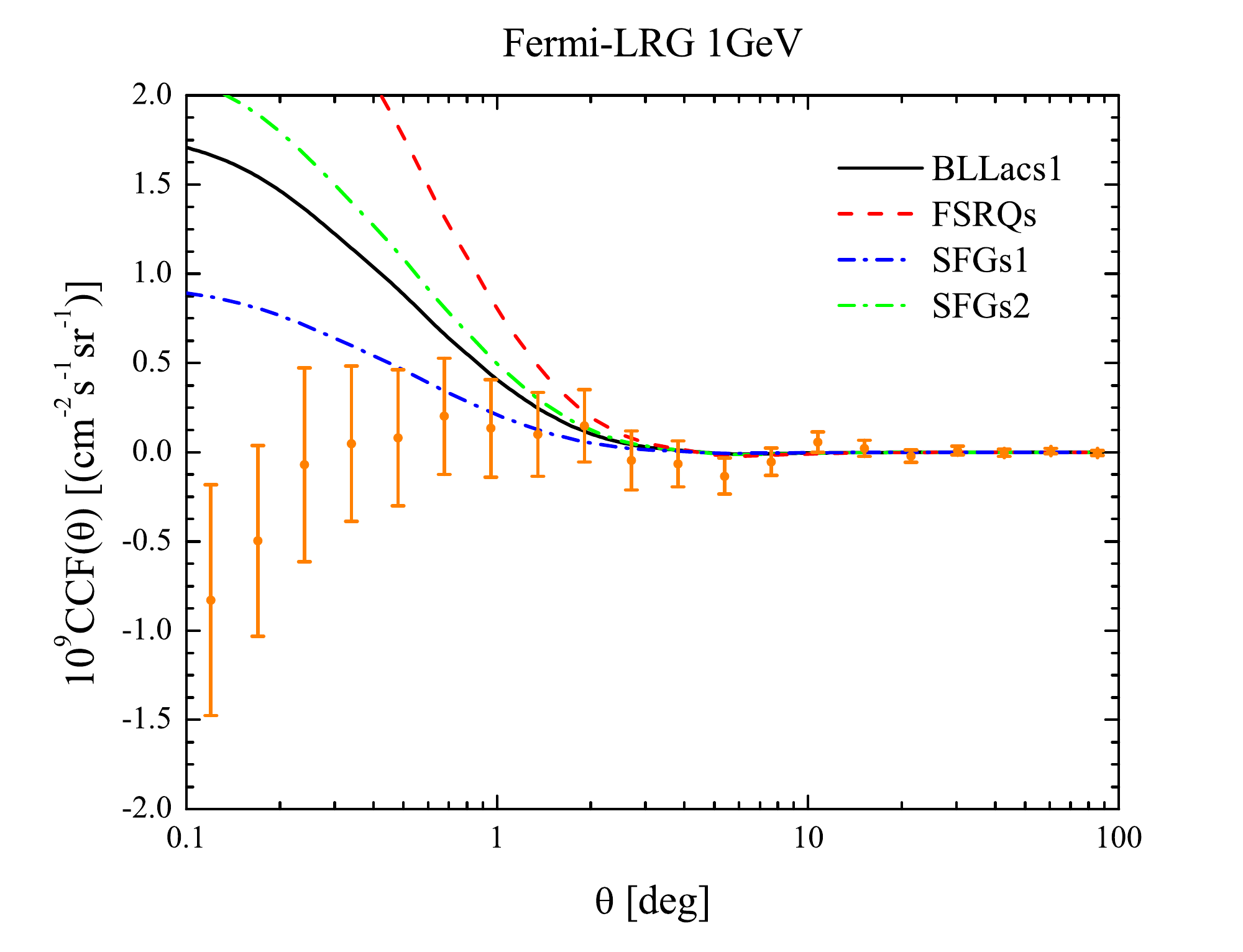}
\includegraphics[width=0.46\linewidth]{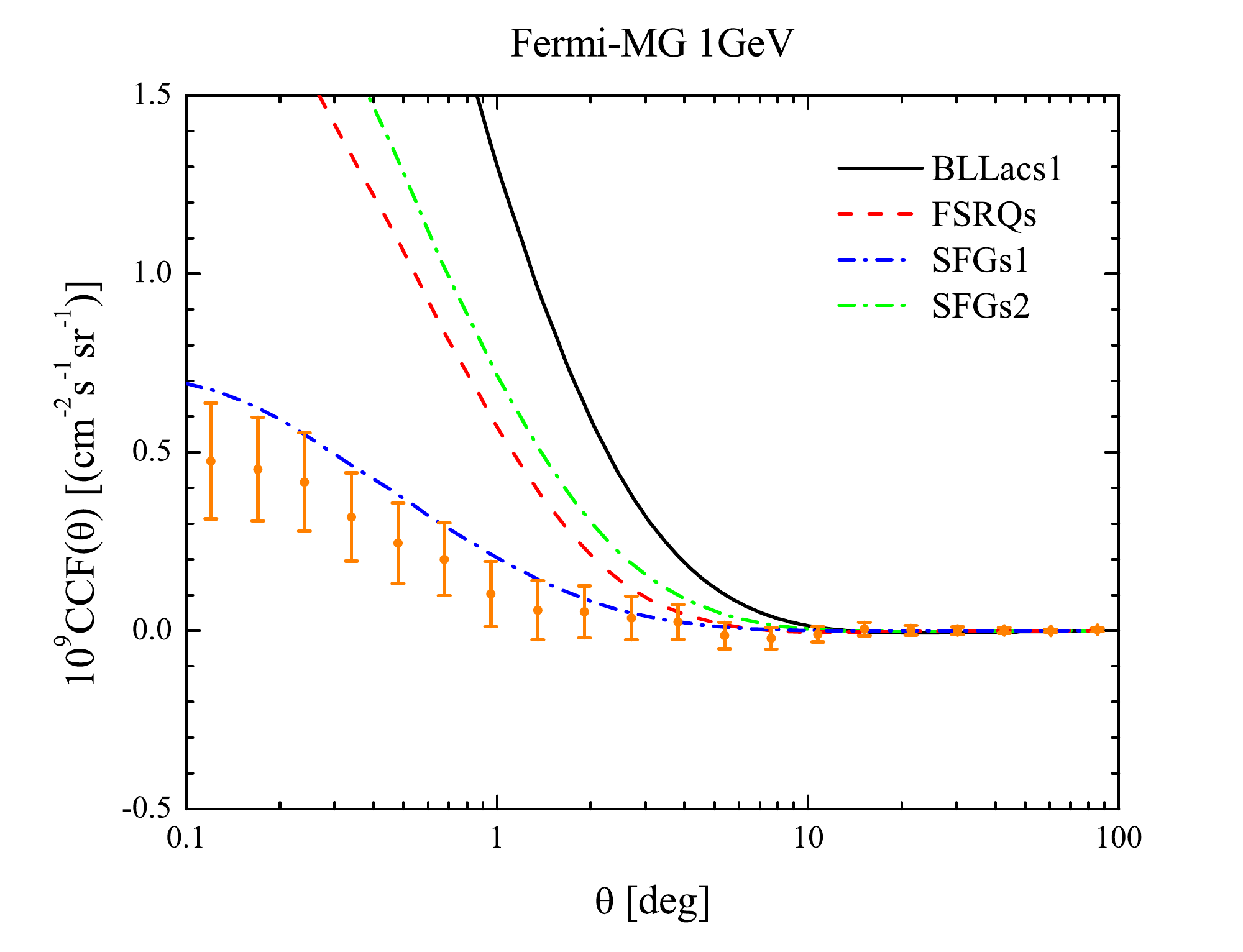}
\caption{\label{fig:cross_correlation_Xia_et_al} {\it Upper left:} 2-point cross-correlation function (orange data points) estimated from the SDSS Data Release 6 optically-selected quasars and the DGRB obtained from 60 months of \fermi LAT data at $|b|>30^\circ$and for $E>500$ MeV. Errors bars represent the diagonal elements of the covariance matrix. Model predictions for different classes of sources are represented by continuous curves: FSRQs (red dashed), BL Lacs (black solid), two models of SFGs (blue and green dot-dashed). All predictions are obtained assuming that each source class contributes 100\% of the DGRB intensity and they do not represent fits to the data. {\it Upper right:} same as in the previous panel but for the cross-correlation with the 2MASS Source Extended Catalog and for energies > 1 GeV. The other panels show the same as in the previous panel but for the NVSS catalog (medium left), luminous red galaxies in the SDSS Data Release 8 (medium right) and the main galaxy sample in the SDSS Data Release 8 (bottom). Note the different scale in the plots. Taken from Ref.~\cite{Xia:2015wka}.}
\end{center}
\end{figure}

The 2-point correlation functions are shown in 
Fig.~\ref{fig:cross_correlation_Xia_et_al} for the 5 different catalogs: a 
cross-correlation signal is evident up to few degrees for all the catalogs 
apart from the SDSS luminous red galaxies (rightmost central panel). The 
significance of these detections is $4.5\sigma$ for the SDSS quasars, 
$3.6\sigma$ for the 2MASS catalog, $3\sigma$ for the SDSS main galaxies and as 
large as $10\sigma$ in the case of NVSS. These numbers refer to the energy 
threshold that maximizes the detection significance, i.e. 500 MeV for the case
of the SDSS quasars and 1 GeV otherwise. These are also the thresholds 
considered in the different panels of 
Fig.~\ref{fig:cross_correlation_Xia_et_al}. The case of the cross-correlation
with NVSS deserves some additional comments: the authors of 
Ref.~\cite{Xia:2015wka} noticed that, for that catalog, the cross-correlation
signal exhibits an angular extension that is consistent with the PSF of \fermi 
LAT. In particular, it decreases as the energy threshold increases. This 
suggests a different origin for the NVSS signal with respect to the one 
observed with the other catalogs. They also noted that a 1-halo component to 
the signal (see Sec.~\ref{sec:implications_APS}) would manifest itself as a 
Dirac delta at $\theta=0$ degrees, smeared up to the angular size of the 
\fermi LAT PSF. Such a 1-halo term would be present, for example, if some of 
the galaxies in NVSS emitted also in the gamma-ray band as well. Indeed, NVSS 
galaxies are standard candidates to be gamma-ray emitters and this catalog is 
routinely searched for counterparts of gamma-ray sources 
\cite{Fermi-LAT:2011iqa,TheFermi-LAT:2015hja}. With all these considerations
in mind, the authors of Ref.~\cite{Xia:2015wka} concluded that the 2-point 
correlation function with NVSS is probably contaminated by a 1-halo term and 
it does trace the LSS. 

In Ref.~\cite{Xia:2015wka}, the measured correlation functions are compared
with the predictions obtained if the DGRB were contributed {\it entirely} by
only one class of astrophysical sources. These theoretical predictions are
included in Fig.~\ref{fig:cross_correlation_Xia_et_al} as colored lines. In
particular, the dashed red line stands for FSRQs, modeled as in the 
luminosity-dependent density evolution scheme of Ref.~\cite{Ajello:2011zi}, 
while the solid black is for BL Lacs, following the results of 
Ref.~\cite{Ajello:2013lka}, see Sec.~\ref{sec:blazars}. The dot-dashed lines
denote the case of SFGs, considering two different models: $i)$ the gamma-ray 
emission of SFGs is assumed to follow the cosmic SFR, as in 
Ref.~\cite{Fields:2010bw}, and $ii)$ the gamma-ray luminosity is related to
the IR one through the $L_\gamma-L_{\rm IR}$ relation obtained in 
Ref.~\cite{Ackermann:2012vca} (see Sec.~\ref{sec:SFGs}). The two scenarios are
represented by the blue and green dot-dashed lines, respectively. Finally, we
note that MAGNs are not included in the analysis of Ref.~\cite{Xia:2015wka} as 
their emission is showed to be very similar (and, therefore, degenerate) with 
the contribution of SFGs. MAGNs, however, are explicitly included in the 
follow-up analysis of Ref.~\cite{Cuoco:2015rfa}.

The 2-point correlation functions are, then, computed assuming that the 
fluctuations in the gamma-ray maps and in the galaxy distributions from the 
catalogs both trace the LSS matter density fluctuations (provide that the
so-called {\it bias factor} is considered). This allows for the correlation 
functions to be determined in terms of the non-linear power spectrum of matter 
fluctuations, which can be obtained, e.g., by means of the public CAMB code 
\cite{Lewis:2002ah} or {\ttfamily Halofit} routine \cite{Smith:2002dz}. 

We note that the cross-correlation expected from the classes of astrophysical 
sources mentioned above (colored lines in 
Fig.~\ref{fig:cross_correlation_Xia_et_al}) is indeed different from zero at
small angles, as in the observed signal. Whether the amplitude of the 
predicted correlation functions is in agreement with the data or not depends 
on the redshift overlap between the gamma-ray emitters and the sources in the 
catalogs. In particular, optically-selected quasars and sources in the NVSS 
catalog have a quite broad redshift distribution, extending to $z\sim$3-4. 
Thus, a large cross-correlation is expected with the emission from unresolved 
SFGs (modeled as in Ref.~\cite{Fields:2010bw}), whose redshift distribution 
peaks at $z \sim 2-3$. On the other hand, the luminous red galaxies and the 
objects in the 2MASS catalog probe the local Universe (respectively, from 
$z\sim0.8$ and $z\sim0.3$, and down to the present time) and they are 
characterized by narrower redshift distributions. Both are expected to 
correlate mainly with BL Lacs. Indeed, by considering several galaxy catalogs 
with different redshift distributions, one can effectively probe different 
redshift ranges, developing a full tomographic approach. Specifically, in 
Ref.~\cite{Xia:2015wka}, the authors build a model of the DGRB that includes 
FSRQs, BL Lacs and SFGs, leaving the normalization of the different components 
free to vary when fitting the cross-correlation data in
Fig.~\ref{fig:cross_correlation_Xia_et_al}. Ref.~\cite{Xia:2015wka} shows that 
including the cross-correlation with SDSS quasars is crucial in deriving a 
lower bound on the contribution of SFGs.\footnote{Note that this lower bound 
is found only if SFGs are modeled according to Ref.~\cite{Fields:2010bw}.} 
These are, indeed, found to be the dominant component in the DGRB, with blazars 
accounting for, at most 10\% of the total DGRB measured by \fermi LAT in 
Ref.~\cite{Ackermann:2014usa} (at $1\sigma$ level). Also, depending on which
description is assumed for the SFGs, the best-fit model to the 
cross-correlation data in Fig.~\ref{fig:cross_correlation_Xia_et_al} can 
account for only 70\% or 20\% of the total DGRB intensity. 

The cross-correlation with galaxy catalogs can also be used to constrain the
DM component of the DGRB. This possibility was studied in 
Refs.~\cite{Ando:2013xwa,Ando:2014aoa}, where the authors considered the 
2MASS Redshift Survey \cite{Huchra:2011ii} and the 2MASS Extended Source 
Catalog from Ref.~\cite{Jarrett:2000qt}. These two catalogs are chosen 
against others because they trace the matter distribution in the local 
Universe and, therefore, they are expected to correlate with any potential 
DM-induced gamma-ray signal. The 2MASS Redshift Survey extends to $z \sim 0.1$, 
while the 2MASS Extended Source Catalog peaks at $z=0.072$ and does not 
contain galaxies beyond $z \sim 0.4$. The \fermi LAT has detected most of the 
blazars in this volume, down to a very low sensitivity. Thus, unresolved 
blazars are not expected to exhibit a large cross-correlation with catalogs of 
the local Universe.\footnote{In other words, the window function of unresolved 
blazars does not overlap significantly with that of the 2MASS catalogs 
considered above.} This means that the cross-correlation APS will be 
potentially very sensitive to other DGRB contributors, emitting mainly at low 
redshift as, e.g., SFGs or DM. 

In Refs.~\cite{Ando:2013xwa,Ando:2014aoa}, galaxies are described by means 
of the so-called Occupation Distribution (HOD) model. This framework 
postulates that each source is embedded into a DM halo of mass $M$ and that
the abundance and distribution of galaxies are related to the properties of 
the host DM halos. The HOD model is a phenomenological formalism, based on 
results from $N$-body simulations and on semi-analytical descriptions of DM 
halos and galaxy formation \cite{Seljak:2000gq,Cooray:2002dia,Zheng:2004id}. 
Under this formalism, Ref.~\cite{Ando:2013xwa} shows that DM dominates the 
cross-correlation with the 2MASS Redshift Survey and that 5 years of \fermi 
LAT data should be able to distinguish a DM scenario from one with purely a
strophysical sources. Ref.~\cite{Ando:2014aoa} also computes the upper limit 
that it would be possible to derive on $\langle \sigma v\rangle$ if the 
cross-correlation were found compatible with an astrophysical interpretation. 
These results, however, largely depend on the model adopted for low-mass DM 
halos and subhalos. In the most optimistic scenario, the cross-correlation 
will be able to exclude thermal cross sections for DM masses up to almost 1 
TeV.

\begin{figure}
\begin{center}
\includegraphics[width=0.9\linewidth]{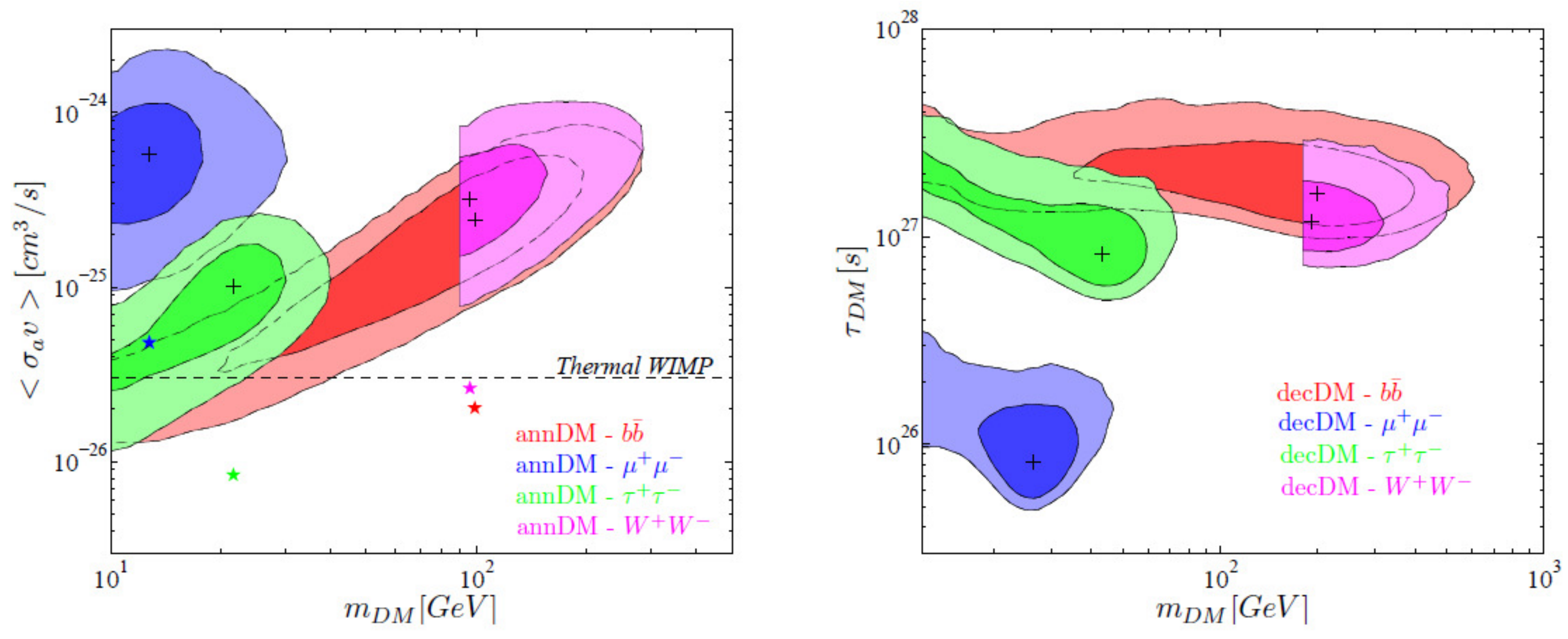}
\caption{\label{fig:DM_interpretation_correlation_2MASS} {\it Left:} $1\sigma$ and $2\sigma$ allowed regions for the DM annihilation rate versus DM mass, for different gamma-ray production channels and assuming the ``LOW'' substructure scheme in Ref.~\cite{Fornasa:2012gu}. Crosses indicate the best-fit models. In the ``HIGH'' scenario of Ref.~\cite{Fornasa:2012gu}, regions remain similar in shape but they shift downward by a factor of $\sim 12$ (star symbols). {\it Right:} The same but for decaying DM, showing the DM particle lifetime as a function of its mass. Taken from Ref.~\cite{Regis:2015zka}.}
\end{center}
\end{figure}

\begin{figure}
\begin{center}
\includegraphics[width=0.9\linewidth]{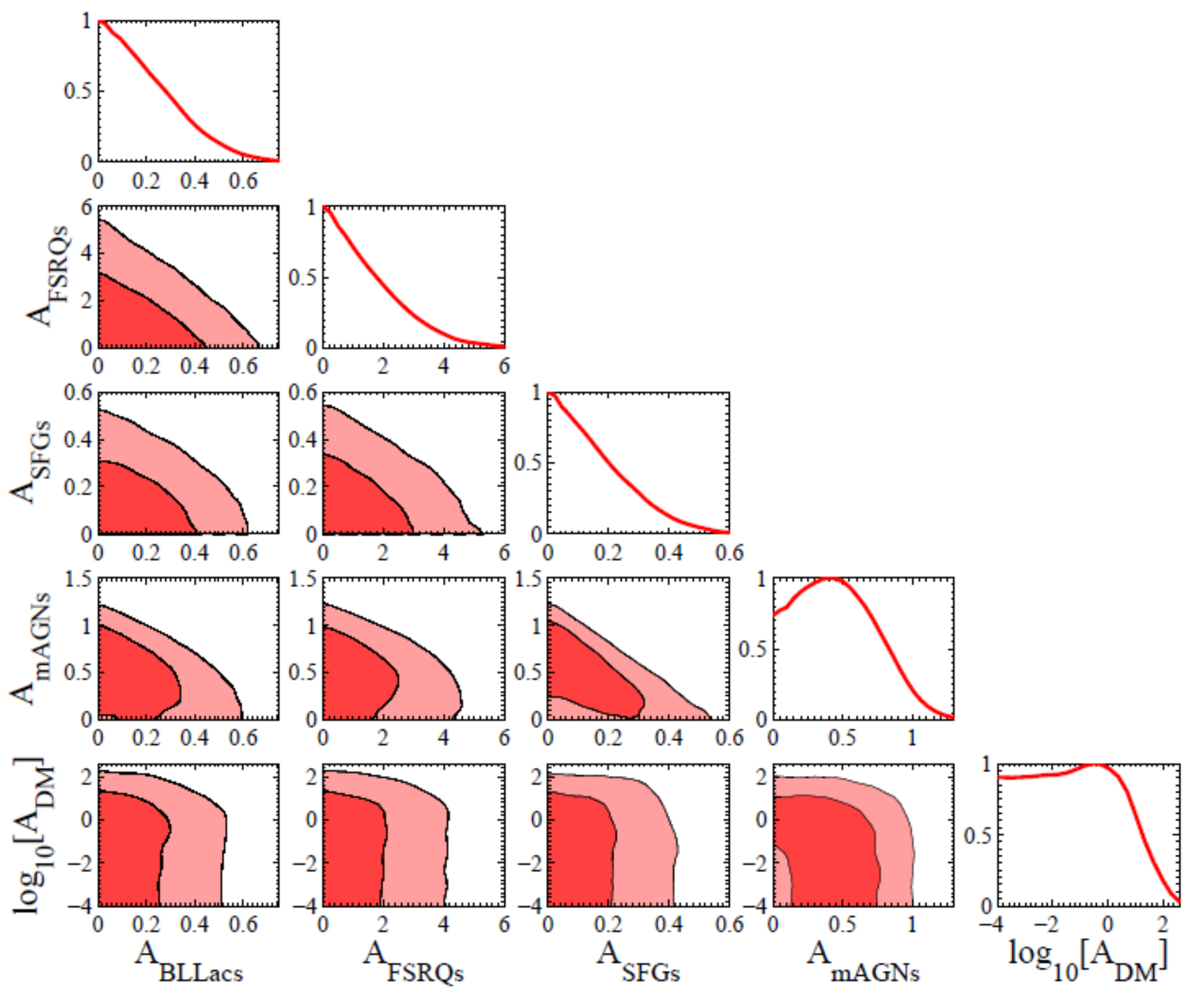}
\caption{\label{fig:interpretation_cross_correlation} Posterior probability distributions for the normalization of the component to the DGRB due to FSRQs ($A_{\rm FSRQs}$), BL Lacs ($A_{\rm BLLacs}$), SFGs ($A_{\rm SFGs}$), MAGNs ($A_{\rm mAGNs}$) and annihilating DM ($A_{rm DM}$). These parameters are defined with respect to a fiducial model introduced in Ref.~\cite{Cuoco:2015rfa}. Panels along the diagonal show the marginalized 1-dimensional probability distribution for each parameters. All the others indicate the $1\sigma$ (darker, innermost) and $2\sigma$ (lighter, outermost) confidence level contours in the probability distributions of the different combinations of parameters. The full model contains a total of 11 free parameters, but only the 5 mentioned above are shown in this figure. Taken from Ref.~\cite{Cuoco:2015rfa}.}
\end{center}
\end{figure}

These predictions were tested against actual data in Ref.~\cite{Regis:2015zka}, 
where the authors explained the measured cross-correlation signal with the 
2MASS Extended Source Catalog in Ref.~\cite{Xia:2015wka} in terms of DM. They 
found that a WIMP DM particle with a mass in the range between 10 and 100 GeV 
(depending on the annihilation channel) and a thermal cross section can 
reproduce both the shape and intensity of the measured cross correlation. This 
is also the case for decaying DM candidates, for a similar range in DM mass
and a decay lifetime between $5 \times 10^{25}$ and $5 \times 10^{27}$ s. The 
colored contours in Fig.~\ref{fig:DM_interpretation_correlation_2MASS} show the
the regions in the $(m_\chi,\langle \sigma v \rangle)$ and $(m_\chi, \tau)$ 
parameter spaces that are compatible with the cross-correlation found with 
2MASS. The figure also shows how, for an annihilating DM candidate, a 
different assumption for the subhalo boost can shift the preferred contours 
by more than one order of magnitude.

In additiona, Ref.~\cite{Regis:2015zka} used the measured cross-correlation to 
derive upper limits on $\langle \sigma v \rangle$, by requiring that the 
DM-induced correlation function do not to over-produce the data. This allowed 
Ref.~\cite{Regis:2015zka} to exclude thermal annihilation cross sections for 
DM masses below 100 GeV (in the case of annihilations into $b$ quarks and a 
``LOW'' subhalo boost model inspired by Ref.~\cite{Fornasa:2012gu}). This 
makes the cross-correlation with {\it local} galaxy catalogs the strongest 
observable up to date to constrain a potential DM contribution to the DGRB, 
compared to the DGRB energy spectrum reported in  Ref.~\cite{Ackermann:2014usa} 
or the auto-correlation APS in Ref.~\cite{Ackermann:2012uf}.

Astrophysical and DM-induced emissions are considered at the same time in 
Ref.~\cite{Cuoco:2015rfa}: the authors define a model of the DGRB that 
includes FSRQs, BL Lacs, SFGs, MAGNs (parametrized according to 
Refs.~\cite{Ajello:2011zi,Ajello:2013lka,Tamborra:2014xia,DiMauro:2013xta},
respectively) and annihilating DM. The galaxy catalogs are described following
the HOM formalism. The normalizations of the emission of the 4 mentioned
astrophysical source classes are left free in the model, as well as the DM 
mass and annihilation cross section. 5 additional parameters are included, one 
for each class of gamma-ray emitters, accounting for possible 1-halo terms in 
the 2-point correlation functions. The model is, then, used to fit the measured 
cross-correlation reported in Ref.~\cite{Xia:2015wka}. As expected, the 
posterior probability distribution function for the intensity of the 1-halo 
term points towards a value different than zero, in the case of the 
cross-correlation with NVSS. The distributions are compatible with zero for
the other catalogs.

\begin{figure}
\begin{center}
\includegraphics[width=0.6\linewidth]{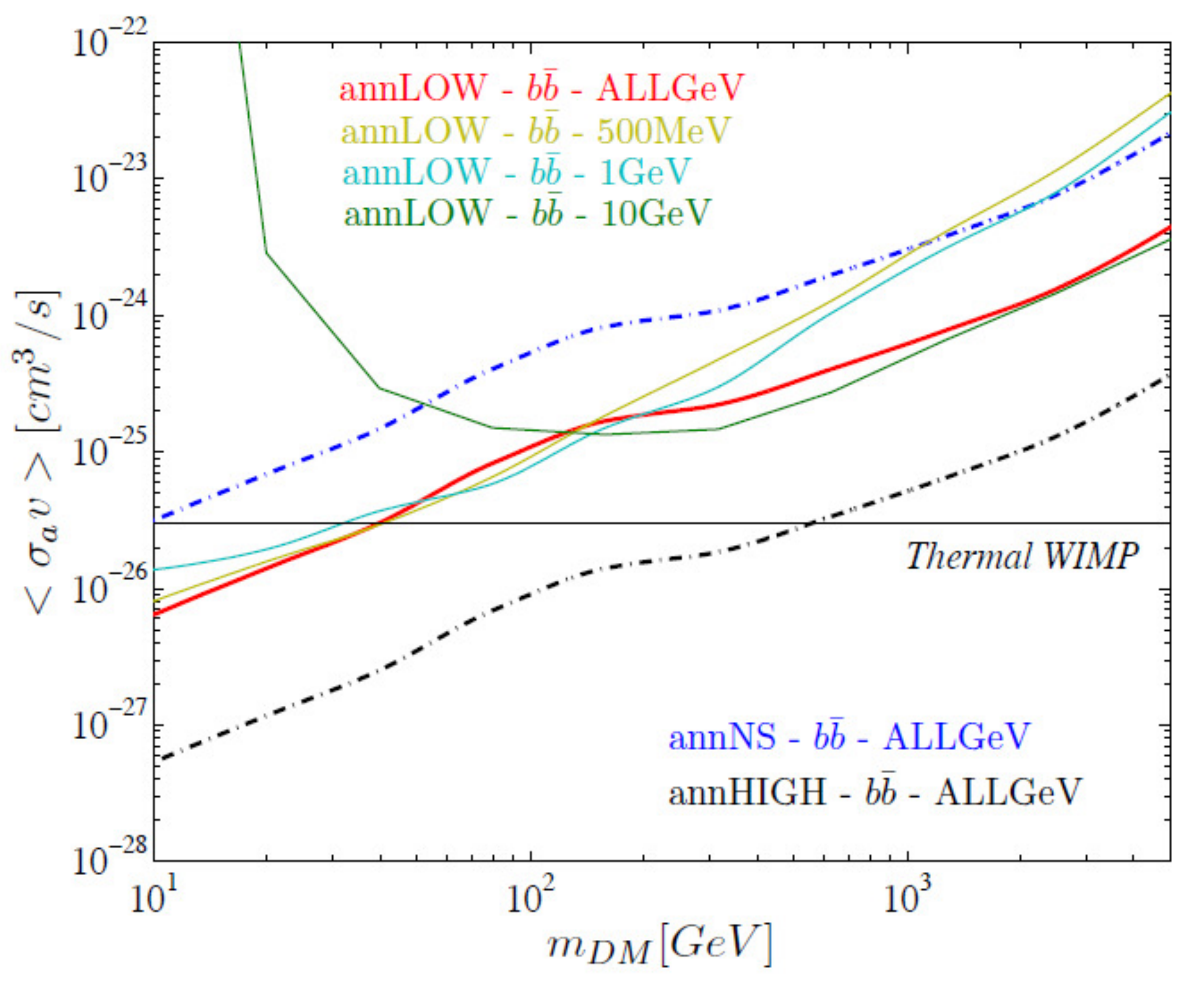}
\caption{\label{fig:cross_correlation_ULs} 95\% confidence limit upper bounds on the DM annihilation rate $\langle \sigma v \rangle$ as a function of the DM mass, for the ``LOW'' substructures model of Ref.~\cite{Fornasa:2012gu} and the reference NVSS-10 $A_{1h}^k \neq 0$ fit (see Ref.~\cite{Cuoco:2015rfa} for details). Solid lines refer to the $b\bar{b}$ annihilation channel: the red line refers to the analysis that combines information from all the three energy bins under consideration (i.e. $E>0.5$, 1 and 10 GeV), while the other three lines refer to the analysis performed on a single energy bin (as stated in the figure label). The upper dot-dashed blue line refers to the ``NS'' substructure model, where halos do not have substructures and $M_{\rm min}=10^7 M_\odot$. The lower dot-dashed black line, instead, representto the ``HIGH'' substructure model, inspired by Ref.~\cite{Fornasa:2012gu}. Taken from Ref.~\cite{Cuoco:2015rfa}.}
\end{center}
\end{figure}

Fig.~\ref{fig:interpretation_cross_correlation} summarizes the probability
distributions of the remaining 5 free parameters (excluding $m_\chi$). An
examination of this figure makes evident that the interpretation of the 
cross-correlation data is affected by significant degeneracies: there is a 
mild indication of a peak in the probability distribution of the 
normalizations of MAGNs ($A_{\rm mAGNs}$ in 
Fig.~\ref{fig:interpretation_cross_correlation}) and of the DM component
($A_{\rm DM}$, proportional to $\langle \sigma v \rangle$), but, otherwise, the
data effectively just enforce upper limits on the other parameters. The 
degeneracy is particularly visible between $A_{\rm mAGNs}$ and $A_{\rm DM}$ in
the fourth panel of the bottom row of 
Fig.~\ref{fig:interpretation_cross_correlation}. Nevertheless, the measured
cross correlations are still able to provide stringent uppers limit on the 
DM component, as it can be seen in Fig.~\ref{fig:cross_correlation_ULs}. The
regions above the solid lines are excluded as they would over-produce the
2-point correlation functions measured above 500 MeV (yellow line), 1 GeV 
(cyan line), 10 GeV (green). The red solid line indicates the upper limit 
obtained from the combined analysis of the data sets for the three mentioned 
energy thresholds. These results are all obtained by adopting the assume a 
``LOW'' subhalo boost. Note that, even for such a moderate description of 
low-mass DM structures, the inferred DM limits are more stringent than those 
obtained in Ref.~\cite{Ackermann:2015gga} by studying the DGRB intensity or 
those derived from the DGRB auto-correlation APS in 
Ref.~\cite{Gomez-Vargas:2014yla}.\footnote{These limits are sensitive to the 
way the 1-halo terms are implemented. See Ref.~\cite{Cuoco:2015rfa} for 
details.} In the case that a ``HIGH'' subhalo boost were adopted instead, the 
upper limit on the DM annihilation cross section would improve by roughly one
order of magnitude (dash-dotted black line in 
Fig.~\ref{fig:cross_correlation_ULs}). On the contrary, in an overly 
conservative scenario in which no substructures are present below 
$M_{\rm min}=10^7 M_\odot$, the exclusion worsen by a factor $\sim 4-5$ (blue
dash-dotted line). 

As already hinted at in Ref.~\cite{Xia:2015wka}, the authors of 
Ref.~\cite{Cuoco:2015rfa} noted that the models that provide a good fit to the
cross-correlation data fall short of accounting for the intensity of the
DGRB: above 1 GeV, in the most likely scenario, the model is able to explain 
only $\sim 30\%$ of the DGRB intensity measured by \fermi LAT in 
Ref.~\cite{Ackermann:2014usa}. Note, though, that multiple reasons can be
invoked to explain this apparent discrepancy, e.g. uncertainties in the 
modeling of the Galactic diffuse foreground when performing the measurement 
of the DGRB intensity, and/or uncertainties in the model predictions. Another
possibility is the presence of a Galactic component in the DGRB that, 
therefore, does not correlate with the LSS. In this regard, Galactic DM would 
be a plausible and interesting candidate (see, e.g.,
Ref.~\cite{Ackermann:2015gga} for further discussion on this issue).

\subsection{The cross-correlation with cosmic shear}
\label{sec:correlation_shear}
Another tracer of LSS is the gravitational lensing effect of cosmic shear. Due
to lensing, the light emitted by distant sources is distorted while it 
propagates towards us by the presence of intervening matter. In the weak 
lensing regime, the effect is very small and it is directly related to the 
distribution of matter at large scale. The signal, referred to as {\it cosmic
shear}, is expected to cross-correlate with the gamma-ray emission since the 
same structures responsible for light bending are also those producing the 
gamma-ray emission, either because they emit light themselves (through DM 
annihilation or decay) or because they host the astrophysical emitters. 

The gravitational distortion can be evaluated on the null-geodesic of the 
unlensed photon. It can be decomposed into the so-called convergence $\kappa$ 
and shear $\gamma$ \cite{Bartelmann:1999yn,Kaiser:1996tp}. In the flat-sky 
approximation (i.e. small distortion angles), $\kappa$ and $\gamma$ share the 
same APS and, for convenience, we focus only on the former from now on. The 
convergence $\kappa$ is a direct estimator of the fluctuations in the Newtonian 
potential of the LSS, integrated along the line of sight. $\kappa$ can be 
estimated via a statistical analysis of the correlations in the ellipticities 
of the images of galaxies. Thanks to Poisson’s equation, which links the 
gravitational potential to the matter distribution, the intensity of the 
cosmic shear signal $I_{\kappa}(\mathbf{n})$ from a direction $\mathbf{n}$ can 
be written as follows:
\begin{equation}
I_{\kappa}(\mathbf{n}) = \int d\chi \, g_{\kappa}(\mathbf{n},\chi) W_\kappa(\chi).
\end{equation}
The decomposition of $I_{\kappa}(\mathbf{n})$ resembles the way we expressed the
gamma-ray emission in Eq.~(\ref{eqn:average_emission_X}): the source field 
$g_{\kappa}(\mathbf{n},\chi)$ indicates directly the distribution of matter and 
it is modulated by the window function. The product of the average source 
field $\langle g_\kappa(\chi) \rangle$ and the window function can be expressed 
as follows:
\begin{equation}
\langle g_\kappa(\chi) \rangle W_{\kappa}(\chi) = \frac{3}{2} H_0^2 \Omega_m 
[1+z(\chi)] \, \chi \, \int_{\chi}^\infty d\chi^\prime 
\frac{\chi^\prime-\chi}{\chi^\prime} \frac{dN_g(\chi^\prime)}{d\chi^\prime},
\end{equation}
where $dN_g/d\chi^\prime$ is the redshift distribution of the background
galaxies, normalized to 1 over the observed redshift range.

As done in Sec.~\ref{sec:angular_spectrum}, the gamma-ray emission associated 
with a certain population $X$ is written in terms of its window function 
$W_X(\chi)$ and of the source field $g_{X}(\chi,\mathbf{n})$ as 
$I_X(\mathbf{n}) = \int d\chi \, g_X(\chi,\mathbf{n}) W_X(\chi)$. The average
source field $\langle g_X(\chi) \rangle$ depends on the abundance of sources
as a function of their $Y$-parameter and of redshift (see 
Eq.~(\ref{eqn:average_source_field})). Similarly to 
Eq.~(\ref{eqn:theory_APS}), the cross-correlation APS between the gamma-ray 
emission produced by population $X$ and the cosmic shear can be written as
\cite{Fornengo:2013rga,Ando:2014aoa}:
\begin{equation}
C_{\ell}^{X,\kappa} = 
\frac{1}{\langle \mathcal{I}_X \rangle\langle \mathcal{I}_\kappa \rangle} 
\int \frac{d\chi}{\chi^2} \langle g_X(\chi) \rangle 
\langle g_\kappa(\chi) \rangle W_X(\chi) W_{\kappa}(\chi) P_{X,\kappa}
\left( k = \frac{\ell}{\chi}, \chi \right).
\end{equation}

\begin{figure}
\begin{center}
\includegraphics[width=0.45\linewidth]{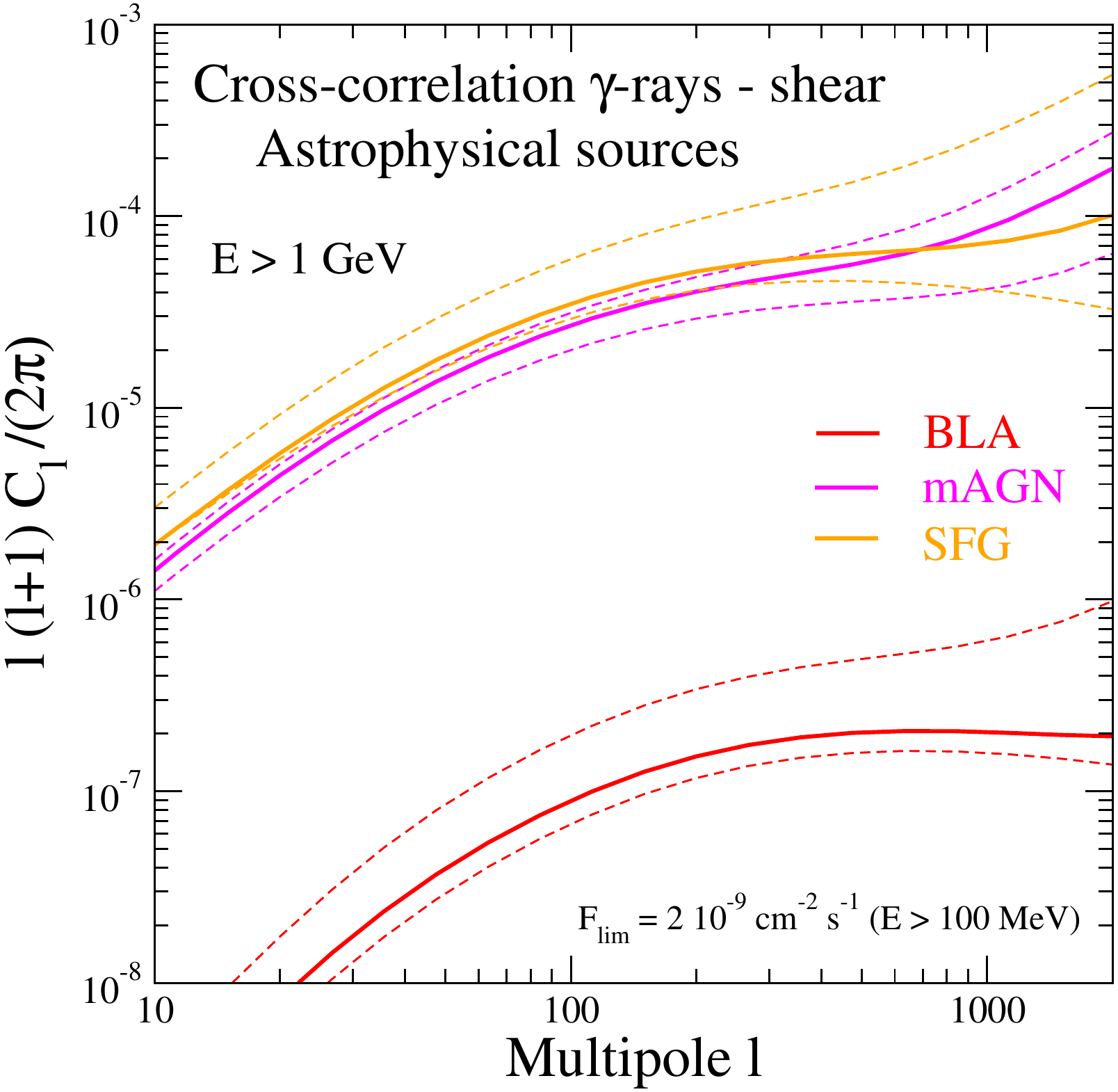}
\hspace{0.1cm}
\includegraphics[width=0.45\linewidth]{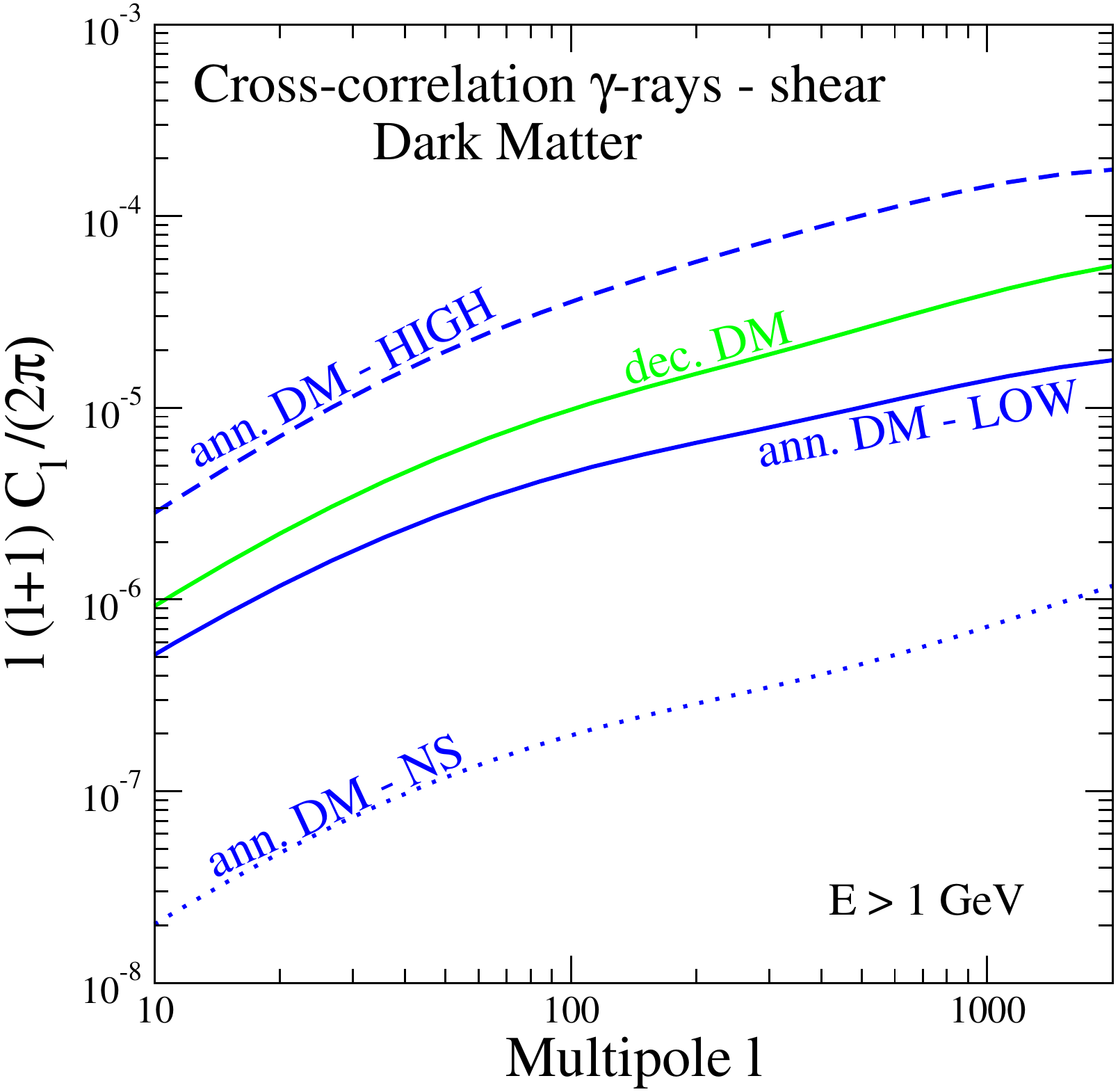}
\caption{\label{fig:cross_correlation_Camera} {\it Left:} Cross-correlation APS between cosmic shear and gamma rays from blazars (red), MAGNs (pink), and SFGs (orange). Among the curves with the same color, different lines correspond to different choices for the $L_\gamma(M)$ relation: solid lines represent the fiducial models considered in Ref.~\cite{Camera:2014rja}, while dashed ones indicate extreme scenarios assumed to bracket the uncertainty on $L_\gamma(M)$. {\it Right}: Cross-correlation APS between cosmic shear and DM-induced gamma-ray emission. Blue lines correspond to an annihilating DM candidate and the green one to a decaying DM candidate. The different blue lines represent different scenarios for low-mass DM (sub)halos (see Ref.~\cite{Camera:2014rja} for details). The mass of the DM particle is taken to be 100 GeV (200 GeV) in the case of annihilating (decaying) DM. The annihilation cross section is fixed at $3 \times 10^{-26} \mbox{cm}^3 \mbox{s}^{-1}$ and the decay lifetime at $3 \times 10^{27} \mbox{s}$. Annihilations and decays entirely into $b\bar{b}$ are assumed. Taken from Ref.~\cite{Camera:2014rja}.}
\end{center}
\end{figure}

The 3-dimensional cross-correlation power spectrum $P_{X,\kappa}$ can be split 
into the following 1-halo and 2-halo terms:
\begin{equation}
P_{1h}(k,z) = \int_{M_{\rm min}}^{M_{\rm max}} dM \, \frac{dn}{dM} \tilde{u}(k|M) 
\frac{\tilde{u}_X(k|Y(M))}{\langle g_X(\chi) \rangle},
\label{eqn:1_halo_lensing}
\end{equation}
\begin{equation}
P_{2h}(k,z) = \left[ \int_{M_{\rm min}}^{M_{\rm max}} dM \, \frac{dn}{dM} b_h(M) 
\tilde{u}(k|M) \right] 
\left[ \int_{Y_{\rm min}}^{Y_{\rm max}} dY  \, \frac{dN}{dY} b_X(Y) 
\frac{\tilde{u}_X(k|Y)}{\langle g_X(\chi) \rangle} \right] P_{\rm lin}(k,z).
\end{equation}
The quantity $\tilde{u}(k|M)$ is the Fourier transform of the radial density 
profile of a DM halo with mass $M$, while $\tilde{u}_{X}(k|Y(M))$ is the 
Fourier transform of the gamma-ray surface brightness profile of the source 
characterized by parameter $Y$. Note that these equations depend on the $Y(M)$ 
relation that links the $Y$-parameter to the mass of the host DM halo. For 
astrophysical sources, $Y$ is usually taken to be the gamma-ray luminosity 
$L_\gamma$. Ref.~\cite{Camera:2014rja} determines the $L_\gamma(M)$ empirically 
for the case of blazars, SFGs and MAGNs, making use of correlations between 
different source properties or of the results of semi-analytical models. 
However, the $L_\gamma(M)$ remains very uncertain for all the source classes 
considered in Ref.~\cite{Camera:2014rja}. Its uncertainty can become an issue 
when estimating the cross-correlation APS $C_\ell^{X,\kappa}$. However, in the 
case of astrophysical sources, $C_\ell^{X,\kappa}$ is dominated by the 2-halo 
term (at least up to multipoles as large as few hundreds), while the 
$L_\gamma(M)$ relation mainly affects the 1-halo term.\footnote{The $L_\gamma(M)$ 
relation enters in the computation of the 2-halo term only through the bias 
factor, $b_X(Y(M))$, which is, generally, of $\mathcal{O}(1)$.} The 
uncertainty on $C_\ell^{X,\kappa}$ generated by the variability of $L_\gamma(M)$ 
can be seen in the left panel of Fig.~\ref{fig:cross_correlation_Camera}. The 
different solid lines show the expected cross-correlation APS for different 
classes of astrophysical sources (red for blazars, orange for SFGs and pink 
for MAGNs), while dashed lines indicate how the cross-correlation APS changes 
for two extreme scenarios bracketing our lack of knowledge on $M(L_\gamma)$. 
The uncertainty bands are within a factor of 2 from the solid lines, at least 
at high multipoles. They increase in size at smaller angular scales (i.e. 
large $\ell$), since the APS becomes more sensitive to the 1-halo term. 

The left panel of Fig.~\ref{fig:cross_correlation_Camera} also demonstrates
that the expected cross-correlation between cross-correlation and the emission 
of unresolved blazars is more than 2 orders of magnitude smaller than the one 
with the other classes of astrophysical sources in Ref.~\cite{Camera:2014rja}, 
i.e. SFGs and MAGNs. As commented in the previous section, this is because 
the \fermi LAT has detected most of the blazars populating the volume probed 
by cosmic shear (i.e. $z \lesssim 2$ \cite{Camera:2014rja}) and, thus, the 
window functions $W_{\rm blazars}(\chi)$ and $W_\kappa(\chi)$ do not overlap 
significantly. On the other hand, only a limited number of MAGNs and SFGs have 
been observed by the \fermi LAT (see also Secs.~\ref{sec:MAGNs} and 
\ref{sec:SFGs}), so that the emission of their unresolved counterparts is 
still characterized by a notable cross-correlation with the lensing signal.

The right panel of Fig.~\ref{fig:cross_correlation_Camera} shows the expected 
cross-correlation APS with the gamma-ray emission produced by annihilating DM 
(blue lines) or decaying DM (green line). Different blue lines corresponds to 
different models for the description of low-mass DM halos. As seen in the 
previous sections, this uncertainty can have a significant impact on the 
properties of the DM-induced emission. Ref.~\cite{Camera:2014rja} estimates 
that different description of the DM (sub)halos below the mass resolution of 
$N$-body simulations can lead to an uncertainty as large as a factor of 100 on 
the expected cross-correlation APS.\footnote{Note, however, that the ``NS'' 
model (dotted blue line in Fig.~\ref{fig:cross_correlation_Camera}) is 
probably underestimating the signal from DM since it assumes no contribution 
from DM subhalos and a quite large $M_{\rm min}=10^7 M_\odot$. On the other hand, 
the ``HIGH'' scenario (dashed blue line) is based on power-law extrapolations 
from Ref.~\cite{Gao:2011rf} and, thus, it likely overestimates the DM signal 
(see Sec.~\ref{sec:DM}).} In the most optimistic scenario considered in 
Ref.~\cite{Camera:2014rja}, the cross-correlation APS from annihilating DM is 
of the same order of that expected from astrophysical sources. This 
demonstrates how effective this observable can be for the detection of the DM 
component in the DGRB.

Ref.~\cite{Camera:2014rja} estimates that the measurement of the 
cross-correlation between the data of the Dark Energy Survey 
\cite{Abbott:2005bi} and those of the \fermi LAT (after 5 years of operation)
has the potential to detect an annihilating DM particle with an annihilation 
cross section smaller that the thermal value of 
$3 \times 10^{-26} \mbox{cm}^3 \mbox{s}^{-1}$ for DM masses up to 300 GeV (for 
annihilations into $b$ quarks). The result refers to a very optimistic subhalo 
boost and the predicted signal decreases by a factor of $\sim 10$ in the case 
of a more realistic description of DM subhalos (see right panel of 
Fig.~\ref{fig:cross_correlation_Camera}. Prospects improve significantly with
the inclusion of data from the forthcoming Euclid mission (expected for 2020) 
\cite{Laureijs:2011gra,Amendola:2012ys}. Indeed, the cross correlation of 
Euclid data with those of a hypothetical future gamma-ray telescope with 
improved performances with respect to the \fermi LAT\footnote{This improved 
version the \fermi LAT is described by a wide energy range, i.e. from 300 MeV 
to 1 TeV. It is also assumed to achieve $\sim 2.5$ times more exposure 
(defined in Ref.~\cite{Camera:2014rja} as the product of the effective area 
and the observation time) than 5 years of \fermi LAT data and to drastically 
improve the angular resolution to a value of 0.027$^\circ$ over the whole energy 
range. Its field of view is similar to that of the \fermi LAT.} has the 
potential to detect a DM component for DM masses up to the TeV scale (assuming 
a thermal annihilation cross section and an optimistic subhalo boost).

\begin{figure}[h]
\begin{center}
\includegraphics[width=0.46\linewidth]{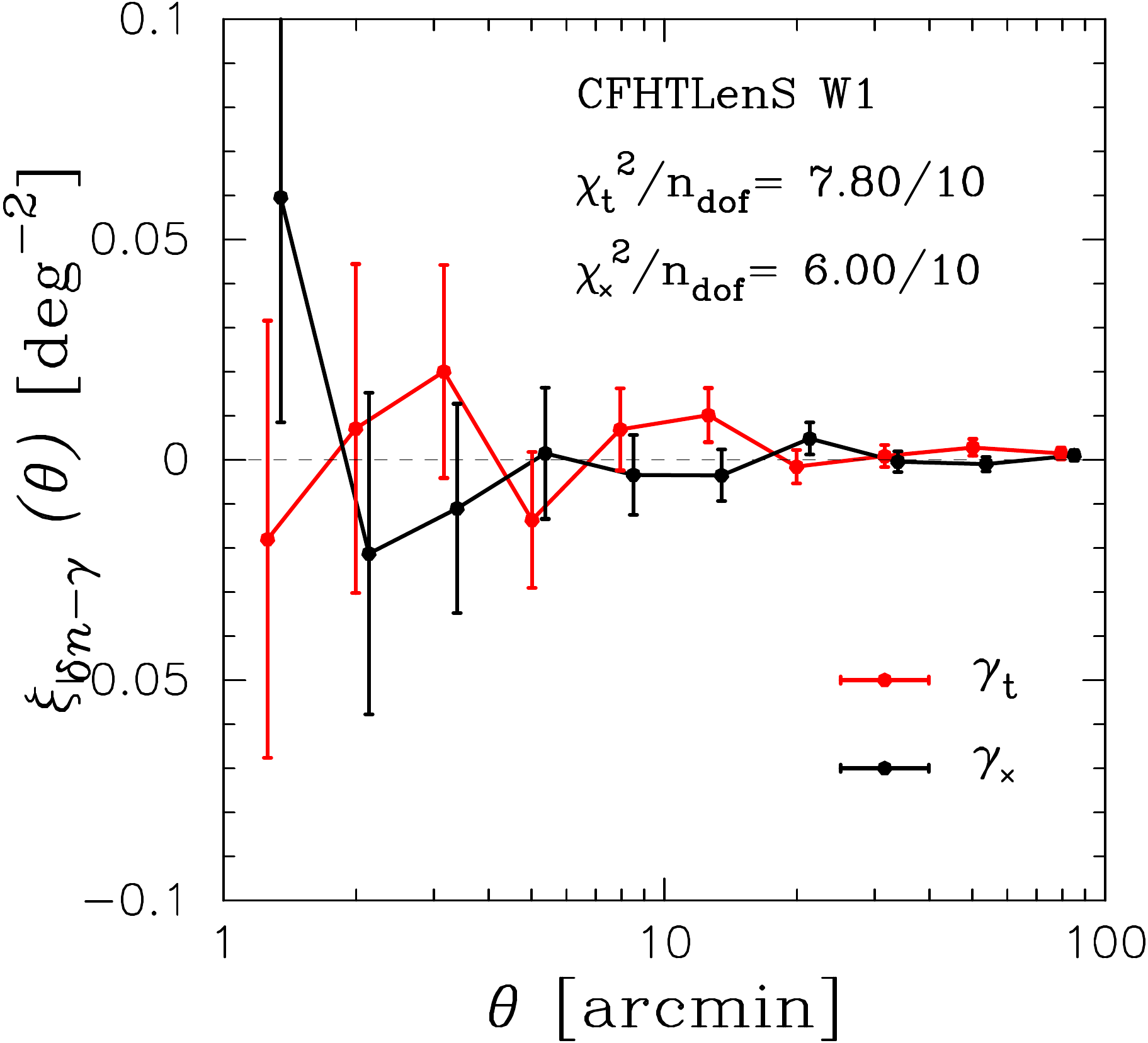}
\includegraphics[width=0.46\linewidth]{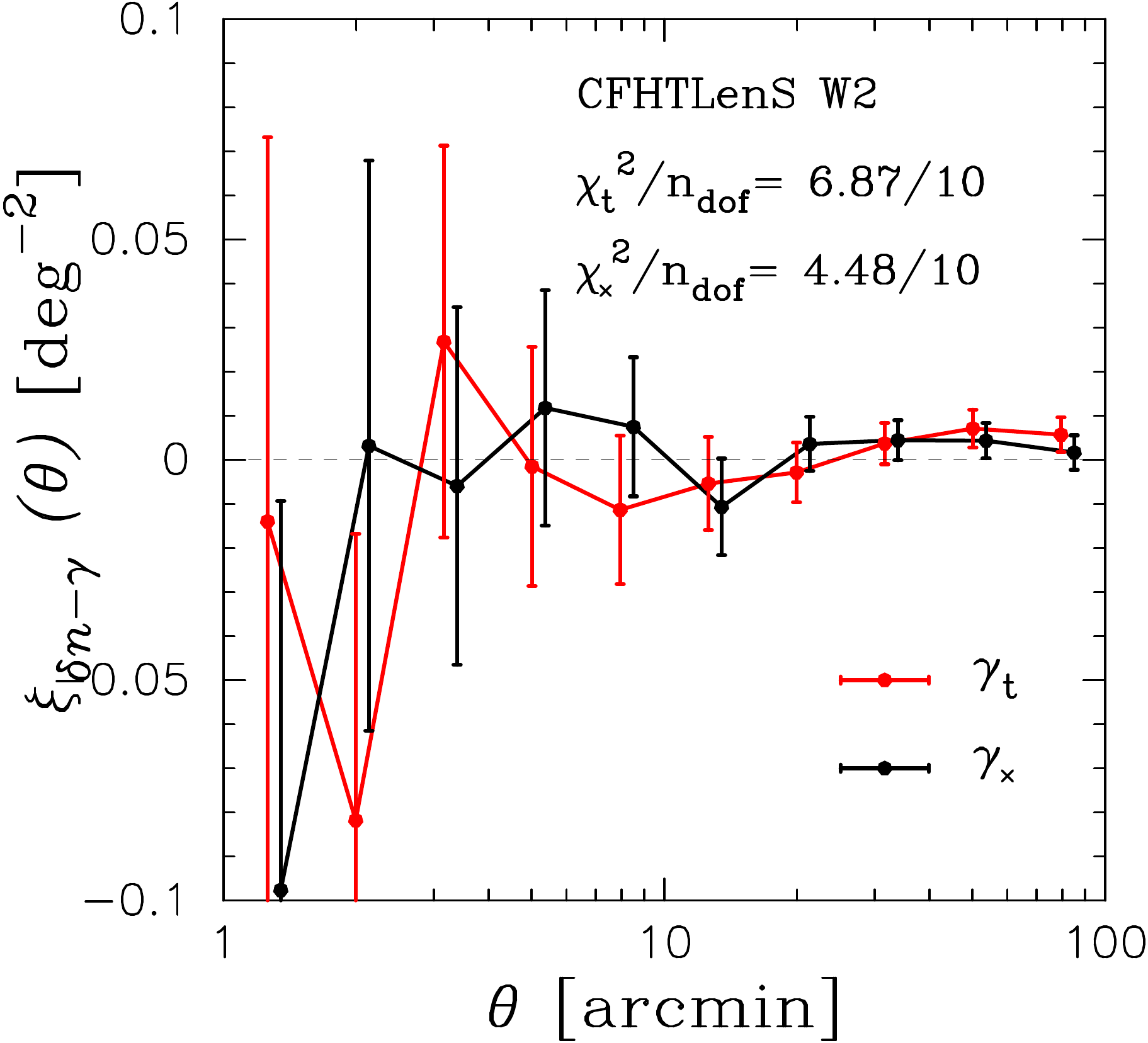}
\includegraphics[width=0.46\linewidth]{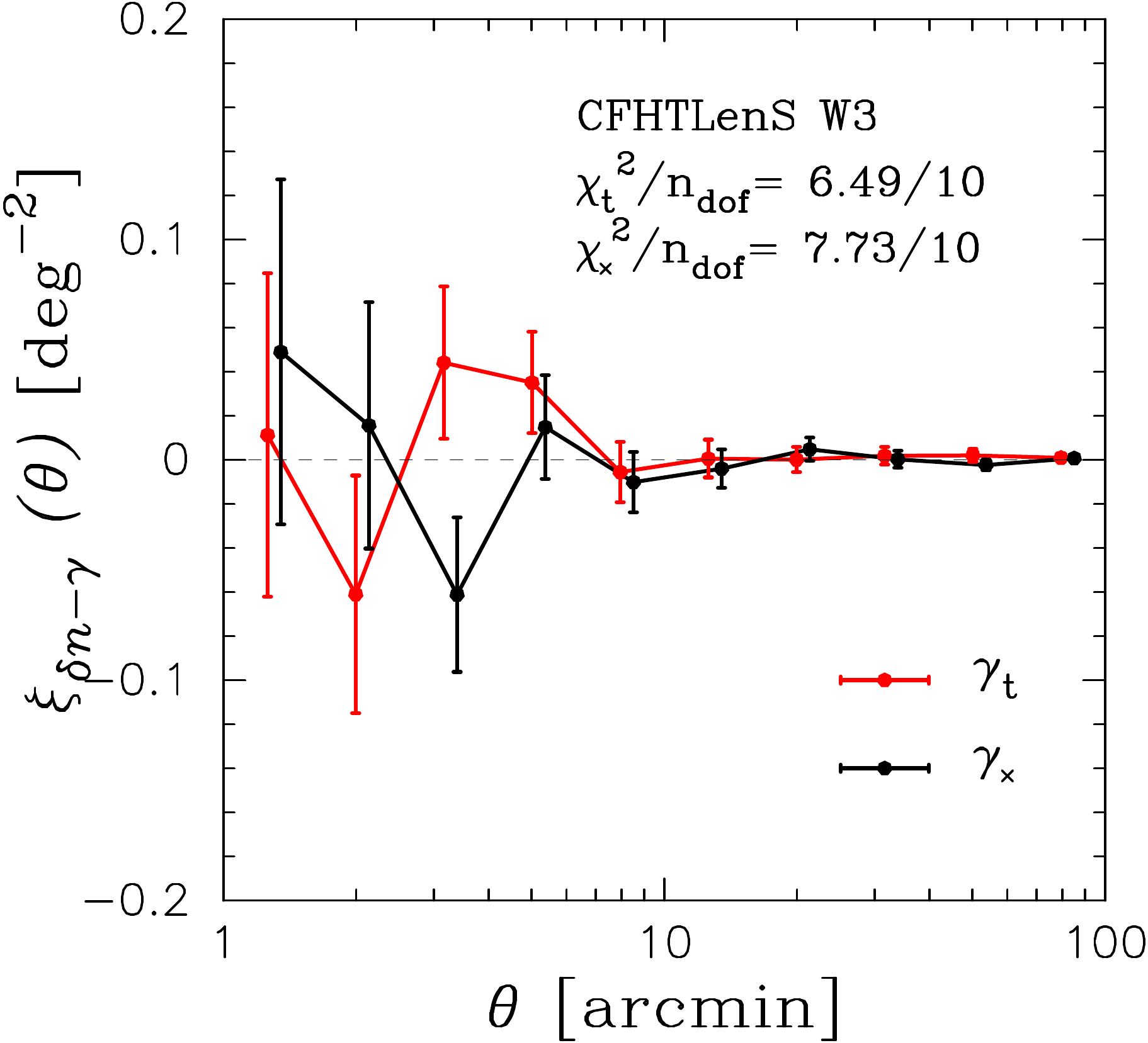}
\includegraphics[width=0.46\linewidth]{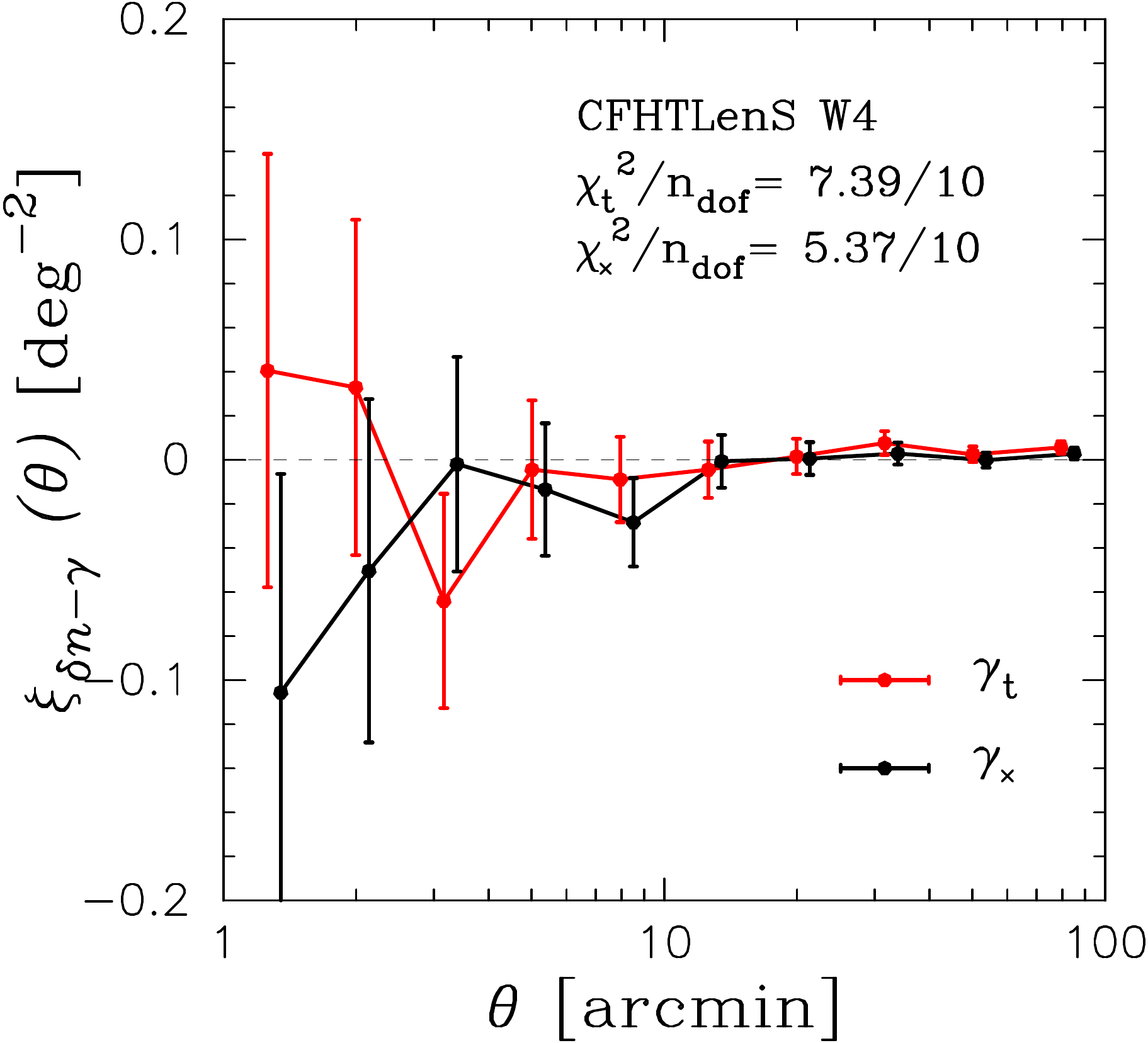}
\caption{\label{fig:cross_correlation_Shirasaki} The cross-correlation signal of cosmic shear and the DGRB. The 4 panels corresponds to the 4 sky-patches W1-W4 covered by CFHTLenS. Red points show the results obtained using tangential shear (indicated as $\gamma_{\rm t}$), while black points are for a component of the shear ($\gamma_\times$ in the legends) rotated by $45^\circ$ with respect to $\gamma_{\rm t}$. The error bars indicate the standard deviation, estimated from 500 randomized shear catalogs. The $\chi^2$ quantifies the significance of the signal with respect to the statistical error. Taken from Ref.~\cite{Shirasaki:2014noa}.}
\end{center}
\end{figure}

Such an optimal prospect for a DM detection is possible thanks to the 
``tomographic spectral approach'' employed in Ref.~\cite{Camera:2014rja}, 
which combines spectral information with the study of the dependence of 
$C_\ell^{X,\kappa}$ on redshift (i.e. tomography). Such a technique provides an 
excellent sensitivity to DM-induced emission, even if the intensity or 
auto-correlation APS of such a component are only subdominant.

The first (and, to date, only) measurement of the cross-correlation between
DGRB and cosmic shear is performed in Ref.~\cite{Shirasaki:2014noa}. The
authors make use of the data of CFHTLens from Ref.~\cite{Heymans:2012gg}. The 
survey detected more than 5 million galaxies in 4 patches, covering an area of 
154 square degrees. The corresponding shear signal is correlated in 
Ref.~\cite{Shirasaki:2014noa} with 65 months of \fermi LAT data from the same 
region in the sky. Only photons between 1 and 500 GeV are considered in the 
analysis. A model for the diffuse Galactic foreground (determined in 
Ref.~\cite{Shirasaki:2014noa} from the gamma-ray data) is subtracted from the 
total gamma-ray emission and 2FGL sources are masked, before computing the 
cross-correlation.

The estimator considered in Ref.~\cite{Shirasaki:2014noa} for the 2-point
correlation function is
\begin{equation}
\xi(\vartheta) = \frac{\sum_{ij} n^\gamma_i(\phi_i) \, w_j 
\epsilon_j(\phi_i + \mathbf{\vartheta}_j)}{(1+K(\vartheta)) \sum_{i,j} w_j},
\label{eqn:cross_correlation_estimator}
\end{equation}
where $n^\gamma_i$ is the number of gamma rays in the pixel centered on 
direction $\phi_i$, and $\epsilon_j$ is the shear-induced tangential 
ellipticity in pixel $\phi_j=\phi_i+\mathbf{\vartheta}_j$. The factors $w_j$ 
and $K(\vartheta)$ depend on the precision in the estimation of $\epsilon_j$ 
(see Refs.~\cite{Miller:2012am,Shirasaki:2014noa}) and the sum runs over all 
the pairs of pixels available. Fig.~\ref{fig:cross_correlation_Shirasaki} 
shows the 2-point cross-correlation function (binned in 10 logarithmic bins 
with $\Delta \log_{10} \vartheta=0.2$) for the 4 CFHTLenS sky patches. The 
measured 2-point correlation function using 
Eq.~(\ref{eqn:cross_correlation_estimator}) are showed in red, while the black 
points are obtained from another shear component rotated by $45^\circ$ with 
respect to the tangential one. In the case of a perfect measurement of the 
shape of the galaxies and of no intrinsic alignment, there should be no 
cross-correlation with this rotated data set. Thus, the black points in 
Fig.~\ref{fig:cross_correlation_Shirasaki} represent a control sample with no 
cross-correlation. Note that black and red points in 
Fig.~\ref{fig:cross_correlation_Shirasaki} are compatible with each other 
within errors, proving that no significant cross-correlation is present 
between CFHTLenS and \fermi LAT. This result can be translated into an upper 
limit on the DM annihilation cross section as a function of its mass. 
Ref.~\cite{Shirasaki:2014noa} excludes cross sections as low as the thermal 
value of $3 \times 10^{-26} \mbox{cm}^3 \mbox{s}^{-1}$ for DM masses smaller 
than $\sim$ 10 GeV (in the case of annihilation into $\tau^+\tau^-$ and of 
the optimistic subhalo boost model of Ref.~\cite{Gao:2011rf}).

\subsection{The cross-correlation with other tracers}
\label{sec:correlation_others}
In addition to galaxy catalogs and cosmic shear, it is also possible to 
consider other observables that trace the LSS of the Universe. Below, we 
highlight some of these studies.

Ref.~\cite{Xia:2011ax} cross-correlates the DGRB inferred from the first 21 
months of \fermi LAT data with the CMB measured by WMAP7 \cite{Komatsu:2010fb}. 
Such a measurement has the potential to probe the properties of dark energy 
through the detection of the so-called integrated Sachs-Wolfe effect 
\cite{Sachs:1967er}. This arises when the LSS gravitational potential changes 
with time during a cosmic era dominated by dark energy, as, e.g., in the local 
Universe. Additional anisotropies are induced in the CMB, which are expected 
to correlate with the LSS and, thus, potentially with the DGRB. In 
Ref.~\cite{Xia:2011ax}, the 2-point correlation function is computed, 
similarly to what was done for galaxy catalogs in 
Sec.~\ref{sec:correlation_galaxies}. Their results are compatible with a null 
detection, due to the large statistical errors.\footnote{The lack of a 
significant detection is compatible with the expectation if the DGRB is 
composed by unresolved sources (parametrized as in Ref.~\cite{Xia:2011ax}).}
Overall, Ref.~\cite{Xia:2011ax} shows that the goal of detecting the 
integrated Sachs-Wolfe effect by cross-correlating the DGRB with the CMB is 
not unrealistic, but beyond the reach of the limited gamma-ray sample 
considered in Ref.~\cite{Xia:2011ax}.

More recently, Ref.~\cite{Fornengo:2014cya} computed, for the first time,
the cross-correlation of the DGRB with the so-called ``lensing potential'' of
the CMB: the gravitational lensing induced by LSS imprints some distortions 
on the anisotropy pattern of the CMB, in such a way that the radiation 
detected by the Planck satellite \cite{Ade:2013hta} today is not exactly the 
one emitted at recombination. A statistical analysis of the non-Gaussianity of 
the CMB allows to reconstruct the lensing potential responsible for such
perturbations \cite{Blanchard:1987,Lewis:2006fu,Okamoto:2003zw}. The first
all-sky map of the CMB lensing potential has been recently reported by the 
Planck collaboration \cite{Ade:2013tyw}. The signal is mainly contributed by
structures at $z \sim 2$ and it exhibits an auto-correlation APS that peaks
at $\ell \sim 20-30$.

\begin{figure}[h]
\begin{center}
\includegraphics[width=0.7\linewidth]{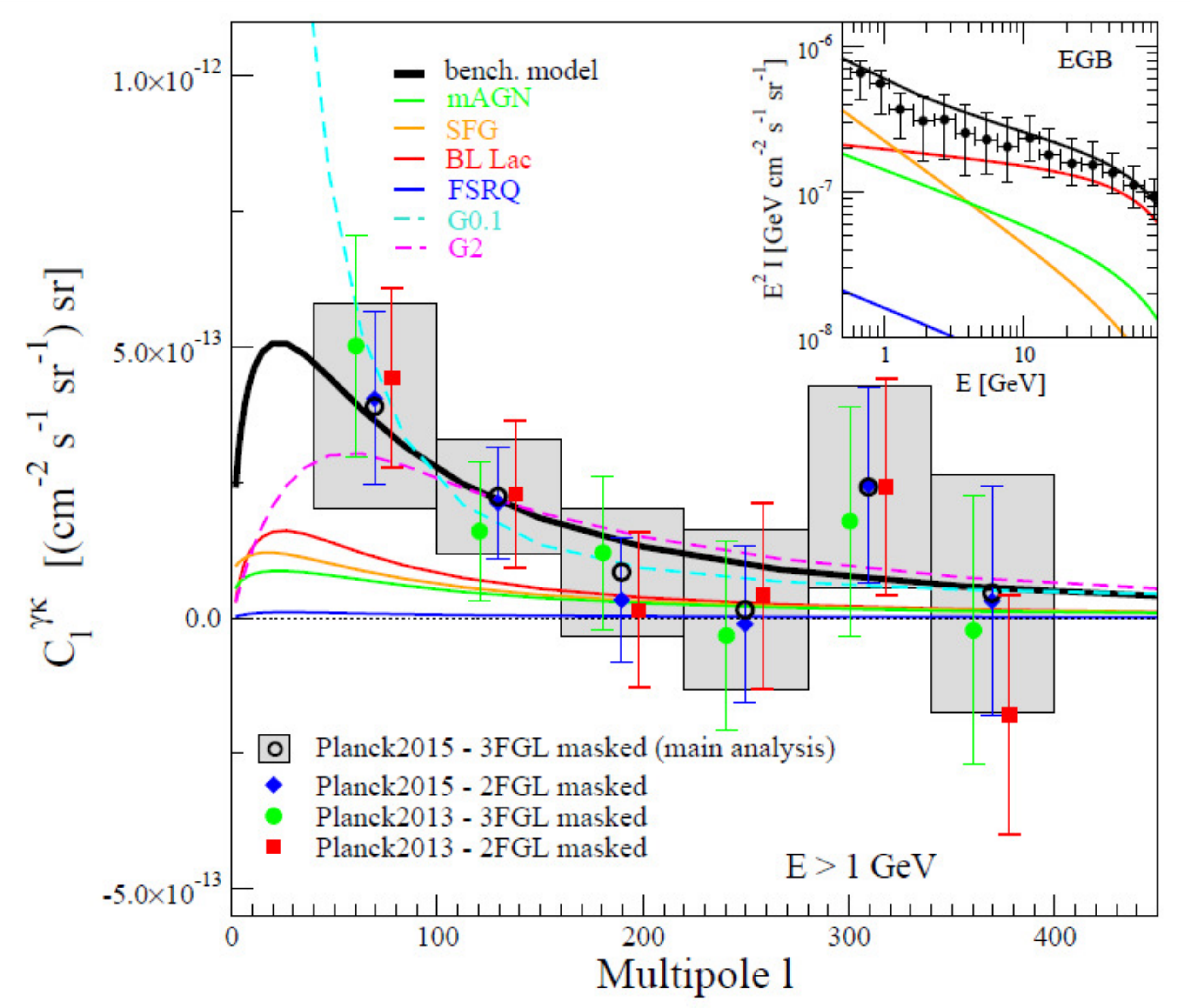}
\caption{\label{fig:cross_correlation_CMB_lensing} Cross-correlation APS $C_\ell^{\gamma\kappa}$ between the DGRB and the CMB lensing potential, as a function of the multipole $\ell$, for gamma-ray energies $E>1$ GeV. The measurement is averaged (linearly in terms of $\ell C_\ell^{\gamma\kappa}$) in multipole bins of $\Delta\ell=60$, starting at $\ell=40$. Points report the minimum-variance combination of the measurement in the individual energy bins (assuming a spectrum $\propto E^{-2.4})$, as described in Ref.~\cite{Fornengo:2014cya}. Four different analyses are shown. They arise from the combination of two lensing maps (from the 2013 and 2015 release of Planck data) and two gamma-ray point-source masks (masking sources in 2FGL or 3FGL). The benchmark theoretical model, shown in black, is the sum of the contributions from BL Lacs (red), FSRQs (blue), MAGNs (magenta) and SFGs (orange), multiplied by a normalization factor $A^{\kappa\gamma}=1.35$. Two generic models (labeled ``G0.1'' and ``G2'') with Gaussian window functions are also shown. The peak of the Gaussian is at $z_0=0.1$ ($z_0=2$), with a dispersion $\sigma_z=0.1$ ($\sigma_z=0.5$) for G0.1 (G2). In the inset, the intensity energy spectrum is shown for the \fermi LAT measurement (black data point, labeled ``EGB'') and for the model predictions. Taken from Ref.~\cite{Fornengo:2014cya}.}
\end{center}
\end{figure}

Ref.~\cite{Fornengo:2014cya} cross-correlates the sky-map of the CMB lensing
potential with 68 months of \fermi LAT data, after having removed the diffuse 
Galactic emission. Six energy bins are considered, between 700 MeV and 300 GeV. 
The region at $|b| < 25^\circ$ along the Galactic plane and a $1^\circ$-circle 
around each source in the 2FGL are masked, together with the {\it baseline} 
70\% Galactic mask from Ref.~\cite{Ade:2013tyw}. The signal region is defined 
between $\ell=40$ and 400. A detection with a significance of $3.2\sigma$ is
reported in the low-multipole region ($\ell < 160$). No signal is present 
above $\ell=160$ (see data points in 
Fig.~\ref{fig:cross_correlation_CMB_lensing}). The signal at low multipoles 
is compared to the predicted cross-correlation between the lensing potential
and the gamma-ray emission from 4 classes of unresolved sources, namely FSRQs, 
BL Lacs, SFGs and MAGNs. The LFs of these populations are fixed to the best-fit 
models from Refs.~\cite{Ajello:2011zi,Ajello:2013lka,DiMauro:2013xta,
Gruppioni:2013jna,Ackermann:2012vca} (see Secs.~\ref{sec:blazars}, 
\ref{sec:MAGNs} and \ref{sec:SFGs}). The combined emission of these 4 source 
classes (solid black line in Fig.~\ref{fig:cross_correlation_CMB_lensing})
reproduces fairly well the experimental data. More precisely, BL Lacs, SFGs 
and MAGNs (red, orange and green lines in 
Fig.~\ref{fig:cross_correlation_CMB_lensing}, respectively) contribute more or 
less in equal parts to the cross-correlation signal, while FSRQs (blue line) 
are subdominant. This model also provides a good fit to the DGRB energy 
spectrum and auto-correlation APS.

\section{The science of the Diffuse Gamma-Ray Background}
\label{sec:conclusions}
In this review we have summarized the current knowledge on the Diffuse 
Gamma-Ray Background (DGRB). The DGRB is what remains in the gamma-ray sky
after the subtraction of the diffuse Galactic foreground and of the resolved
sources. It is interpreted as the cumulative emission of the objects that are
not bright enough to be detected individually.

Since its first detection in 1972 by the OSO-3 satellite \cite{Kraushaar:1972}, 
this emission has been deeply investigated in an attempt to understand its 
composition. The \fermi LAT satellite, in operation since 2008, has greatly 
improved our understanding of the DGRB. This emission is measured as an 
isotropic template in a multi-component fit to the \fermi LAT data. The fit
also includes a model for the resolved sources and for the diffuse Galactic 
foreground. The most recent measurement of the DGRB energy spectrum is 
reported in Ref.~\cite{Ackermann:2014usa} and it covers almost 4 orders of 
magnitude in energy, from 100 MeV to 820 GeV. Our imperfect knowledge of the 
Galactic foregrounds represents the main source of uncertainty in the analysis 
and it induces a systematic error on the DGRB intensity of $\sim 15-30\%$, 
depending on the energy range considered.

Before the \fermi LAT, the energy spectrum was the only source of information 
available on the DGRB. However, the scenario drastically changed in 2012 when, 
for the first time, the \fermi LAT measured also the angular power spectrum 
(APS) of anisotropies in the DGRB \cite{Ackermann:2012uf}. The emission was
found to exhibit a Poissonian APS in the multipole range between $\ell=155$ and 
504, with a significance as large as 7.2$\sigma$ between 1.99 and 5.0 GeV. 
This signal is independent of energy between 1 and 50 GeV.

Recently, cross-correlations of the DGRB with different data sets have also 
revealed to be a powerful tool to unveil the composition of the DGRB. 
Ref.~\cite{Xia:2015wka} reported a significant cross-correlation of the DGRB 
with 4 out of the 5 galaxy catalogs considered, namely the optically-selected 
quasares of the Sloan Digital Sky Survey (SDSS) Data Release 6 in 
Ref.~\cite{Richards:2008eq}, the IR galaxies of the 2 Micron All-Sky Survey 
(2MASS) Extended Source Catalog from Ref.~\cite{Jarrett:2000qt}, the main 
galaxy sample of SDSS Data Release 8 from Ref.~\cite{Aihara:2011sj} and the 
radio sources from the NRAO VLA Sky Survey (NVSS) \cite{Condon:1998iy}. These 
cross-correlation signals, obtained after the analysis of 60 months of \fermi 
LAT data, are localized at small angles (below few degrees) with significances 
that range between $3\sigma$ (in the case of SDSS Data Release 8 main galaxies) 
and more than $10\sigma$ (for the NVSS catalog), for gamma-ray energies above 
1 GeV. The cross-correlation with NVSS, however, is most likely contaminated 
by a 1-halo term not related to the Large Scale Structure (LSS) of the 
Universe.

In Ref.~\cite{Shirasaki:2014noa}, the authors measured, for the first time, 
the cross-correlation of the DGRB with the cosmic shear induced by the 
gravitational lensing detected in the Canada-France-Hawaii Telescope Lensing
Survey (CFHTLenS) \cite{Heymans:2012gg}. Their results are compatible with a 
null cross-correlation signal in the angular range between 1 and 100 arcmin. 
Ref.~\cite{Shirasaki:2014noa} proves, though, that measuring the 
cross-correlation with the cosmic shear signal is not an unrealistic goal. 
Indeed, more promising results are expected from the cross-correlation between
the DGRB and the data expected from the Dark Energy Survey (DES) or from 
Euclid \cite{Camera:2014rja}. Finally, Ref.~\cite{Fornengo:2014cya} reported a 
$3.2\sigma$ detection of the cross-correlation between the DGRB and the 
lensing potential of the CMB measured by the Planck Collaboration 
\cite{Ade:2013tyw}. The signal is localized in the multipole region between 
$\ell=40$ and 160, for gamma-ray energies between 700 MeV and 300 GeV. 

The increased amount of observational data on the DGRB has allowed significant 
progress in the modeling of its composition. The DGRB is interpreted as the 
cumulative emission of unresolved sources, e.g. blazars, misaligned Active 
Galactic Nuclei (MAGNs), star-forming galaxies (SFGs) and MilliSecond Pulsars 
(MSPs). It, therefore, represents a reservoir of invaluable information on 
these astrophysical sources as it may be the only way to study the emission of 
objects that are too faint to be detected individually. In particular, the 
DGRB can potentially determine the faint end of the luminosity function of 
the aforementioned populations, i.e. a goal that would be quite difficult, if 
not impossible, to achieve otherwise.

Before the \fermi LAT, the predictions for the contribution of unresolved 
blazars to the DGRB were affected by large uncertainties \cite{Padovani:1993,
Stecker:1993ni,Stecker:1996ma,Muecke:1998cs,Narumoto:2006qg,Pavlidou:2007dv,
Bhattacharya:2009yv,Inoue:2008pk}. Nowadays, the wealth of new information on 
the DGRB, combined with the population studies of resolved blazars performed 
by the \fermi LAT Collaboration, have established that this source class 
cannot account for more than $(20 \pm 4)\%$ of the DGRB in the energy range 
between 0.1 and 100 GeV \cite{Ajello:2015mfa}. The subclass of blazars 
responsible for the bulk of the blazar contribution varies depending on the 
energies considered. Indeed, Refs.~\cite{Ajello:2013lka,DiMauro:2013zfa,
Ajello:2015mfa} showed that unresolved high-synchrotron-peaked BL Lacs can 
explain the whole DGRB at energies above $\sim 100$ GeV. At lower energies, 
astrophysical populations other than blazars are required. Yet, the modeling 
of these other gamma-ray emitters, such as namely MAGNs, SFGs and MSPs, is not 
as robust as that of blazars. This is caused by difficulty in performing 
reliable population studies with the limited sample of resolved sources 
currently available in the gamma-ray range. Generally, it is useful to assume 
a correlation between luminosities at different wavelengths (i.e. gamma-ray 
with radio frequencies \cite{Inoue:2011bm,DiMauro:2013xta} in the case of 
MAGNs, or gamma rays and infra-red light \cite{Ackermann:2012vca} for SFGs). 

The overall picture indicates that the 4 classes of astrophysical sources
mentioned above are enough to explain the totality of the DGRB energy spectrum 
measured in Ref.~\cite{Ackermann:2014usa} (see 
Fig.~\ref{fig:summary_astrophysics}). As for the APS of DGRB anisotropies,
Refs.~\cite{Cuoco:2012yf,Harding:2012gk,DiMauro:2014wha} prove that unresolved 
blazars alone (more specifically, high-synchrotron-peak BL Lacs) can account 
for the whole APS reported in Ref.~\cite{Ackermann:2012uf}. These two 
important results can be reconciled by the fact that the blazar component is 
produced by a relatively small number of bright but unresolved objects and, 
thus, it gives rise to significant anisotropies. On the contrary, the other
source classes produce fairly isotropic cumulative emission, as their members
are more numerous and fainter.

The picture gets more complicated, though, when considering also the measured 
cross-correlation with LSS tracers: the sum of unresolved blazars, MAGNs and 
SFGs provides a good fit to the cross-correlation APS detected between the DGRB 
and the CMB lensing potential \cite{Fornengo:2014cya}. It is also compatible 
with the lack of significant cross-correlation with the CFHTLenS cosmic shear 
\cite{Shirasaki:2014noa}. However, Ref.~\cite{Cuoco:2015rfa} finds that the 
model that best fits the two-point correlation functions measured in 
Ref.~\cite{Xia:2015wka} with 5 galaxy catalogs can only account for $\sim 30\%$ 
of the DGRB intensity reported in Ref.~\cite{Ackermann:2014usa}. It is still 
unclear if such a limitation of the astrophysical interpretation of the DGRB 
can be alleviated by a more sophisticated modeling of its components. Also, it 
will be interesting to verify whether a similar scenario will be confirmed by 
new surveys or by more complete data releases of the catalogs currently 
employed. Alternatively, one will be forced to supplement the model with 
another component, which does not correlate with the LSS, such as gamma-ray 
emission associated with the DM halo of the MW or with its DM substructures.

Quantifying the contribution of known astrophysical populations automatically 
constrains the intensity of other potential contributors to the DGRB. We 
briefly presented, among others, the case of clusters of galaxies, Type Ia 
supernovae and of Ultra-High-Energy Cosmic Rays interacting with background 
radiation. However, the most studied scenario is that of a potential gamma-ray 
emission from Dark Matter (DM) annihilation or decay. Since no DM signal has 
been undoubtedly detected in the gamma-ray sky so far, it is expected that 
most of this hypothetical gamma-ray emission will contribute to the DGRB. 
Searching for DM in the DGRB has the advantage that the DM-induced component 
is sourced by the emission coming from {\it all} DM halos and subhalos around 
us. It will, thus, depend on the ensemble-averaged properties of the DM halo 
population, which can be inferred from $N$-body cosmological simulations 
\cite{Springel:2005mi,BoylanKolchin:2009nc,Prada:2011jf,Angulo:2012ep,
Alimi:2012be} or predicted by the theory of structure formation 
\cite{Press:1973iz,Cooray:2002dia}. This is intrinsically different from the
observation of a specific target, e.g. the Galactic Center or a dwarf 
Spheroidal satellite galaxy. Indeed, each of these targets could be very 
peculiar and deviate considerably from the ensemble average, potentially 
hamper the interpretation of any data. Another benefit of using the DGRB to 
search for DM is that a potential DM signal contributing to the DGRB would be 
sensitive to the process of assembly of DM halos and their subsequent 
evolution. This kind of information would be difficult (if not impossible) to 
extract by observing individual targets in the sky. In this regard, the DGRB 
may be the only cosmological non-gravitational probe of DM. In addition, it 
represents a fundamental source of complementary information in the study of 
any claimed DM signal.

The intensity of this DM-induced cosmological emission depends on the 
properties of the DM particle, e.g. its mass, annihilation cross section or 
decay lifetime. Also, it rests on the abundance and properties of DM 
structures. Our understanding of DM (sub)halos heavily relies on the results
of $N$-body cosmological simulations. Yet, as of today, even the simulations
with the highest resolution are far from resolving the whole DM halo hierarchy
down to the predicted $M_{\rm min}$. The properties of low-mass DM structures,
thus, need to be inferred by extrapolating the characteristics of their more 
massive counterparts, that are well resolved in the simulations. Heuristic 
power-law extrapolations, which predict very bright low-mass DM halos, have 
been commonly used in the literature \cite{Pieri:2007ir,Zavala:2009zr,
Pinzke:2011ek,Gao:2011rf,Ando:2013ff,Cholis:2013ena}. However, recent 
high-resolution simulations of the smallest DM halos \cite{Ishiyama:2014uoa,
Anderhalden:2013wd} suggest that those extrapolations are not well motivated, 
favoring, instead, models that yield a more moderate DM-induced gamma-ray
emission from low-mass structures \cite{Prada:2011jf,Sanchez-Conde:2013yxa,
Ludlow:2013vxa}. Overall, this results in a DM-induced cosmological signal
which is substantially weaker than the one obtained when assuming power-law
extrapolations. Also, and perhaps even more importantly, the improved 
knowledge provided by Refs.~\cite{Prada:2011jf,Sanchez-Conde:2013yxa,
Ludlow:2013vxa} on the structural properties of the smallest DM halos has
considerably reduced the theoretical uncertainty associated with the DM
contribution to the DGRB down to a factor of $\sim 20$ \cite{Ackermann:2015gga}.

Not unexpectedly, the measured DGRB energy spectrum can be used to constrain 
the intensity of the DM-induced emission and, thus, to derive upper (lower) 
limits on the annihilation cross section (decay lifetime). 
Figs.~\ref{fig:DM_annihilation_limits} and \ref{fig:summary_decay} summarize
some of these results. For annihilating DM, the so-called ``sensitivity-reach''
limits derived in Ref.~\cite{Ackermann:2015gga} from their fiducial Halo Model
exclude thermal annihilation cross sections of 
$3 \times 10^{-26} \mbox{cm}^2 \mbox{s}^{-1}$ for masses below $\sim 100$ GeV in
the case of annihilations into $b$ quarks. When compared to other indirect 
searches for DM, the upper limits inferred from the DGRB represent the 
strongest constraints on $\langle \sigma v \rangle$ currently available, for
DM masses below $\sim 1$ TeV. A more conservative statistical analysis would 
increase the upper limits by a factor of $\sim 10$.\footnote{Note also that 
both the fiducial sensitivity-reach limits and the more conservative ones in 
Ref.~\cite{Ackermann:2015gga} are affected by an uncertainty of a factor of 
$\sim 3$.} In the case of decaying DM, the lower limits on $\tau$ obtained in 
Ref.~\cite{Cirelli:2012ut} from the DGRB measurement in Ref.~\cite{Abdo:2010nz} 
exclude decay lifetimes smaller than $\sim 3 \times 10^{27}$ s for 
$m_\chi \lesssim 2$ TeV and decays into $\mu^+\mu^-$. At larger masses,
the strongest lower limit comes from Ref.~\cite{Ando:2015qda}, which excludes
lifetimes as large as $10^{28}$~s. When compared to other searches of decaying 
DM, these limits represent the most constraining information available on 
$\tau$, up to DM masses of 20 TeV.

A similar strategy can be used when comparing the APS measured in 
Ref.~\cite{Ackermann:2012uf} to the predicted anisotropies in the DM-induced
emission. Nevertheless, the latter turns out to be quite isotropic 
\cite{Zavala:2009zr,Fornasa:2012gu,Ando:2013ff} and, thus, the corresponding
upper limits, although competitive with other indirect DM searches, are less 
stringent than those inferred from the DGRB \cite{Ando:2013ff,
Gomez-Vargas:2014yla}. The cross-correlation with galaxy catalogs and with 
cosmic shear are also very promising strategies to constrain (or even to
detect) a potential DM signal in the DGRB \cite{Ando:2013xwa,Ando:2014aoa,
Camera:2012cj,Camera:2014rja}. This is possible mainly thanks to the fact 
that both cross-correlations are particularly sensitive to the way the matter
is distributed in the local Universe and, in particular, to the most massive 
DM halos. We remind that unresolved blazars, the main contributors to the 
auto-correlation APS, do not populate neither the local volume nor the largest 
DM halo masses. A model of the DGRB including both astrophysical sources and 
annihilating DM is used in Ref.~\cite{Cuoco:2015rfa} to describe measurement 
of the 2-point correlation function reported in Ref.~\cite{Xia:2015wka}. The 
analysis excludes DM candidates with annihilation cross sections larger than 
the thermal value for masses below 40 GeV, for annihilations into $b$ quarks 
and a moderate value of the subhalo boost. When assuming the same value of 
the subhalo boost, the obtained  upper limit is currently the strongest one 
among all those derived from DGRB data, including measurements of the DGRB 
intensity and of the auto-correlation APS.

The energy spectrum, auto-correlation and cross-correlation APS of the DGRB
are sensitive to different characteristics of the sources contributing to the 
emission. Considering {\it at the same time} all three observables provide a 
very powerful handle to reconstruct the composition of the DGRB. Needless to 
say, such an ambitious goal requires ample data sets, including information 
from wavelengths other than the gamma-ray energy range. Fortunately, a great 
wealth of new observational information is expected in the near future: the 
\fermi LAT will continue gathering data until, at least, 2016. The Cherenkov 
Telescope Array (CTA), expected to be in operation by 2020, will improve by a 
factor $\sim 10$ the sensitivity of current Cherenkov telescopes in the energy 
range between a few dozens of GeV and a few dozens of TeV. Its complementarity
with respect to the \fermi LAT will allow an improved precision in the
determination of the DGRB energy spectrum in the sub-TeV energy range, as well
as the extension of the measurement beyond the TeV. In addition, CTA will 
perform the first survey of a significant portion of the sky at these 
very-high energies \cite{Dubus:2012hm,Ripken:2012db}. Combined with the \fermi 
LAT data gathered since Ref.~\cite{Ackermann:2012uf}, it will be possible to 
extend the data on the auto-correlation APS to higher energies and to decrease 
the size of the bins in energy. On the other hand, the cross-correlation of 
the DGRB with LSS tracers will hugely benefit for the imminent release of the 
data gathered by the DES \cite{Abbott:2005bi} during its first year of 
operation. The near future will also see the advent of the next generation of 
galaxy catalogs, e.g. the extended Baryon Oscillation Spectroscopy Survey 
(eBOSS)\footnote{https://www.sdss3.org/future/eboss.php} and the Dark Energy
Spectroscopic Instrument (DESI, formerly BigBoSS) \cite{Schlegel:2011zz}. On a 
longer scale, Euclid will offer weak lensing measurements with unprecedented 
precision after 2020 \cite{Laureijs:2011gra}. 

The increased observational data available on the DGRB will also be 
accompanied by significant progress in the modeling of its contributors. 
Improving our understanding of the emission of blazars, SFGs and MAGNs will
alleviate the degeneracies currently affecting our interpretation and will
reduce the uncertainty associated with each component. A fully multi-wavelength 
approach is required to achieve such a goal: in the X-ray band the Nuclear
Spectroscopic Telescope Array (NuSTAR) \cite{Harrison:2013md} has been 
recently launched and ASTRO-H \cite{Takahashi:2012jn,Kitayama:2014fda} will 
follow within this year (2015). Infra-red data from Herschel and the 
Wide-field Infrared Survey Explorer 
(WISE)\footnote{http://wise.ssl.berkeley.edu/} are already public and the
James Webb Space Telescope (JWST)\footnote{http://www.jwst.nasa.gov/} is 
expected to be launched in 2018. Regular observation in radio has started with 
Low-Frequency Array (LOFAR) \footnote{http://www.lofar.org/}, a pathfinder for 
the Square Kilometer Array (SKA) \footnote{http://www.skatelescope.org/} 
planned for 2020.

We end the review by noting that the IceCube Collaboration has recently 
reported the detection of the first extraterrestrial neutrinos 
\cite{Aartsen:2013bka,Aartsen:2013jdh,Aartsen:2014gkd}. The 37 observed 
neutrino events represent an excess over the atmospheric background extending 
to the PeV scale, with a significance of more than $5\sigma$ 
\cite{Aartsen:2014gkd}. Even if the origin of these neutrinos is not clearly 
established yet (see Refs.~\cite{Waxman:2013zda,Anchordoqui:2013dnh} and 
references therein), a possibility is that they originate from a diffuse 
neutrino flux similar to the DGRB.\footnote{To stress the similarity with the
DGRB, we propose to call this neutrino emission, the Diffuse Neutrino 
Background (DNB).} If this interpretation was confirmed, many of the 
techniques discussed in this review could also be adopted to investigate this 
diffuse neutrino flux. Indeed, many of the sources contributing to the DGRB 
are expected to emit neutrinos as well \cite{Loeb:2006tw,Stecker:2013fxa,
Murase:2014foa,Padovani:2014bha}. This implies that any constraint on their 
neutrino emission would indirectly constrain also their contribution to the 
DGRB (and viceversa). The development of a fully multi-messenger approach is 
certainly a very tantalizing possibility that has been already considered, 
e.g., in Refs.~\cite{Murase:2013rfa,Ahlers:2013xia,Liu:2013wia,
Kashiyama:2014rza,Anchordoqui:2014yva,Tamborra:2014xia}.

In conclusion, the DGRB is a fundamental component of the gamma-ray sky,
whose exact composition still remains unveiled. The recent rapid growth of 
available data on the DGRB (mostly thanks to the outstanding performance of
the \fermi LAT) has triggered an increased attention from the scientific 
community. Astrophysicists and astroparticle physicists aim at reconstructing 
the composition of the DGRB to infer novel information on the sources 
contributing to the emission, especially in their low-luminosity regime. New 
data sets are already available (or will be soon) which can provide a 
significant progress to the common goal of dissecting the true nature of the 
DGRB. By summarizing where we stand on our current understanding of this 
emission, with this review we hope to have offered a useful reference to 
those who will analyze and interpret the data to come, as well as to help 
finding new avenues and opportunities for further research.

\section{Acknowledgments}
MF gratefully acknowledges support of the Leverhulme Trust. MASC is 
Wenner-Gren Fellow and acknowledges the support of the Wenner-Gren foundation
to develop his research. We also acknowledge the project MultiDark 
CSD2009-00064. We thank M. Ajello, S. Ando, K. Bechtol, M. Di Mauro, A. M. 
Green, G. Zaharijas and J. Zavala for useful comments on the manuscript.

\bibliographystyle{elsarticle-num}
\bibliography{Review.bib}

\end{document}